\documentclass[twocolumn,showpacs,preprintnumbers,amsmath,amssymb,floatfix]{revtex4}

\usepackage{graphicx}
\usepackage{dcolumn}
\usepackage{bm}
\usepackage{hyperref}

\begin{document}


\title{Depinning and dynamics of vortices confined in \\
mesoscopic flow channels.}

\author{R.~Besseling}
\altaffiliation[Presently at ]{ the School of Physics, University
of Edinburgh, Kings Buildings, Mayfield Road, Edinburgh EH9 3JZ,
United Kingdom}

\author{P.H.~Kes}%
\affiliation{Kamerlingh Onnes Laboratorium, Leiden University,
P.O. Box 9504, 2300 RA Leiden, the Netherlands.}

\author{T.~Dr\"{o}se}
\affiliation{I. Institut f\"ur Theoretische Physik, Universit\"at
Hamburg, Jungiusstrasse 9, D-20355 Hamburg, Germany.}
\altaffiliation{Presently at Siemens AG, St.-Martin-Str. 76,
D-81617 M\"{u}nchen, Germany}

\author{V.M. Vinokur}
\affiliation{Materials Science Division, Argonne National
Laboratory, Argonne, Illinois 60439}

\date{\today}

\begin{abstract}
We study numerically and analytically the behavior of vortex
matter in artificial flow channels confined by pinned vortices in
the channel edges (CE's). The critical current density $J_s$ for
channel flow is governed by the interaction with the static
vortices in the CE's. Motivated by early experiments which showed
oscillations of $J_s$ on changing (in)commensurability between the
channel width $w$ and the natural vortex row spacing $b_0$, we
study structural changes associated with (in)commensurability and
their effect on $J_s$ and the dynamics. The behavior depends
crucially on the presence of disorder in the arrays in the CE's.
For ordered CE's, maxima in $J_s$ occur at commensurability
$w=nb_0$ ($n$ integer), while for $w\neq nb_0$ defects {\it along}
the CE's cause a vanishing $J_s$. For weak disorder, the sharp
peaks in $J_s$ are reduced in height and broadened via nucleation
and pinning of defects. The corresponding structures in the
channels (for zero or weak disorder) are quasi-1D $n$ row
configurations, which can be adequately described by a
(disordered)sine-Gordon model. For larger disorder, matching
between the longitudinal vortex spacings in and outside the
channel becomes irrelevant and, for $w\simeq nb_0$, the shear
current $J_s$ levels at $\sim 30 \%$ of the value $J_s^0$ for the
ideal commensurate lattice. Around 'half filling' ($w/b_0 \simeq
n\pm 1/2$) the disorder leads to new phenomena, namely
stabilization and pinning of {\it misaligned} dislocations and
coexistence of $n$ and $n \pm 1$ rows in the channel. At
sufficient disorder, these quasi-2D structures cause a {\it
maximum} in $J_s$ around mismatch, while $J_s$ smoothly decreases
towards matching due to annealing of the misaligned regions. Near
threshold, motion inside the channel is always plastic. We study
the evolution of static and dynamic structures on changing
$w/b_0$, the relation between the $J_s$ modulations and transverse
fluctuations in the channels and find dynamic ordering of the
arrays at a velocity with a matching dependence similar to $J_s$.
We finally compare our numerical findings at strong disorder with
recent mode-locking experiments, and find good qualitative
agreement.

\end{abstract}

\maketitle

\tableofcontents

\section{Introduction}
\label{secintro}

The depinning and dynamics of the vortex lattice (VL) in type II
superconductors is exemplary for the behavior of driven, periodic
media in presence of a pinning potential \cite{Blatterbible}.
Other examples range from sliding surfaces exhibiting static and
dynamic friction and absorbed monolayers \cite{Perssonbook} to
charge density waves (CDW's) \cite{Gruner}, Wigner crystals
\cite{Wigner} and (magnetic) bubble arrays \cite{Seshadri}. Vortex
matter offers the advantage that the periodicity $a_0$ of the
hexagonal lattice can be tuned by changing the magnetic induction
$B$. In addition, the effect of various types of pinning
potentials can be studied. This pinning potential, arising from
inhomogeneities in the host material, can be completely random, as
in most natural materials, or can be arranged in periodic arrays
using nano-fabrication techniques
\cite{BaertperpinPRL95,MartinperpinPRL99}. In a variety of cases
correlated inhomogeneities occur naturally in a material, such as
twin boundaries and the layered structure of the high $T_c$
superconductors \cite{twinHTC}.

Depinning of the VL in a {\it random potential} generally involves
regions of plastic deformations
\cite{JensenPRLPRB88,ShiPRL91_def,MarchevskyPRB99_depinning,PardoPRL97_topdef,pardonatureBragg},
i.e. coexistence of (temporarily) pinned domains with moving
domains. For very weak pinning the typical domain size can exceed
the correlation length $R_c$ of the VL (see
\cite{MarchevskyPRB99_depinning}) and the weak collective pinning
theory \cite{Larkin} can be successfully used to estimate the
critical current density $J_c$ \cite{Kes2DCP,WordenweberPRB86}.
However, as either the ratio of the VL shear modulus $c_{66}$ and
the elementary pinning strength or the number of pins per
correlated volume decreases, plastically deformed regions start to
have a noticeable effect on $J_c$. Recent imaging experiments
\cite{TroyanovskiPRL02_PE} have shown directly that the rise in
$J_c$ in weak pinning materials near the upper critical field
$B_{c2}$, known as the peak effect
\cite{Kes2DCP,WordenweberPRB86}, originates from such, rather
sudden, enhancement of the defect density. This strong reduction
of the VL correlation length is also accompanied by a qualitative
change in the nature of depinning: for strong pinning, depinning
proceeds through a dense network of quasi-static flow channels
(filaments) such that the typical width of both static and moving
'domains' has approached the lattice spacing
\cite{JensenPRLPRB88,ShiPRL91_def,MehtaReichOlson,RyuPRL96_2ddynord}.
Depinning transitions via a sequence of static, channel like
structures have also been observed experimentally via transport
experiments \cite{HellerDanckwerts}.

In superconductors with {\it periodic} pinning arrays (PPA's),
matching effects between the lattice and the PPA become important.
As shown first by Daldini and Martinoli
\cite{DaldiniPRL74,MartinoliPRB78}, when the vortex spacing
coincides with the periodicity of the potential, pronounced maxima
can occur in $J_c$, while at mismatch defects (discommensurations)
appear which gives rise to a reduced $J_c$. The last decade, many
more studies of VL's in PPA's have appeared, both experimentally
and numerically. Pronounced commensurability effects were found in
films with 2-D periodic pinning
\cite{BaertperpinPRL95,MartinperpinPRL99,matching2dperpin_exp} for
flux densities equal to (integers of) the density of dots. In
these systems, vortex chains at interstitial positions of the
periodic arrays (e.g. at the second matching field of a square
pinning array) can exhibit quasi one-dimensional motion under the
influence of the interaction with neighboring, pinned vortices
\cite{RosseelLook}, as has also been observed in numerical
simulations \cite{Reichhardtperpin}. In addition these simulations
have revealed that, depending on the vortex interactions and the
symmetry or strength of the PPA, a rich variety of other states
and dynamic transitions can occur, often leading to peculiar
transport characteristics.

\begin{figure}
\scalebox{0.40}{\includegraphics{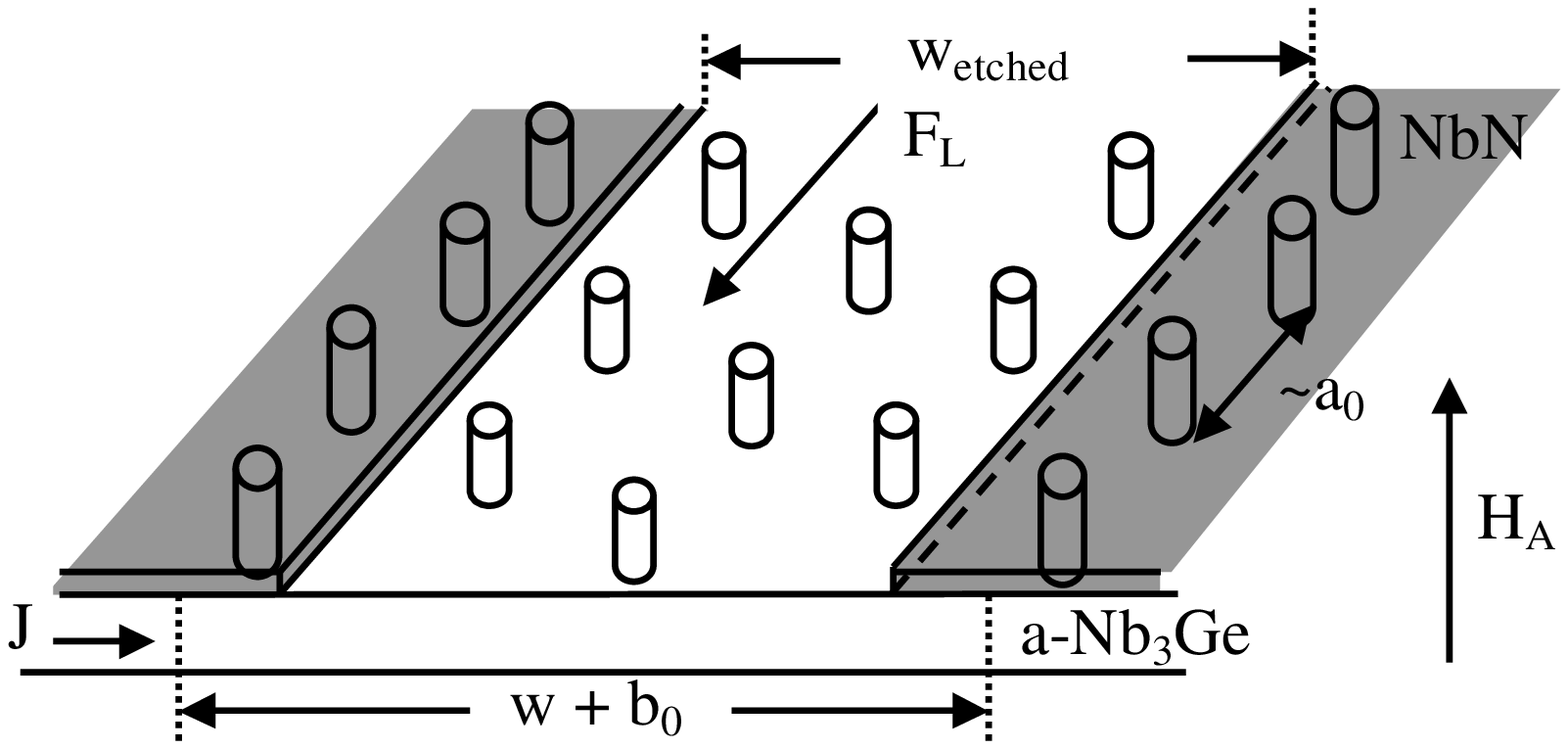}}

\vspace{0cm} \caption{Sketch of the artificial channel geometry.
In the gray areas vortices are pinned by the strong-pinning NbN
layer while inside the channels pinning due to material
inhomogeneities is negligible. The etched channel width
$w_{etched}$ (of the order of a few row spacings $b_0$) and the
effective width $w$ are indicated.} \label{plot_sketch}
\end{figure}

Besides the above examples, the phenomenon of vortex channelling
can also arise from the presence of grain boundaries in the
sample. Historically, the 'shear' depinning of vortices in grain
boundaries in low $T_c$ materials received considerable attention
\cite{Dewhughes87_shearinGB,Pruymboomthesis} because it could
explain the quadratic decrease of $J_c$ near $B_{c2}$ in
practically relevant poly-crystalline superconductors. More
recently, the interesting issue of channelling of mixed
Abrikosov-Josephson vortices in grain boundaries in high $T_c$
materials was addressed in detail by Gurevich \cite{Gurevich}.

\begin{figure}
\scalebox{0.40}{\includegraphics{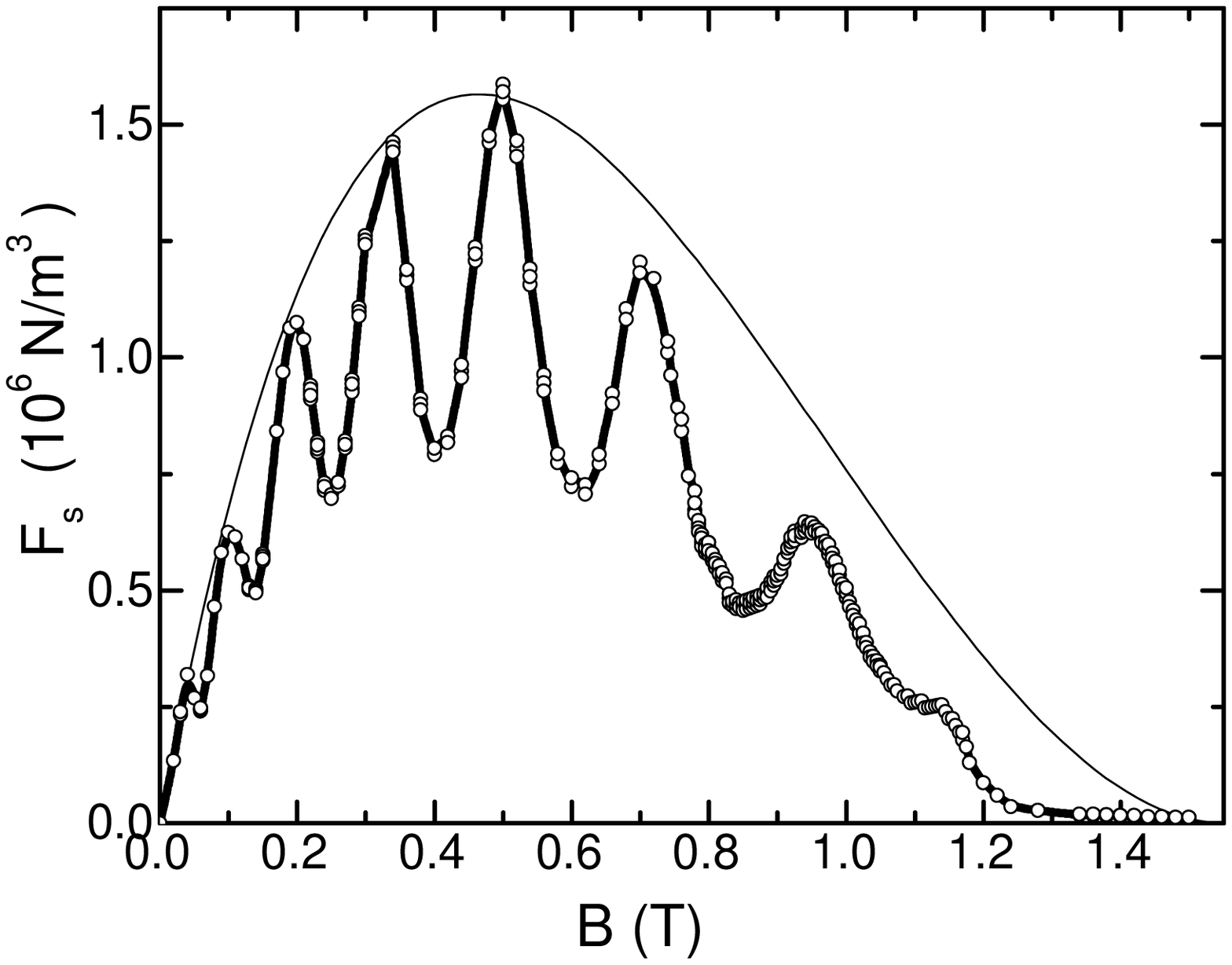}}

\vspace{0cm} \caption{Data: critical shear force density
$F_s=J_sB$, determined using a velocity criterion $v/a_0 \approx
1$ MHz, versus applied field for a channel sample with
$w_{etched}\approx 230$ nm at $T=1.94$ K. Drawn line:
Eq.(\ref{phenomenological}) with $A=0.05$, $B_{c2}=1.55$ T,
$\lambda(T)=1.13$ $\mu$m and effective width $w=300$ nm.}
\label{plot_introFs}
\end{figure}

A system in which channel motion and its dependence on the
structural properties of vortex matter can be studied
systematically is that of narrow, weak pinning flow channels in a
superconducting film \cite{Pruymboom}, see Fig. \ref{plot_sketch}.
The samples are fabricated by etching straight channels of width
$w_{etched}\gtrsim 100$ nm through the top layer of an a-NbGe/NbN
double layer. With a magnetic field applied perpendicular to the
film, vortices penetrate both the strong pinning NbN in the
channel edges (CE's) and the remaining NbGe weak pinning channels.
The strongly pinned CE vortices provide confinement to the
vortices inside the channel as well as the pinning (shear)
potential which opposes the Lorentz force from a transport current
$J$ applied perpendicular to the channel. By changing the applied
field $H$ one can tune the commensurability between the vortex
lattice constants and the channel width, allowing a detailed study
of the shear response and threshold for plastic flow as a function
of the mismatch and the actual microstructure in the channel.

Phenomenologically, plastic flow in the channel occurs when the
force density $F=JB$ (with $B/\Phi_0$ the vortex density) exceeds
$2\tau_{max}/w$, where $\tau_{max}=Ac_{66}$ is the flow stress at
the edge (the factor 2 is due to both CE's) and $w$ is the
effective width between the first pinned vortex rows, defined in
Fig. \ref{plot_sketch}. Thus, the critical force density is given
by:
\begin{eqnarray}
F_s=J_s B= 2Ac_{66}/w. \label{phenomenological}
\end{eqnarray}
The parameter $A$ describes microscopic details of the system: it
depends on lattice orientation, (an)harmonicity of the shear
potential, details of the vortex structure in the CE's and the
microstructure of the array inside the channel. Critical current
measurements as function of applied field reflected this change in
microstructure through oscillations of $F_s$, shown in Fig.
\ref{plot_introFs} for a channel with $w_{etched}\approx 230$ nm.
Note that in such a narrow channel, the pinning strength due to
intrinsic disorder in the a-NbGe is at most $10 \%$ of $F_s$
(except for $B\lesssim 50$ mT) and does not affect the
oscillations. Since the natural row spacing of the VL is
$b_0=\sqrt{3}a_0/2$, with $a_0^2=2\Phi_0/\sqrt{3}B$, and in our
geometry $B\simeq \mu_0 H$ one can check that the periodicity of
the oscillations corresponds to transitions from $w=nb_0$ to
$w=(n\pm 1)b_0$ with $n$ integer, i.e. the principal lattice
vector $\vec{a}_0$ is oriented along the channel (Fig.
\ref{plot_sketch}). The envelope curve represents
Eq.(\ref{phenomenological}) with Brandt's expression for the VL
shear modulus \cite{Brandtelas}:
\begin{eqnarray}
c_{66}=\frac{\Phi_0 B_{c2}}{16\pi \mu_0 \lambda^2}
b(1-b)^2(1-0.58b+0.29b^2),\label{c66}
\end{eqnarray}
($b=B/B_{c2}$ is the reduced field and $\lambda$ is the
penetration depth) and a value $A= 0.05$. This value for $A$ is
close to the value $\sqrt {\langle u^2 \rangle}/a_0=0.047$ for the
relative displacements at the crossover from elastic to plastic
deformations as obtained from measurements on the peak effect
\cite{WordenweberPRB86,TroyanovskiPRL02_PE}. This led to a
qualitative interpretation of the reduction of $F_s$ at minima as
being due to defects in the channel, which develop at
incommensurability. However, recent developments
\cite{KokuboPRL02,KokuboPRB_chanMLvcB,BesselingEPL2003,BaarleAPL03}
have shown that (strong) structural disorder may be present in the
CE arrays, in which case the interpretation can drastically
differ.

In this article we present numerical and analytical studies of the
threshold force and dynamics of vortices in the channel system for
various degrees of edge disorder. In an early paper
\cite{BesselingPRL99} we studied the commensurability effects in
the idealized case with periodic arrays in the CE's. In this
situation $F_s$ at matching ($w=nb_0$) is equal to the ideal
lattice strength $2A^0 c_{66}/w$ (the value $A^0=1/(\pi \sqrt{3})$
follows from Frenkels considerations \cite{Frenkel1926}), while at
mismatch dislocations develop, leading to $A\simeq 0$. The
resulting series of delta-like peaks in $F_s$ versus matching
parameter differed considerably from the experimental results,
which could not be explained by thermal fluctuations or intrinsic
disorder inside the channel. Therefore we investigated the effect
of positional disorder in the CE arrays on $F_s$ near
commensurability ($w \approx nb_0$) \cite{BesselingEPL2003}. In
this regime the behavior is dominated by the longitudinal
displacements of vortices in the chains, i.e. quasi
one-dimensional ($1$D), and $F_s$ is controlled by defects with
Burgers vector along the channel. At weak disorder, we found a
clear reduction of $F_s$ at commensurability caused by nucleation
of defects at threshold, while the existing defects at
incommensurability become pinned by disorder, leading to an
increase of $F_s$ in the mismatching case.

The present paper first describes in detail these quasi 1D
phenomena near commensurability and/or for weak disorder. Using a
generalized sine-Gordon model, we quantitatively describe how the
structure and transport properties depend on the vortex
interaction range and on weak disorder in the CE's. Besides the
connection to our system, these results also provide a background
for understanding quasi-$1$D vortex states and matching effects in
artificial PPA's, including the effects of disorder which these
PPA's may contain due to fabrication uncertainties.

The 1D model shows that, above a certain disorder strength,
spontaneously (disorder) induced defects along the CE's dominate
over incommensurability induced defects. The commensurability peak
in $F_s$ is then completely smeared out with a value of $F_s$ at
matching ($w=nb_0$) saturating at $\sim 30 \%$ of the ideal
lattice strength. In the more general case of wider channels, the
{\it transverse} degrees of freedom, especially away from matching
($w/b_0\simeq n \pm 1/2$), lead to new phenomena: under the
influence of disorder, the channel array may split up in regions
with $n$ and $n\pm 1$ rows, involving dislocations with Burgers
vector strongly {\it misaligned} with the CE's. At sufficient
disorder strength, such dislocations lead to a more effective
pinning of the array then the 'aligned' dislocations around
matching. $F_s$ then exhibits a smooth oscillation as function of
$w/b_0$, similar to Fig. \ref{plot_introFs}, with yield strength
maxima occurring {\it around mismatch}. This behavior resembles
the classical peak effect, i.e. at mismatch the enhanced ability
of the arrays to sample configuration space allows better
adjustment to the random CE's. In the last part of the paper we
show detailed simulations of both static and dynamical aspects of
this behavior, including a study of reordering phenomena at large
drive. We find an ordering velocity of the arrays with a channel
width dependence similar to that of the threshold force. Using a
modified version of the dynamic ordering theory in
\cite{Koshelevrecryst} it is shown that such behavior can be
explained by a reduction of the energy for formation of misaligned
defect pairs away from matching. The numerical results at strong
disorder are also in good qualitative agreement with recent
mode-locking experiments on the channel system
\cite{KokuboPRL02,KokuboPRB_chanMLvcB}.

The outline of the paper is as follows. In Sec.\ref{secmodel} the
channel geometry and the simulation procedure are discussed. The
first part of the paper deals with channels having hexagonal,
ordered arrays in the CE's: in Sec.\ref{secord1D} we present the
sine-Gordon description and numerical results for a single 1D
vortex chain in an ordered channel; in section \ref{secord2D} we
show how the 1D behavior extends to wider channels with multiple
rows and ordered CE's. The second part of the paper deals with
disordered channels: section \ref{secdis1D} describes the effects
of weak CE disorder on the behavior of a 1D chain, both
analytically and using numerical simulations. The effects of weak
disorder in wider channels are discussed in
Sec.\ref{secweakdis2D}. Section \ref{secstrongdis2D} describes the
static and dynamic properties of wider channels in presence of
strong disorder, including an analysis of the reordering phenomena
in this situation. A comparison with the dynamic ordering theory,
the confrontation with experiments and a summary of the results
are presented in Secs.\ref{secdiscuss} and \ref{secsum}.

\section{Model and numerical procedure}
\label{secmodel}

\begin{figure}
\scalebox{0.40}{\includegraphics{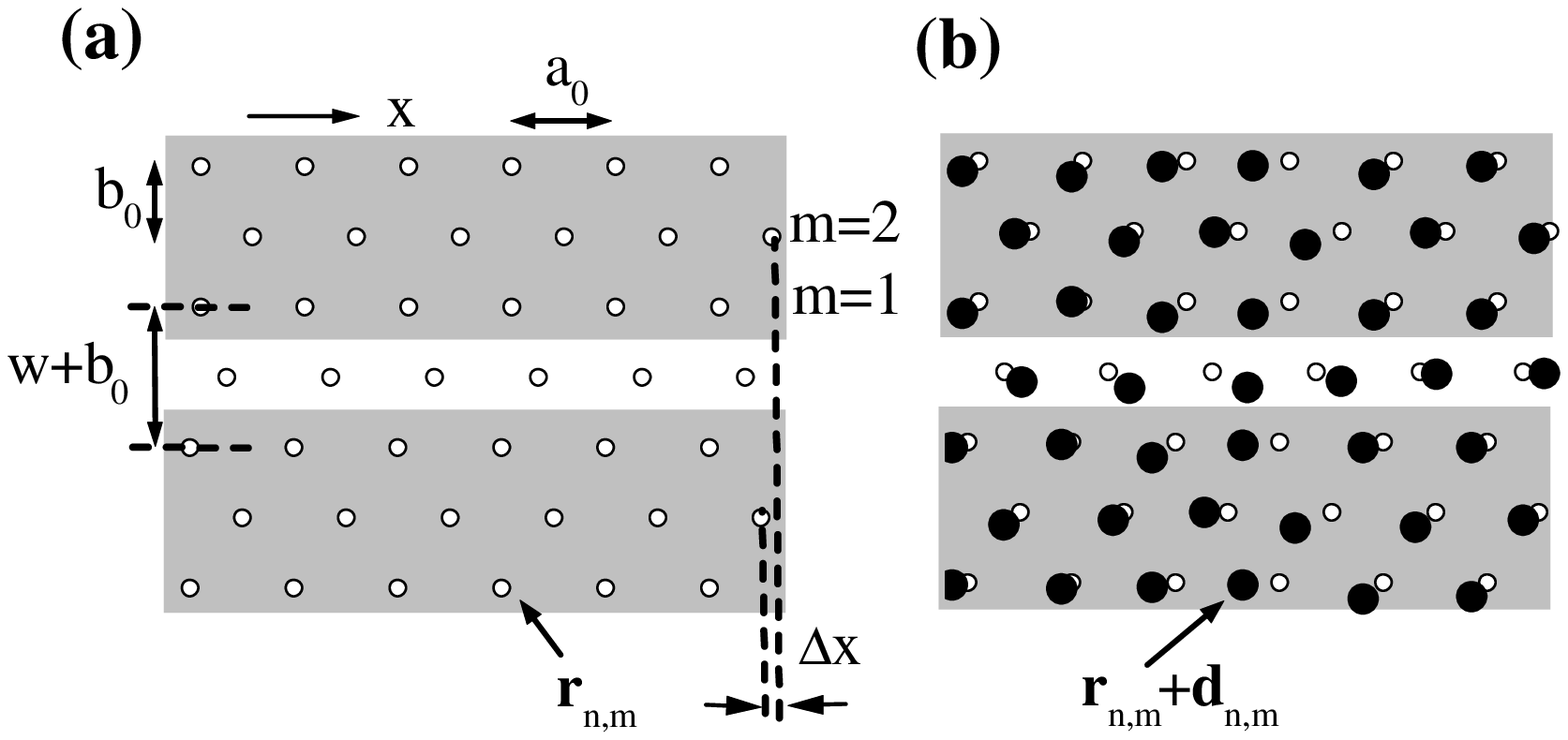}}

\vspace{0cm} \caption{Channel geometry with pinned vortices in the
gray areas. The specific case of $w \simeq b_0$, i.e. with $1$ row
in the channel is illustrated. (a) Ordered situation: the
equilibrium positions ${\bf r}_{n,m}$ of pinned vortices in the
CE's are denoted by ($\circ$). The effective width $w$ and
relative shift $\Delta x$ are indicated. (b) Disordered case. The
randomized vortex positions are denoted by ($\bullet$).}
\label{plot_geometry}
\end{figure}

We consider straight vortices at $T=0$ in the geometry as
illustrated in Fig. \ref{plot_geometry} for the case of 1 row in
the channel. The approximation $T=0$ is well justified over a
considerable range of the experiments (see Sec.\ref{secdiscuss}).
The CE's are formed by two semi-infinite static arrays. The
distance between the first vortex rows on both sides of the
channel is $w+b_0$, with $w$ the {\em effective} channel width.
The vortices are assumed to be fixed by columnar pins in the CE's.
The principal axis of the pinned arrays is along the channel
direction $x$. A relative shift $\Delta x$ is allowed between the
arrays. In (a) the simplest configuration is shown, where CE
vortices form a perfect triangular lattice. For $\Delta x=0$,
their coordinates are:
\begin{equation}
{\bf r}_{n,m}=([n+frac(m/2)]a_{0},m[b_0+(w-b_0)/2|m|]),
\label{ordedgecoord}
\end{equation}
for $m\neq 0$ and $frac(m/2)$ denotes the remainder of $m/2$.

Disorder is incorporated in the model by adding random shifts
${\bf d}$ to the coordinates of the ordered arrays:
\begin{eqnarray}
{\bf R}_{n,m}={\bf r}_{n,m}+{\bf d}_{n,m}, \label{disordedge}
\end{eqnarray}
as shown in Fig. \ref{plot_geometry}(b). The amplitudes of the
random shifts are characterized by disorder parameters $\Delta_x$
and $\Delta_y$ as follows: transverse relative displacements
$d^y/a_0$ are chosen independently from a box distribution
$[-\Delta_y ,\Delta_y]$. The longitudinal shifts $d^x_{n,m}$ are
chosen such that the {\it strain} $(d_{n+1}-d_{n})/a_0$ along the
channel is uniformly distributed in the interval
$[-\Delta_x,\Delta_x]$. The latter provides a simple way of
implementing loss of long range order along the CE's. For
$\Delta_x$ and $\Delta_y$ we study the following specific cases:
$\Delta_x,\Delta_y=0$ in Secs. \ref{secord1D},\ref{secord2D},
$\Delta_x \equiv \Delta,\Delta_y=0$ in Sec.\ref{secdis1D} and
$\Delta_x=\Delta_y \equiv \Delta$ in
Secs.\ref{secweakdis2D},\ref{secstrongdis2D}.

To study the commensurability effects, the effective width of the
channel is varied from a value $w/b_0\sim 1-10$. We assume that
the vortex density in and outside the channel are the same. The
number of vortices in the channel is then given by
$N_{ch}=(L/a_0)(w/b_0)$ with $L$ the channel length. In a
commensurate situation one has $w=pb_0$ and both the row spacing
and (average) longitudinal vortex spacing in the channel match
with the vortex configuration in the CE's. When $w\neq nb_0$ these
spacings become different, leading to generation of topological
defects. While this model differs from the experimental case where
the applied field drives the incommensurability, the method offers
a simple way of introducing geometrical frustration and study
various (mis)matching configurations.

With a uniform transport current $J$ applied perpendicular to the
channel, the equation of motion for vortex $i$ in the channel
reads (in units of N/m):
\begin{equation}
\gamma \partial_t{\bf r} _i = f -\sum_{j\neq i}\nabla V({\bf
r}_i-{\bf r}_j)-\sum_{n,m}\nabla V({\bf r}_i-{\bf R}_{n,m}).
\label{eqm}
\end{equation}
$V({\bf r})$ is the vortex-vortex interaction potential, $j$
labels other vortices inside the channel, the damping parameter
$\gamma$ is given by $\gamma=B\Phi_0/\rho_f$ with $\rho_f$ the
flux flow resistivity, and $f=J\Phi_0$ is the drive along the
channel. For films which are not too thin compared to the
penetration depth $\lambda$ and magnetic fields small compared to
the upper critical field $B_{c2}$, the interaction $V({\bf r})$ is
given by the London potential:
\begin{equation}
V({\bf r})=U_0K_0({\bf |r|}/\lambda), \label{VLondon}
\end{equation}
where $U_0=\Phi_0^2/2\pi\mu_0\lambda^2$ and $\Phi_0$ is the flux
quantum.

In the simulations we integrate Eq.(\ref{eqm}) numerically for all
vortices in the channel. We use a Runge-Kutta method with variable
time steps such that the maximum vortex displacement in one
iteration was $a_0/50$. Distances were measured in units $a_0$
(${\bf \bar{r}}={\bf r}/a_0$), forces in units $U_0/a_0$ and time
in units $\gamma a_0^2/U_0$. Following Ref.
\cite{Koshelevrecryst}, the London potential was approximated by:
\begin{eqnarray}
V({\bf \bar{r}})=\ln\left(\frac{r_{c}}{|{\bf
\bar{r}}|}\right)+\left(\frac{|{\bf
\bar{r}}|}{r_c}\right)^2-\frac{1}{4}\left(\frac{|{\bf
\bar{r}}}{r_c}\right)^4-0.75, \label{siminteraction}
\end{eqnarray}
with a cut-off radius $r_c$ corresponding to $r_c \simeq 3
\lambda/a_0$. We performed most simulations for $r_c=3.33$.
Periodic boundary conditions in the channel direction were
employed. For each $w/b_0$ we relaxed the system to the ground
state for $f=0$. We found that this is best achieved by starting
from a uniformly stretched or compressed $n$ or $n\pm 1$
configuration. For an initial configuration with $N_{ch}$ vortices
distributed randomly in the channel, relaxation resulted in
(slightly) metastable structures, even when employing a finite
temperature, simulated annealing method. Some peculiarities
associated with such structures are mentioned in
Secs.\ref{secord2D} and \ref{secweakdis2D}. After the $f=0$
relaxation, the average velocity versus force ({\it v-f}) curve
was recorded by stepwise varying the force from large $f_{max}
\rightarrow 0$ (occasionally $f=0 \rightarrow f_{max} \rightarrow
0$ was used to check for hysteresis). At each force we measured
$v(f)=\langle\dot{x_i}\rangle_{i,t}$ after the temporal variations
in $v$ became $< 0.5 \%$ (ignoring transients by discarding the
data within the first $3a_0$). In addition, at each force we
measured several other quantities, e.g. the temporal evolution of
$\bf{r}_i$ and the time dependent velocity
$v(t)=\langle\dot{x_i}\rangle_{i}$.

\section{Single chain in an ordered channel}
\label{secord1D}

The first relevant issue for plastic flow and commensurability
effects in the channel is to understand the influence of
periodically organized vortices in the CE's (see Fig.
\ref{plot_geometry}a). The characteristic differences between
commensurate and incommensurate behavior can be well understood by
focusing on a 1-D model in which only a single vortex chain is
present in the channel. Therefore the CE's are assumed to be
symmetric with respect to $y=0$ (i.e. $\Delta x=0$ in Fig.
\ref{plot_geometry}) and only the longitudinal degrees of freedom
of the chain are retained. At commensurability, $w=b_0$, the
longitudinal vortex spacing $a=a_0$. For $w\neq b_0$ the average
spacing $a=\Phi_0/(Bw)=a_0b_0/w$ does not match with the period
$a_0$ in the edges and interstitials or vacancies develop in the
channel. Their density $c_d$ is given by
$c_d=|a_0^{-1}-a^{-1}|=(1/a_0)|1-(w/b_0)|$.

\subsection{Continuum sine-Gordon description}
\label{subord1Dana}

We first consider the interaction of a vortex in the channel with
the periodic arrays in the CE's. As shown in App.\ref{appA}, when
$B \lesssim 0.2 B_{c2}$ and $\lambda \gtrsim a_0$ the edge
potential arising from this interaction is:
\begin{eqnarray}
V_{ce,0}(x,y)=-2U_0e^{-k_0(w+b_0)/2 }\cosh(k_0y) \cos{k_0 x},
\label{OrdVrlowfield}
\end{eqnarray}
where $k_0=2\pi/a_0$. For $w=b_0$ and $y=0$, the associated
sinusoidal force caused by the edge has an amplitude which we
denote by $\mu$:
\begin{eqnarray}
\mu=(4\pi U_0/a_0)e^{-\pi\sqrt{3}}\simeq U_0/(6\pi a_0).
\label{mudef}
\end{eqnarray}
Next we consider a static chain of vortices {\em inside} the
channel. The chain is most easily described in terms of a
continuous displacement field $u(x)$, representing the deviations
of vortices in the chain with respect to the commensurate
positions, i.e. $u(ia_0)=u_i=x_i-ia_0$. The edge force is then
given as $f_p=-\mu \sin(k_0 u)$. To describe the interaction
between vortices {\it within} the chain, we assume that their
relative displacements are small, $\partial_x u\ll 1$. Then one
can use linear elasticity theory. Taking into account that the
interaction range $\lambda> a_0$, the elastic force at $x=ia_0$
is:
\begin{equation}
f_{el}=\sum_{s=ja_0>ia_0} \partial_{s}^2
V(s)[u(x+s)+u(x-s)-2u(x)]. \label{semidiscretefel}
\end{equation}
Using the Fourier transform of $V$ the force due to a displacement
$u_q(x)=$Re$(u_q e^{iqx})$ with wave vector $q$ is:
\begin{equation}
f_{el}=\int \frac{dk}{2\pi} \frac{U_0\pi k^2
}{\sqrt{k^2+\lambda^{-2}}}\sum_{s>x} 2e^{iks}[1-\cos(qs)]u(x).
\label{kappaderive}
\end{equation}
Recasting this into a sum over reciprocal vectors $lk_0\pm q$ and
retaining only the $l=0$ term, one obtains the following
dispersive elastic modulus of the chain:
\begin{eqnarray}
\kappa_q=\frac{U_0\pi\lambda/a_0}{\sqrt{1+\lambda^2q^2}}.
\label{kappaexpression}
\end{eqnarray}
For deformations of scale $> 2\pi \lambda$, the elastic force is
$f_{el}=\kappa_0\partial_x^2 u$ with a long wavelength stiffness
$\kappa_0=U_0\pi (\lambda/a_0)$.

The equation of motion for $u$ for a uniformly driven chain, is
obtained by adding the driving force $f$ to the edge force and the
intra-chain interactions resulting in: $\gamma\partial_t
u=f+f_p+f_{el}$. Assuming for the moment that the long wavelength
description is valid, the evolution of $u$ is given by the
following sine Gordon (s-G) equation:
\begin{eqnarray}
\gamma\partial_t u=f-\mu \sin(k_0 u) +\kappa_0\partial_x^2 u
\label{ordconteqm}
\end{eqnarray}
A useful visual representation of Eq.(\ref{ordconteqm}) is an
elastic string of stiffness $\kappa_0$ with transverse coordinate
$u(x)$ in a tilted washboard potential $(\mu/k_0)\cos(k_0 u)-fu$.

\begin{figure}
\scalebox{0.40}{\includegraphics{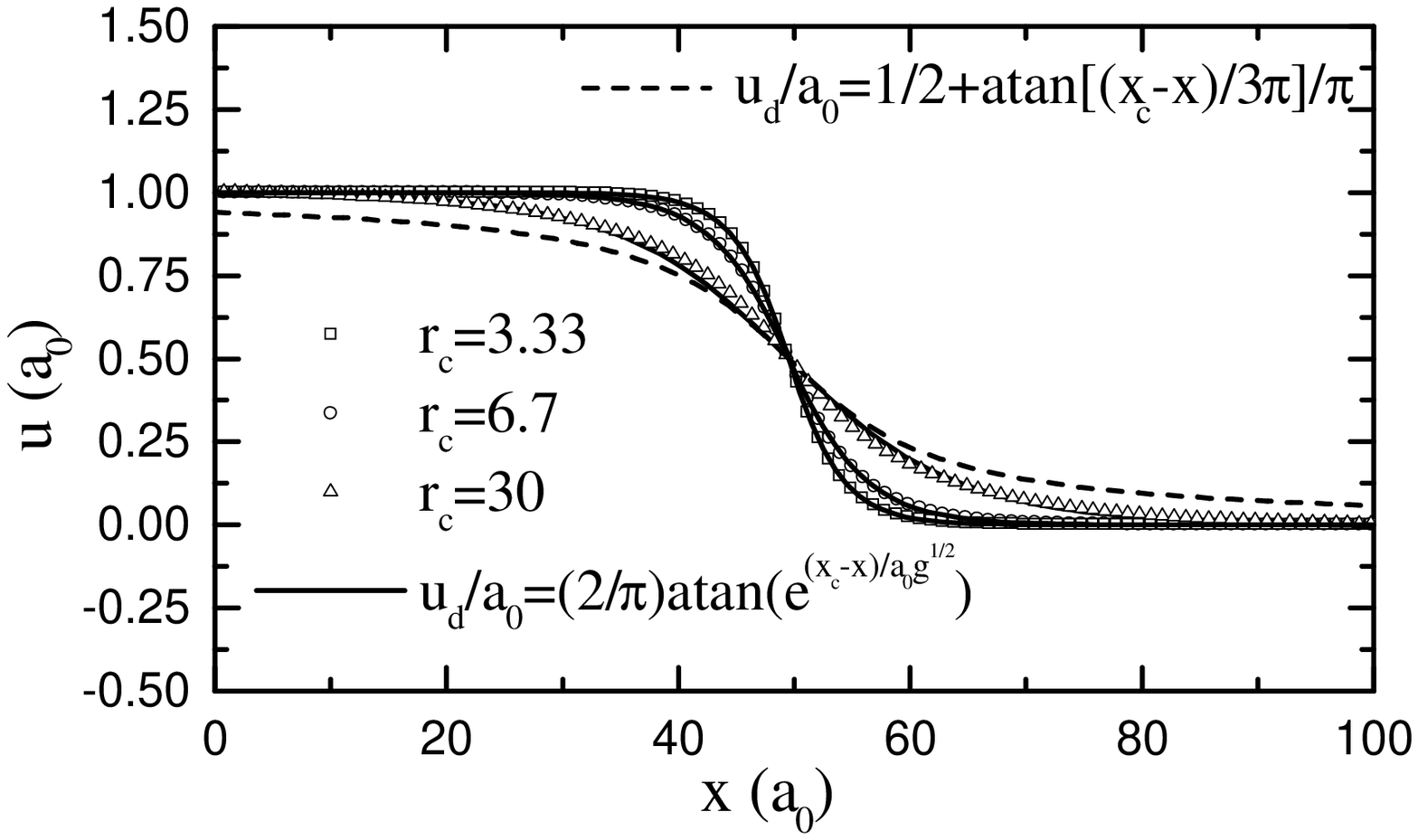}}

\vspace{0cm} \caption{Drawn lines: the anti-kink solution
Eq.(\ref{kinkforms}) for $\lambda/a_0=1$, $2$ and $9$ (most
extended line). Symbols: numerically obtained displacement field
for an isolated interstitial for the corresponding $r_c$. Dashed
line: defect shape Eq.(\ref{nonlocalkink}) in the nonlocal limit.}
\label{plot_sgkink}
\end{figure}

The s-G equation (\ref{ordconteqm}) has been thoroughly studied in
different contexts (e.g.
\cite{OwenScalapinoPR67_LJJ,ButtikLandauerPRA81,BraunKivsharPhysRep98}).
In the static case ($f=0$), it has the trivial solution $u=0$,
corresponding to a commensurate chain, or kinked, incommensurate,
solutions in which $u(x)$ periodically jumps by $\pm a_0$, each
jump representing a point defect in the channel. In the context of
long Josephson junctions (LJJs, \cite{OwenScalapinoPR67_LJJ}), a
kink corresponds to a Josephson vortex where the phase difference
across the junction changes by $2 \pi$. An {\em isolated} defect
is represented by the familiar 'soliton' solution of the s-G
model:
\begin{eqnarray}
u_d(x)=2a_0\arctan[\exp(\pm 2\pi (x-x_c)/l_d)]/\pi.
\label{kinkforms}
\end{eqnarray}
Here $x_c$ denotes the center of the defect and the $+$ ($-$) sign
denotes a vacancy or interstitial (kink or antikink). The length
$l_d$ represent the core size of the defect:
\begin{eqnarray}
l_d=2\pi a_0 \sqrt{g}, \label{defectwidth}
\end{eqnarray}
with $g$ the dimensionless ratio between the chain stiffness and
maximum curvature of the pinning potential:
\begin{eqnarray}
g=\kappa_0/2\pi\mu a_0=3\pi(\lambda/a_0), \label{g_definition}
\end{eqnarray}
as follows from Eq.(\ref{mudef}),(\ref{kappaexpression}). For
$\lambda/a_0\gtrsim 1$, $l_d$ thus considerably exceeds the
lattice spacing. The continuum approach is validated since
$\partial_x u_d\lesssim 2a_0/l_d\ll 1$. In Fig. \ref{plot_sgkink}
we have illustrated the characteristic defect shape
Eq.(\ref{kinkforms}), along with numerical data from a later
section.

The long wavelength limit is only valid when $l_d$ considerably
exceeds $\lambda$. Since $l_d$ grows only as $\sqrt{\lambda/a_0}$
the dispersion in the elastic interactions becomes important
beyond a certain value of $\lambda/a_0$. This value is estimated
by setting $\lambda q_d=2\pi \lambda/l_d=1$ in
Eq.(\ref{kappaexpression}), resulting in $\lambda/a_0 \simeq 9$,
in which case $l_d\simeq 54 a_0$.

For larger interaction range one employs the following approach,
first derived by Gurevich \cite{Gurevich} for mixed
Abrikosov-Josephson vortices in grain boundaries. Expression
(\ref{semidiscretefel}) for the elastic force can be written as an
integral $f_{el}=\int (ds/a_0)
\partial^2_s V(s)u(x+s)$. For defects on a scale
$<\lambda$, only the short distance behavior of $V$ has to be
retained: $V(x)=U_0K_0(x/\lambda)\approx -U_0\ln(|x/ \lambda|)$.
Integrating the expression for $f_{el}$ by parts and adding the
edge force and the drive, the equation of motion becomes:
\begin{eqnarray}
\gamma\partial_t u=f-\mu \sin(k_0 u)
+(U_0/a_0)\int_{-\infty}^{\infty} ds \frac{\partial_s u}{s-x}.
\label{nonlocalordconteqm}
\end{eqnarray}
A static solution of Eq.(\ref{nonlocalordconteqm}) for a single
defect is \cite{Gurevich}:
\begin{eqnarray}
k_0u=\pi+\arctan(\pm 2\pi x/l_d^{nl}), \label{nonlocalkink}
\end{eqnarray}
with $l_d^{nl}=6\pi^2a_0$, which is valid when $\lambda>l_d^{nl}$.
The value for $l_d^{nl}$ is nearly the same as the s-G core size
$l_d$ for $\lambda/a_0=9$. This means that upon approaching the
nonlocal regime, the increase of core size saturates at $\sim
60a_0$, while only the tails of the defect are affected according
to Eq.(\ref{nonlocalkink}), see Fig. \ref{plot_sgkink}. A more
accurate calculation of the onset of the nonlocal field regime
using Brandt's field-dependent vortex interaction (App.\ref{appA})
shows that non-locality is only relevant for a channel in a
superconductor with $\lambda/\xi \gtrsim 50$.

So far, we discussed isolated defects. For finite defect density,
the repulsive interaction between defects of the same 'sign'
causes a periodic superstructure in the chain. When $c_d$ grows to
$\sim 1/l_d$, the defects start to overlap significantly. For the
(local) s-G model, explicit solutions for $u$ have been obtained
in terms of the Jacobi elliptic functions, for which we refer to
\cite{OwenScalapinoPR67_LJJ,BraunKivsharPhysRep98}. Recently, also
in the non-local limit where $l_d^{nl}>\lambda$, the 'soliton'
chain has been described analytically \cite{Gurevich}, which we
will not repeat here.

\subsection{Transport properties}
\label{subord1Dfv}

With a uniform drive $f$, the transport properties strongly depend
on the presence and density of defects in the channel. At
commensurability ($a=a_0$, $c_d=0$), a threshold force $f_s=\mu$
is required, above which all vortices start moving uniformly.
Their velocity is identical to that of an overdamped particle in a
sinusoidal potential: $v=\sqrt{f^2-\mu^2}$ \cite{Tinkham}. The
threshold $\mu$ coincides with the well known relation between
shear strength and shear modulus of an ideal lattice by Frenkel
\cite{Frenkel1926}: for a harmonic shear interaction, a value
$A=A^0=a_0/(2\pi b_0)=1/\pi\sqrt{3}$ applies in
Eq.(\ref{phenomenological}). Identifying $F_s a_0b_0=f_s=\mu$ for
$w=b_0$, one finds:
\begin{eqnarray}
c_{66}=\pi\sqrt{3}\mu/(2a_0)=U_0/(8a_0b_0), \label{Londonc66}
\end{eqnarray}
which coincides with the familiar expression for the shear modulus
in the London limit: $c_{66}=\Phi_0B/(16\pi\mu_0\lambda^2)$. In
App.\ref{appA} we generalize the expression for the ordered
channel potential to higher field and show that also in that case
the potential is harmonic and that $A=A^0$ holds for a
commensurate channel.

At incommensurability, depinning of the chain is governed by the
threshold force to move a defect. In the present continuum
approach such threshold is absent. However, taking into account
the discreteness of the chain, in which case Eq.(\ref{ordconteqm})
turns into a Frenkel-Kontorova (FK) model \cite{BesselingPRL99}, a
finite Peierls-Nabarro (PN) barrier exists to move a defect over
one lattice spacing (see e.g. \cite{BraunKivsharPhysRep98}). The
magnitude of the PN barrier has been studied for a variety of
cases, including FK-models with anharmonic and/or long-range
interactions \cite{BraunKivsharPhysRep98,BraunBraunMingaleev}. For
$g<1$, $f_{PN}$ can amount to a considerable fraction of $\mu$.
Additionally, in this regime anharmonicity may renormalize $g$
\cite{BraunKivsharPhysRep98,BraunBraunMingaleev} and cause
pronounced differences between the properties of kinks (vacancies)
and antikinks (interstitials). In our limit $g \gg 1$, where
$\partial_x u<< 1$ and harmonic elastic theory applies, these
differences are small and the pinning force vanishes as $f_{PN}=32
\pi^2 g \mu \exp{(-\pi^2\sqrt{g})}$. Hence, defects in an ordered
channel give rise to an essentially vanishing plastic depinning
current $J_s$ \cite{fn_discreteness}.

Considering the dynamics, for small drive $f<\mu$ the motion of
defects, each carrying a flux quantum, provides the flux transport
through the channel. When defects are well separated, for
$c_d<l_d^{-1}$, the mobility of the chain is drastically reduced
compared to free flux flow and the average velocity $v$ is
proportional to the defect density: $v=c_d v_d a_0$. Here $v_d$ is
the velocity of an isolated defect at small drive. It can be
calculated from the general requirement that the input power must
equal the average dissipation rate:
\begin{eqnarray}
f v=\gamma \langle (\partial_t u)^2 \rangle_{L,t}=(\gamma/l_d)
\int^{l_d} (\partial_t u)^2 dx. \label{dissiprate}
\end{eqnarray}
The last step arises from the space and time periodicity of $u$.
Using $\partial_t u=v_d
\partial_x u$ and the kinkshape Eq.(\ref{kinkforms}), one obtains
the 'flux flow resistivity' at small defect density:
\begin{eqnarray}
dv/df=c_d a_0 (\pi^2 \sqrt{g}/2\gamma), \label{linearkinkresist}
\end{eqnarray}
where $\pi^2 \sqrt{g}/2\gamma=M_d$ is the kink mobility in the s-G
model \cite{ButtikLandauerPRA81}. For larger defect density, where
defects start to overlap this relation changes. The linear
response for $f\lesssim \mu$ may then be obtained from the
solutions for $u$ based on elliptic integrals
\cite{Gurevich,OwenScalapinoPR67_LJJ,BraunKivsharPhysRep98}.

For larger drive $f\gtrsim \mu$, the 'tilt' induced reduction of
the (washboard) edge potential becomes important. This leads to an
expansion of the cores of the sliding defects and causes a
nonlinear upturn in the {\it v-f} curves. Exact solutions of
Eq.\ref{ordconteqm} describing this behavior do not exist.
Therefore we use a perturbative method similar to that in
\cite{MartinoliPRB78,StrunzPRE98} which is able to describe the
full {\it v-f} curve over a wide range of defect densities. It is
convenient to define the displacements $h(x,t)=u(x,t)-s(x,t)$,
where $s(x,t)=(q/k_0) x+ vt$, with $(q/k_0)=c_da_0$, is the
continuous field describing the displacements of an undeformed
incommensurate chain (i.e. straight misoriented string in the
washboard potential) moving with velocity $v$. In terms of $h$,
the equation of motion (\ref{ordconteqm}) and
Eq.(\ref{dissiprate}) can be written as:
\begin{equation}
\gamma v(1+\partial_s h)=f+\mu\sin(k_0h+qx+k_0vt)+\kappa(q/k_0)^2
\partial^2_s h \label{conthulleqm}
\end{equation}
\begin{eqnarray}
f =\gamma v+(\gamma v/a_0)\int^{a_0} (\partial_s h)^2 ds.
\label{fvdispl}
\end{eqnarray}
The last term in Eq.(\ref{fvdispl}) describes additional
dissipation due to internal degrees of freedom in the chain. Under
influence of the potential, $h$ acquires modulations with period
$1/c_d$ in $x$, i.e. period $a_0$ in $s$. These modulations are
then expressed as a Fourier series of modes with wavelength
$1/(mc_d)$ ($m$ integer $\geq 1$) and amplitude $h_m$:
\begin{eqnarray}
h(x,t)=\sum_m h_m \exp[imk_0 s]+c.c. \label{persolansatz}
\end{eqnarray}
The overlap of defects and the core expansion for $f\geq \mu$
appears in the $q$ and $v$ dependence of $h$. Both effects cause a
reduction of the relative displacements $h$. An approximate
solution for $h(v)$ is obtained by substituting
Eq.(\ref{persolansatz}) into Eq.(\ref{ordconteqm}), yielding the
coefficients $h_m$ (details of the solution are deferred to
App.\ref{appB}). The {\it v-f} relation Eq.(\ref{fvdispl}) attains
the form:
\begin{eqnarray}
f =\gamma v \left[1+
\frac{\omega_p^2}{2[\omega_0^2+\omega_r^2]}\right].
\label{fvresultana}
\end{eqnarray}
The additional 'friction' force is represented in terms of the
pinning frequency $\omega_p=\mu k_0/\gamma$, the washboard
frequency $\omega_0=k_0 v$ and $\omega_r=K_{eff}^2(c_d)/\gamma$
which is the effective relaxation frequency for nonlinear
deformations associated with a defect density $c_d=q/(2\pi)$, with
$K_{eff}^2(c_d)$ given in App.\ref{appB}. At small $v$, the
elastic relaxation time $1/\omega_r$ for the chain to relax is
much smaller than the timescale $1/\omega_0$ between passage of
maxima in the edge potential. This corresponds to the linear
sliding response of the static structure of (overlapping) defects.
For large $v$, $1/\omega_r\gg 1/\omega_0$ meaning that the
incommensurate chain is not given enough time to deform. This
leads to expanded defects described by a sinusoidal variation of
$h$ with reduced amplitude (see App.\ref{appB}). The {\it v-f}
curve then approaches free flux flow according to $f-\gamma v \sim
v^{-1}$ as for a single particle.

Recently, exact solutions describing the nonlinear dynamics and
core expansion of mixed Abrikosov-Josephson vortices based on the
nonlocal Eq.(\ref{nonlocalordconteqm}) have been derived in
\cite{Gurevich}. The resulting transport curves are very similar
to those obtained from Eq.(\ref{fvresultana}), see Fig.
\ref{plot_fvord1d}. We also note the similarity with the $IV$
curves obtained from a model for kinked Josepson strings
\cite{Ziswiler} in high $T_c$ superconductors with the field under
an angle with respect to the insulating layers.

\subsection{Numerical results}
\label{subord1Dnum}

The simulations of symmetric channels ($\Delta x=0$) for $w\sim
b_0$ fully support the above findings. The interaction with the
CE's for $w=b_0$ provides a maximum restoring force with a value
$0.054$, independent of the interaction cut-off $r_c$ used in the
numerics. This value is in agreement with the dimensionless values
for $\mu$ and $c_{66}$ in Eq.(\ref{mudef}) and (\ref{Londonc66}):
$a_0\mu/U_0=1/(6\pi)$ and $c_{66}a_0^2/U_0=1/(4\sqrt{3}))$.

The data points in Fig. {\ref{plot_sgkink}} show the displacement
field of a single defect (obtained by adding one vortex to a
commensurate chain) for three values of the cut-off $r_c$ (i.e.
various $\lambda/a_0$). We conclude that up to $r_c=30$
($\lambda/a_0=9$) the s-G kink shape Eq.(\ref{kinkforms}) forms a
good description of a defect in the chain.

\begin{figure}
\scalebox{0.40}{\includegraphics{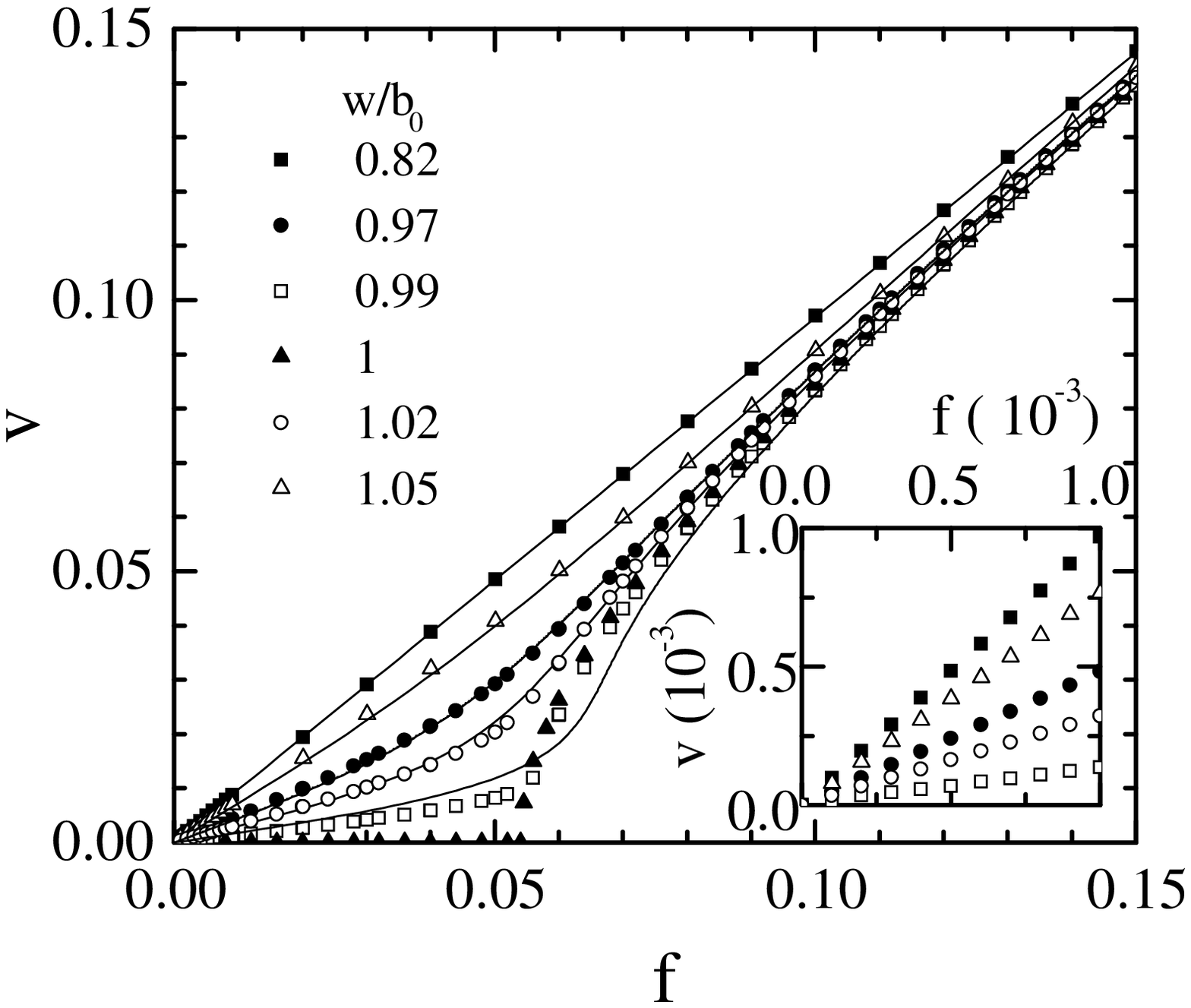}} \vspace{0cm}
\caption{{\it v-f} characteristics for ordered vortex channels
with $w \approx b_0$ and $\Delta x=0$. Symbols are simulation
results, drawn lines are obtained with Eq.(\ref{fvresultana}). The
inset shows an expanded view of the small velocity regime.}
\label{plot_fvord1d}
\end{figure}

\begin{figure}
\scalebox{0.40}{\includegraphics{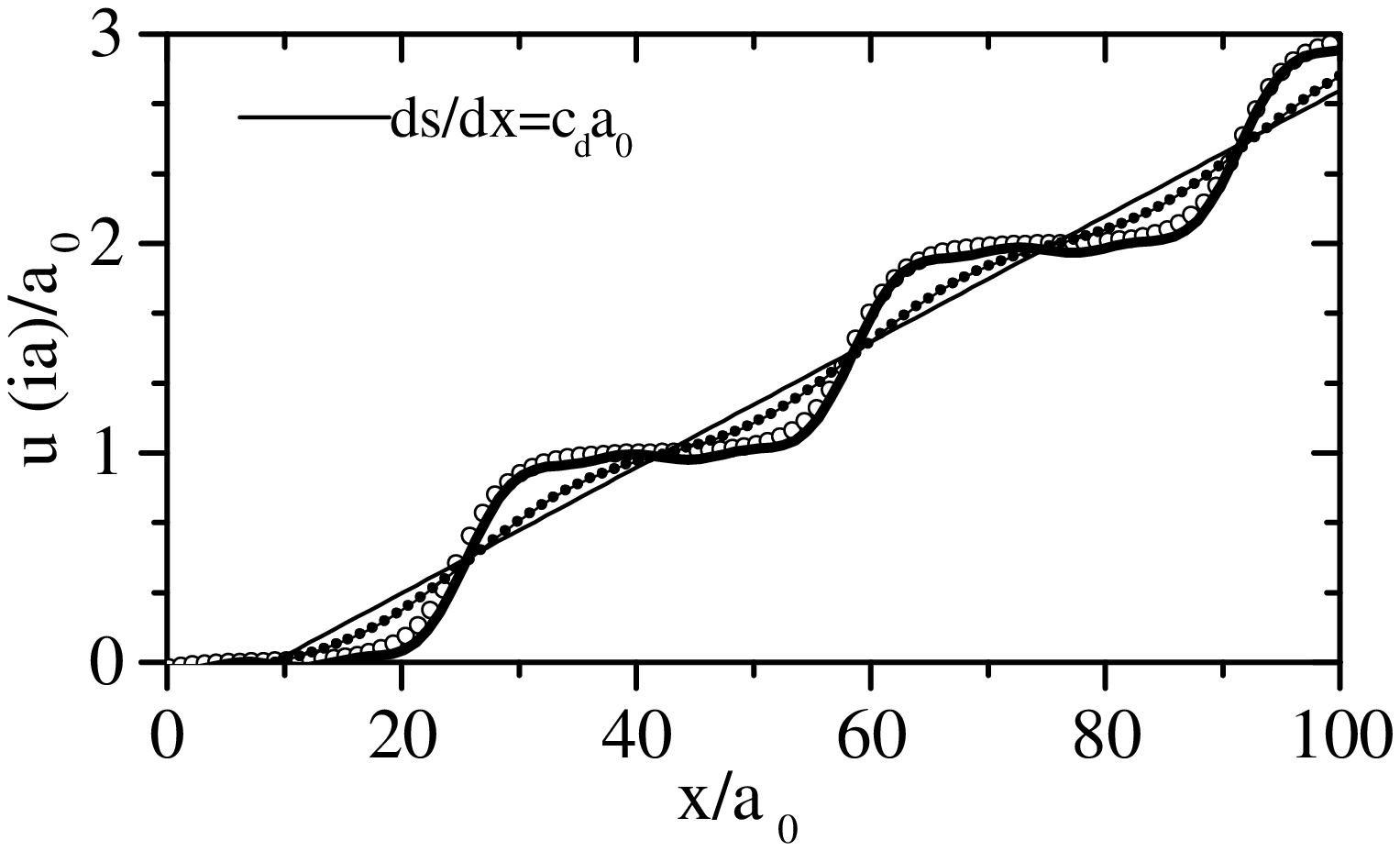}}

\caption{Displacements $u(x)$ along the channel for
$w=0.97b_0$:($\circ$) Numerical result for $f=0$ also representing
a snapshot of the moving chain at low drive, $f=0.01$. The thick
drawn line shows the result for $u$ as calculated from the Fourier
modes given in App.\ref{appB}. ($\bullet$) Displacement field for
$v=0.09$ ($f=0.1$). The data mask a drawn line which is obtained
from Eq.(\ref{largevelrel}) in App.\ref{appB}. The straight drawn
line shows the displacement field $s(x)$ in absence of the
periodic potential.} \label{displ97ord}
\end{figure}

The data points in Fig. \ref{plot_fvord1d} show numerical results
for the transport of a single chain in channels of various widths
and $r_c=3.33$. The features discussed previously, i.e. the
vanishing PN barrier and nonlinear transport, clearly appear in
the data for incommensurate chains. We also plotted the results
according to Eq.(\ref{fvresultana}), with $K_{eff}^2(c_d)$
evaluated using the results in App.\ref{appB} for $\lambda/a_0=1$
and taking into account that $\mu$ slightly depends on $w$. The
analytical treatment gives a very reasonable description of the
data. Finally, we show in Fig. \ref{displ97ord} the numerical
results and analytical results of App.\ref{appB} for the
quasi-static and dynamic shape of the chain for $w/b_0=0.97$
($c_d=0.03/a_0$). The numerical results closely mimic the analytic
results, both for the kinked shape at small $v$ and the core
expansion with the associated reduction of $h$ for large $v$.

To conclude this section we mention that, at incommensurability,
due to the vanishing barrier for defect motion, the average
velocity $\langle \dot{x}_i(t) \rangle_i$ has a vanishing
ac-component. Only at commensurability the washboard modulation is
retained, the velocity at large drive being $\langle v(t)\rangle =
v+(\mu/\gamma)\sin(\omega_0 t)$.

\section{Ordered CE's and multiple chains}
\label{secord2D}

We now turn to the results for channels containing multiple vortex
rows and ordered CE's. The simulations are performed with the full
$2$D degrees of freedom and $r_c=3.33$. We implemented an edge
shift $\Delta x(w)$ with a saw tooth shape ($0\leq \Delta x\leq
a_0/2$). This assures that, as we vary $w$, a perfect hexagonal
structure is retained for $w=pb_0$ with $p$ an integer. However,
for $w\neq pb_0$, the qualitative behavior did not depend on
$\Delta x$.

\begin{figure}
\scalebox{0.40}{\includegraphics{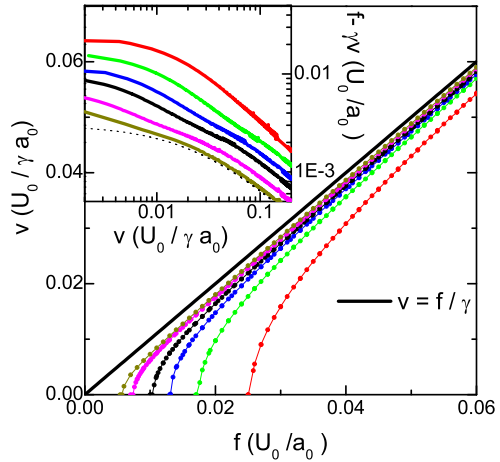}} \vspace{0cm}
\caption{Transport curves for commensurate channels with
$w/b_0=2,3,4,5,7,9$ from right to left. The thick drawn line
represent free flux flow. Inset: friction force $f-\gamma v$ vs.
$v$ for $w/b_0=2,3,4,5,7,9$ from top to bottom. The dotted line
represents Eq.(\ref{widecomfvexpression}) for $n=9$.}
\label{widecompurefv}
\end{figure}

Figure \ref{widecompurefv} shows {\it v-f} curves of commensurate
channels, $w/b_0=n$ with integer $n \geq 2$. In these cases the
arrays are perfectly crystalline and have a shear strength
$f_s=\mu b_0/w$, inversely proportional to the channel width and
in accord with Eq.(\ref{phenomenological}) with
$A=A^0=1/\pi\sqrt{3}$. This is consistent with the fact that only
the first mobile chains within a distance $\sim b_0$ from both
CE's experience the periodic edge potential (see
Eq.(\ref{ordVmb})) while the other chains provide an additional
pulling force via the elastic interaction. This interaction brings
an additional feature to the dynamics, namely shear waves. The
shear displacements of rows $n$ in the bulk of the channel can be
described in continuum form, $u_n(t)\rightarrow u(y,t)$, by the
following equation of motion:
\begin{eqnarray}
\gamma\partial_t u(y,t)=f+c_{66}a_0b_0\partial^2_y u(y,t).
\label{elasheareqm}
\end{eqnarray}
At large $v$ the CE interaction can be represented by oscillating
boundary conditions. As shown in App.\ref{appC}, this causes an
oscillatory velocity component $dh/dt$ with $y$-dependent
amplitude and phase describing periodic lagging or advancing of
chains with respect to each other:
\begin{eqnarray}
\partial_t h(y,t) \sim -f(y)\sin(\omega_0 t) -g(y)\cos(\omega_0 t).
\label{comuhprimeyt}
\end{eqnarray}
Here $\omega_0$ is the washboard frequency $k_0v$,
$f(y)=\cos(y/l_{\perp,v})\cosh(y/l_{\perp,v})$,
$g(y)=\sin(y/l_{\perp,v})\sinh(y/l_{\perp,v})$. The length scale
$l_{\perp,v}=\sqrt{(\mu/\gamma v)}b_0$ explicitly depends on $v$
and represents the distance over which the amplitude and phase
difference decay away from the CE's. Although in principle
Eq.(\ref{comuhprimeyt}) is only valid for $\gamma v/\mu \gtrsim
0.25$, it provides useful qualitative insight in the dynamics at
all velocities: at small velocity $l_{\perp,v}$ is large, meaning
that for all rows the velocity modulation and phase become
similar. Hence, for $v \rightarrow 0$ the array may be described
as a single vortex chain, which is the underlying origin of the
fact that close to threshold the curves approach the 1D
commensurate behavior $v=\sqrt{f^2-f_s^2}$ with reduced threshold
$f_s=\mu/n$. At large velocity, $l_{\perp,v}$ eventually becomes
less than the row spacing. In that limit only the two chains
closest to the CE experience a significant modulation. In
App.\ref{appC} we quantitatively analyze the friction force in
this regime with the result:
\begin{eqnarray}
f-\gamma v = \frac{\mu^2}{2n(2\gamma v+\mu)}.
\label{widecomfvexpression}
\end{eqnarray}
In the inset of Fig. \ref{widecompurefv} this behavior is
displayed for $n=9$ by the dotted line. In the high velocity
regime the result agrees well with the numerical data, at lower
velocities Eq.(\ref{widecomfvexpression}) underestimates the true
friction.

\begin{figure}
\scalebox{0.40}{\includegraphics{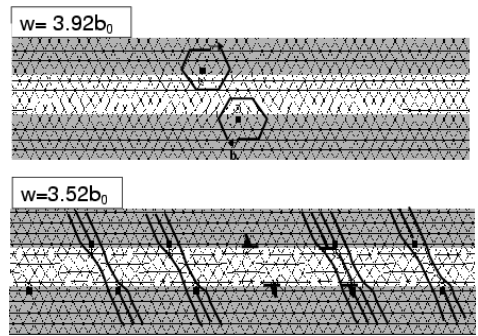}} \vspace{0cm}
\caption{Delauney triangulation of the static structure for two
incommensurate channels: $w/b_0=3.92$ and $w/b_0=3.52$. Open
circles and filled squares denote $7$ and $5$-fold coordinated
points, respectively. The construction for the Burgers vector is
shown for $w/b_0=3.92$; the drawn lines for $w/b_0=3.52$ mark the
transverse stacking faults.} \label{392tria352stat}
\end{figure}

Next we discuss the behavior of incommensurate channels. The
static vortex configuration for a channel of width $w/b_0=3.92$ is
shown in the upper part of Fig. \ref{392tria352stat}. A Delauney
triangulation shows that the array consists of 4 rows with two
pairs of 5,7-fold coordinated vortices at the CE constituting two
misfit dislocations of opposite Burgers vector $\vec{b}$ and glide
planes along $x$. Due to their mutual attraction, dislocations at
the upper and lower CE are situated along a line with an angle of
$\sim 60^{\circ}$ with $\vec{x}$. The two edge dislocations thus
form a 'transverse' stacking fault (TSF). In the lower part of
Fig. \ref{392tria352stat} the structure for a channel with
$w/b_0=3.52$ is shown. Here the density of stacks, given by
$c_{TSF}=|(1/a_0)-(1/a)|$, is enhanced.  The dislocations at one
side of the CE repel each other and are equally spaced, like the
periodic superstructure for a single chain in Fig.
\ref{displ97ord}. The slight misalignement between the 'upper' and
'lower' dislocations of a pair is due to the relative shift
between the CE's: the exact orientation of the pairs is determined
by the choice of $\Delta x$. For channel widths in the regime
$n<w/b_0\lesssim n+1/2$ with integer $n$, we find very similar
structures but instead of TSF's consisting of vacancies, we now
have TSF's consisting of {\it interstitial vortices}, again
arranged in a periodic superstructure.

\begin{figure}
\scalebox{0.40}{\includegraphics{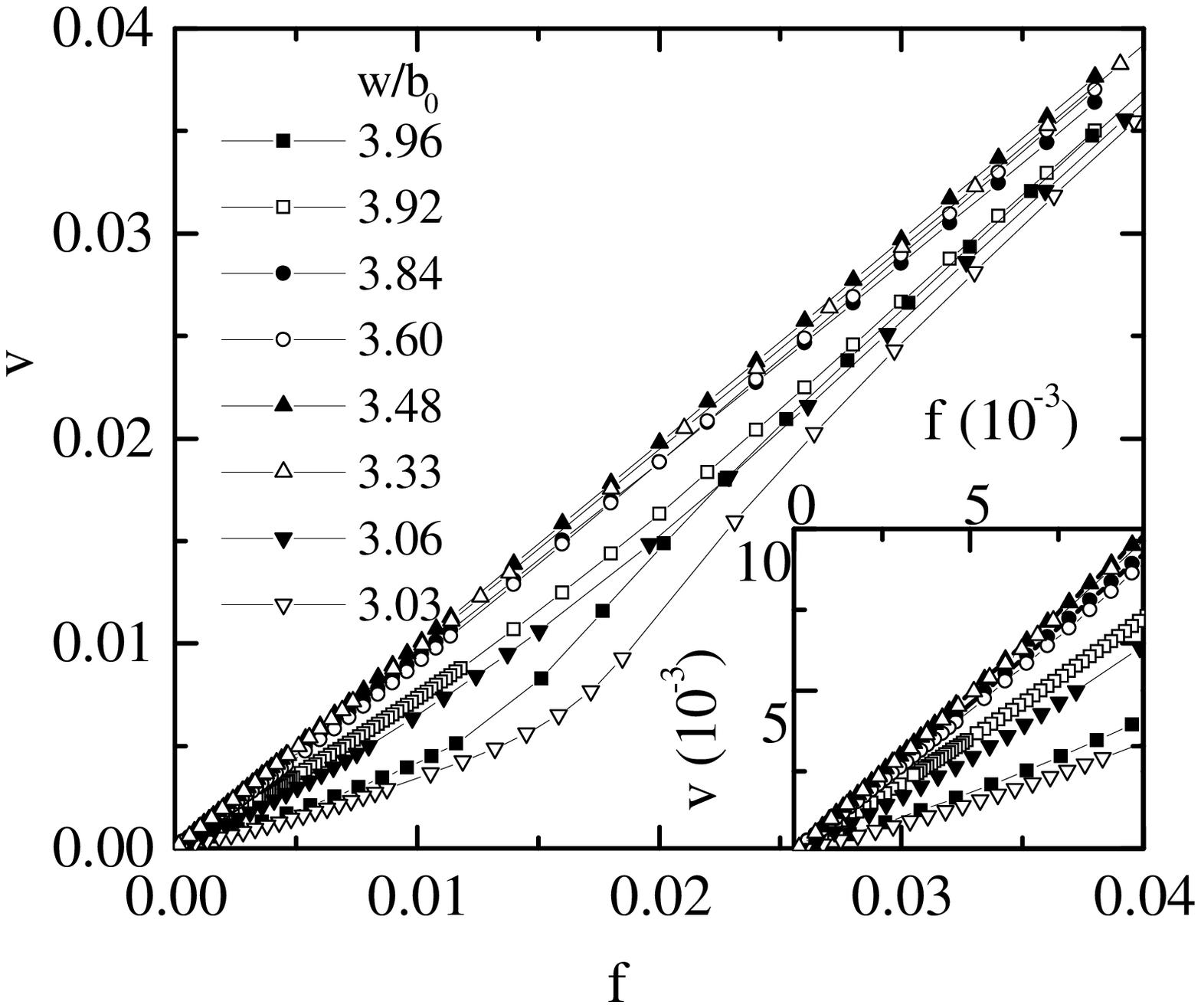}} \vspace{0cm}
\caption{{\it v-f} curves for incommensurate channels. Inset:
regime of small drive.} \label{fvwideincom}
\end{figure}

Figure \ref{fvwideincom} shows the transport curves associated
with these structures. As for the single chain, the presence of
misfit defects causes an essentially vanishing threshold force.
For small drive, $f< \mu b_0/w$, a low mobility regime occurs
associated with glide of the edge dislocation pairs along the CE.
This allows for elastic motion of a complete TSF, i.e. the
vortices in the 'bulk' of the channel remain $6$-fold coordinated.

\begin{figure}
\scalebox{0.40}{\includegraphics{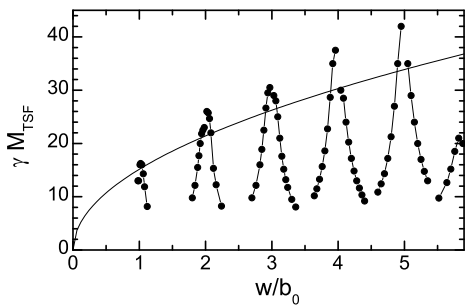}}

\vspace{0cm} \caption{The mobility per stack
$M_{TSF}=(dv/df)/c_{stack}$ versus $w/b_0$. The drawn line shows
the predicted form Eq.(\ref{M_TSFformula}) for $\lambda/a_0=1$.}
\label{plot_M_TSF}
\end{figure}

It is interesting to study how the mobility due to the TSF's
changes on increasing the number of rows. In Fig. \ref{plot_M_TSF}
we plot the mobility per stack, $M_{TSF}=(dv/df)_{f\rightarrow
0}/c_{TSF}$ versus channel width. $M_{TSF}$ around each peak
decreases with increasing $c_{TSF}$. This is caused by overlap of
the strain fields of the defects, in analogy with the behavior for
a single chain. The overall increase of the peak value of
$M_{TSF}$ is related to a change in the size of an isolated TSF.
An extension of the analysis in Sec.\ref{secord1D} allows to
describe this change quantitatively. For small $n$ the
longitudinal deformations do not vary strongly over the channel
width. This can be understood by considering shear and compression
deformations related by the equation $\kappa\partial_x^2
u_x+c_{66}a_0b_0
\partial_y^2 u_x=0$. It follows that a longitudinal
deformation on a scale $l_{\parallel}$ along the channel varies
over a scale $l_{\perp}=l_{\parallel}\sqrt{c_{66}a_0b_0/\kappa}$
perpendicular to the channel. In case $l_{\perp} \gtrsim w$, the
transverse variation of $u_x(y)$ is small and can be neglected so
that $\kappa_0$ in Eq.(\ref{ordconteqm}) can be replaced by an
effective stiffness $n\kappa_0$ due to $n$ rows. Similarly, the
driving force is replaced by $f\rightarrow nf$. This results in
the same equation (\ref{ordconteqm}) with a rescaled edge force
$\mu \rightarrow \mu/n$. Accordingly, the longitudinal size of a
defect (TSF) is given by $l_{TSF}=2\pi a_0 \sqrt{ng}$ and the
mobility of an isolated TSF by (compare $M_d$ below
Eq.(\ref{linearkinkresist})):
\begin{eqnarray}
M_{TSF}\simeq \pi^2 \sqrt{ng}/2\gamma. \label{M_TSFformula}
\end{eqnarray}
As shown by the drawn line in Fig. \ref{plot_M_TSF}, this form
gives a reasonable description of the data up to $n=3$. Working
out the condition $l_{\perp} \gtrsim w$ given above for the
validity of Eq.(\ref{M_TSFformula}), one obtains $w \lesssim
(l_d/2) \sqrt{c_{66}a_0b_0/\kappa}\simeq 3 b_0$, in agreement with
the data. At larger $n$, 'bulk-mediated' elasticity
\cite{CuleHwaPRB98} leads to decay of the longitudinal
deformations towards the channel center. We also note that, due to
the increase of $l_{TSF}$ with $n$, the density $c_{TSF}$ for
which defects are non-overlapping, decreases on increasing $n$.

In the {\it v-f} curves of Fig. \ref{fvwideincom} we observe at
larger velocity features very similar to the transport of the $1$D
chain: for $f\gtrsim \mu/n$ the effective barrier is reduced,
leading to core expansion of the TSF's. Accordingly, the curves
approach free flux flow behavior. As in the commensurate case,
this approach is initially more slowly then $f-\gamma v\sim 1/v$
due to additional oscillating shear deformations in the channel
for $f \gtrsim \mu/n$.

\begin{figure}
\scalebox{0.40}{\includegraphics{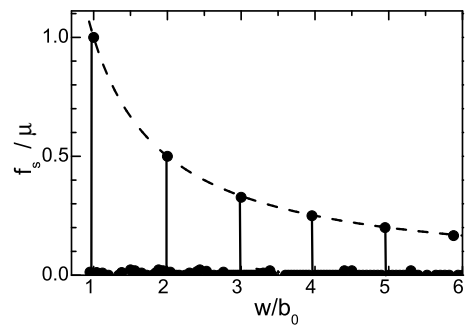}}

\vspace{0cm} \caption{Threshold force versus $w/b_0$ for ordered
channels. The dashed line represents Frenkels prediction for an
ideal lattice in the continuum limit.} \label{fcwpure}
\end{figure}

Figure \ref{fcwpure} summarizes the behavior of the shear force
$f_s$, taken at a velocity criterion $v \approx 0.01 \mu/\gamma$,
versus the matching parameter. At integer $w/b_0=n$, the threshold
is $f_s=\mu/n$, but we note that it can be reduced due to a finite
edge shift $\Delta x$. At mismatch $f_s$ is essentially vanishing.
Near 'half filling', $w/b_0\simeq n \pm 1/2$, where the arrays
switch from $n$ to $n \pm 1$ chains, a small enhancement of $f_s$
is observed. In this regime, the static ($f=0$) structure was
obtained by annealing from a random initial configuration,
attempting to determine the exact switching point. This results in
{\it metastable} structures with coexisting $n$ and $n\pm 1$ row
regions (or longitudinal stacking faults, LSF's) bordered by
dislocations with misoriented Burgers vector, see
\cite{BesselingPRL99}. The increase in $f_s$ is caused by the
finite barrier for climb-like motion of these dislocations, via
which an LSF can move as 'giant' defect through the channel. For
sufficiently large drive, (part of) the LSF's are annealed which
may result in hysteresis for up/down cycled {\it v-f} curves. We
will discuss these 'mixed' $n/n\pm1$ structures in more detail in
Secs.\ref{secweakdis2D} and \ref{secstrongdis2D} in the context of
disordered CE's. We also note that the integer chain structures
with TSF's away from half filling differ from the results in
\cite{BesselingPRL99}. The structures there, obtained from a
random initial configuration, contained point defects unequally
distributed among rows, yielding 'gliding' dislocations within the
channel. Such structures are also slightly metastable but the
conclusion of vanishing $f_s$ for incommensurate, integer chain
structures, drawn in \cite{BesselingPRL99}, remains unaltered.

\section{Single chain in a disordered channel}
\label{secdis1D}

We will now consider the influence of disorder in the CE arrays on
transport in the channels, focussing in this section on the
characteristics of a single chain for $w/b_0 \sim 1$ with only
longitudinal degrees of freedom. The CE disorder is implemented
with longitudinal random shifts as described in
Sec.\ref{secmodel}. We note that both CE's remain 'in phase'; the
effect of quenched phase slips or dislocations between the CE's
will be treated in the discussion in Sec.\ref{secdiscuss}.

\subsection{Disordered sine-Gordon equation}
\label{subdis1deqm}

First we consider the form of the channel potential in presence of
weak disorder. To that end we generalize Eq.(\ref{Vedge}) in
App.\ref{appA} and express the CE potential at ${\bf r_0}=(x,y=0)$
in terms of the vortex density $\rho_e$ in the CE's:
\begin{eqnarray}
V_{ce}({\bf r_0})=(2\pi)^{-2}\int d{\bf k}V({\bf k})\rho_e({\bf
k})e^{i{\bf k}\cdot{\bf r_0}}, \label{Vedgeformal}
\end{eqnarray}
with $\rho_e({\bf k})$ the Fourier transform of $\rho_e$. For weak
disorder ($\nabla\cdot {\bf d}\ll 1$), this density can be
expressed in terms of the displacement field ${\bf d}$ in the CE
as follows \cite{GiamarchiPRB95}: $\rho_e({\bf r}_e,{\bf d})\simeq
(B/\Phi_0)(1-\nabla\cdot{\bf d}+\delta \rho_e)$ where $\delta
\rho_e=\sum_i\cos[{\bf K}_i({\bf r}_e-{\bf d}({\bf r}_e))]$
represents the microscopic modulation due to the lattice (${\bf
K}_i$ spans the reciprocal lattice) while $\nabla\cdot{\bf d}$
reflects density modulations. As described in App.\ref{appD}, this
decomposition of $\rho_e$ leads to two contributions to the
potential:
\begin{eqnarray}
V_{ce}=&&V_l(x)+V_p(x)\nonumber\\
=&&-(B/\Phi_0)\int d{\bf r}_eV({\bf r_0}-{\bf r}_e)\nabla \cdot
{\bf d}({\bf r}_e)\nonumber\\
&&-[\mu+\delta \mu(x)]\cos[k_0(x-d)]/k_0, \label{1Ddisordpotform}
\end{eqnarray}
where in the second term $\delta \mu(x)/\mu=\pi \sqrt{3}\partial_x
d$. The term $V_l$ represents long range potential fluctuations
and is smooth on the scale $\sim a_0$. Its correlator
$\Gamma_l(s)=\langle V_l(x)V_l(x+s) \rangle$ is derived in
App.\ref{appD}. Assuming that $\partial_x d$ has short range
correlations (on the scale $\sim a_0/2$) and a variance $\langle
(\partial_x d)^2 \rangle=\Delta^2/3$ as in the simulations,
$\Gamma_l$ can be written as:
\begin{eqnarray}
\Gamma_l(s)\simeq C_{\alpha}\Delta^2U_0^2(\lambda/a_0)^{1+\alpha}
e^{-(\frac{s}{\lambda})^2}. \label{correlator}
\end{eqnarray}
The exponent $\alpha$ and the prefactor $C_{\alpha}$ depend on the
disorder correlations between rows in the edge: $\alpha=2$ when
the strain $\partial_x d(x)$ is identical for all rows and
$\alpha=1$ when the strain is uncorrelated between rows. The term
$V_p$ in Eq.(\ref{1Ddisordpotform}) is the quasi-periodic
potential arising from $\delta \rho_e$ of the vortex rows nearest
to the CE's. The amplitude fluctuations $\delta\mu /k_0$ are
characterized by (see App.\ref{appD}):
\begin{eqnarray}
\Gamma_a(s)=\frac{\langle \delta \mu (x) \delta \mu
(x+s)\rangle_x}{k_0^{2}} \simeq (\mu \Delta a_0/2)^2
e^{-(\frac{2s}{a_0})^2} \label{deltamucorrel}
\end{eqnarray}

To obtain the energy of the vortex chain and the equation of
motion, the vortex density inside the channel, $\rho_c$, is
decomposed similar to $\rho_e$: $\rho_c(x,u)=a_0^{-1}[1-\partial_x
u+\delta\rho_c(x,u)]$, where $u$ is the displacement field of the
chain. As shown in App.\ref{appD}, in the limit $\lambda> a_0$ the
resulting interaction with the CE's can be written as
$H=H_{SG}+H_a+H_s$ where $H_{SG}=a_0^{-1}\int dx
[(\kappa_0/2)(\partial_x u)^2-(\mu/k_0) \cos(k_0 u)]$ represents
the sine-Gordon functional for an ordered channel, and $H_a$,
$H_s$ are the disorder contributions due to amplitude fluctuations
and random coupling to the strain:
\begin{eqnarray}
H_a=-\int \frac{dx}{a_0}\frac{\delta \mu(x)}{k_0}\cos(k_0 u) \nonumber\\
H_s=-\int \frac{dx}{a_0} V_s(x)\partial_x u.
\label{textformHdistotal}
\end{eqnarray}
The term $V_s(x)=V_l(x)-\kappa_0
\partial_x d(x)$ contains contributions from local and nonlocal strains. The
latter dominates for $\lambda > a_0$ (see App.\ref{appD}). Hence
$\Gamma_s(s)=\langle V_s(x)V_s(x+s) \rangle \simeq \Gamma_l(s)$.

The model described by $H=H_{SG}+H_{a}+H_{s}$ is also used to
describe LJJs or commensurate CDW's with weak disorder, however
with different disorder correlations. In the former case, the term
$H_a$ in Eq.(\ref{textformHdistotal}) describes local variations
in the junctions critical current
\cite{MalomedPRB89_JosCDWrapid,VinokurJETP90_disordJosJun}. For
CDW's, a disorder contribution of the form $H_a$ arises from so
called backward scattering impurities, while the term $H_s$
originates from 'forward' scattering impurities
\cite{FeigelVinCDW}. We also note that our model differs from the
usual Fukuyama-Lee-Rice model for CDW's \cite{FukuLeeRice}, in
which commensurability is ignored either due to strong direct
random coupling to $u$ ($\delta \mu(x) \gg \mu$) or due to large
mismatch.

In principle, the equation of motion for the chain is given by
$\gamma \partial_t u=-\delta H/\delta u$. However, it has been
shown in previous studies
\cite{CuleHwaPRB98,ChenPRB96,KrugPRL95,BalentstempPRL95} that in
the moving state a convective term $-\gamma v\partial_x u$ should
be included. While irrelevant for the depinning process, such term
can be important for the dynamics and for completeness we include
it \cite{fn_convective term}. The resulting equation of motion is:
\begin{eqnarray}
\gamma\partial_t u=&& f+\kappa_0\partial_x^2u -[\mu+\delta \mu(x)] \sin(k_{0}u) \nonumber \\
&& -\partial_xV_s-\gamma v\partial_x u. \label{contdiseqm}
\end{eqnarray}
In writing Eq.(\ref{contdiseqm}) we have assumed, for simplicity,
that the elastic deformations in presence of disorder can be
described by the long wavelength stiffness $\kappa_0$. Ignoring
the last term, Eq.(\ref{contdiseqm}) describes the transverse
displacements $u(x)$ of an elastic string in a tilted 'washboard'
potential with random amplitude $\mu(x)/k_0$ and random phase
$\phi(x)=-\int_{-\infty}^x dx'V_s(x')/\kappa_0$. The latter
represents a $u$ independent random deformation of the chain.

\subsection{Numerical results}
\label{subdis1Dnum}

The influence of disorder on the threshold force and the dynamics
of the chain are directly visible in numerical simulations. The
simulations were performed using $r_c=3.33a_0$ and channels of
length $L \geq 1000 a_0$.

\begin{figure}
\scalebox{0.40}{\includegraphics{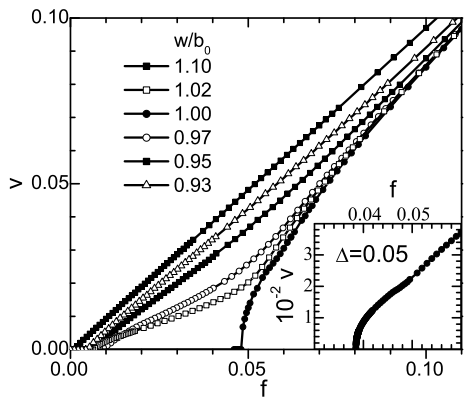}} \vspace{0cm}
\caption{{\it v-f} curves of a commensurate and various
incommensurate chains for weak disorder $\Delta=0.025$. Inset:
{\it v-f} curve of a commensurate channel for $\Delta=0.05$}
\label{plot_fv1Dweakdis}
\end{figure}

\begin{figure}
\scalebox{0.5}{\includegraphics{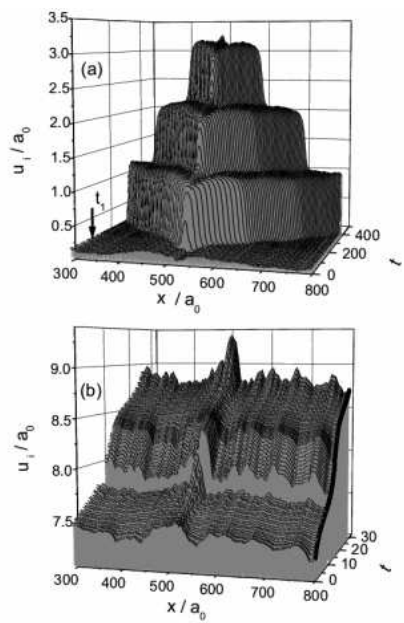}}

\caption{(a) Evolution of {\em longitudinal} displacements
$u_i(t)$ for the commensurate chain in Fig.
\ref{plot_fv1Dweakdis}($\Delta=0.025$), plotted for clarity in a
transverse way vs. $x$. At $t=t_1$ the force is increased above
threshold. (b) Stationary evolution of $u_i(t)$ for large drive
($f=0.08$) showing the motion over a distance $\sim1.8a_0$ (the
$t$ and $u$ axis have arbitrary offset, and a few frames around
$t=10$ were omitted for clarity). } \label{plot_nuccomdisord}
\end{figure}

Figure \ref{plot_fv1Dweakdis} shows several {\it v-f} curves for
channels with $0.93<w/b_0<1.1$ at a disorder strength
$\Delta=0.025$. We first focus on the result for the commensurate
case $w=b_0$. The disorder leads to a significant reduction of the
threshold $f_s$ with respect to the pure value $\mu=0.054$. The
reduction is enhanced on increasing $\Delta$, as shown for
$\Delta=0.05$ in the inset. The origin of the reduction is that
disorder lowers the energy barrier for formation of
vacancy/interstitial (kink/antikink) pairs in the chain. Figure
\ref{plot_nuccomdisord}a shows the time evolution of the
displacements $u_i=x_i-ia_0$ upon a sudden increase of $f$ to a
value $f=0.049>f_s$ at $t_1$. For $t<t_1$ $u$ is 'flat' and the 2D
crystal formed by the chain and the CE's is topologically ordered.
At $t=t_1$, the motion starts at an unstable site (at $x/a_0
\simeq 500$) by nucleation of a vacancy/interstitial pair visible
as steps of $\pm a_0$ in $u$. We henceforth denote the force at
which this local nucleation occurs by $f_n$. The defects are
driven apart by the applied force and when their spacing becomes
$\sim l_d$ a new pair nucleates at the same site. This process
occurs periodically with rate $R_n$, leading to the formation of a
domain with defect density $c_d=R_n/\langle v_d \rangle$ and a net
velocity $v=c_d \langle v_d \rangle a_0=R_na_0$ with $\langle v_d
\rangle$ the average defect velocity. In the present case of weak
disorder $\langle v_d \rangle$ is essentially the same as for
$\Delta=0$. For a further increase of the force to $f=0.053$ an
increase of the nucleation rate is observed. In
Ref.\cite{BesselingEPL2003} we showed that in larger systems
coarsening occurs in the initial stage of depinning due to a
distribution of unstable sites. However, after sufficiently long
times the stationary state consists of one domain around the site
with the largest nucleation rate (smallest local threshold
$f_n^{min}$) with vacancies travelling to the left and
interstitials to the right. It is interesting to compare this to a
study of CDW's with competing disorder and commensurability
pinning \cite{ChenPRB96}. Using a coarse grained version of
Eq.(\ref{contdiseqm}) it was found in \cite{ChenPRB96} that in the
pinning dominated, low velocity regime, the so-called interface
width $W(L)=\sqrt{\langle (u(x)-\langle u\rangle)^2\rangle_x}$
grows linearly with the system size $L$. The mechanism of defect
nucleation which we observe naturally explains this phenomenon. In
addition, we found that at depinning the average velocity
$v=R_n^{max}a_0$ can be described by $R_n^{max}\propto
(f-f_n^{min})^\beta$ with a depinning exponent $\beta=0.46\pm
0.04$, similar as previously reported for 1D periodic media
\cite{MyersPRB93}.

The defective flow profile does not persist up to a arbitrary
large forces. In the commensurate {\it v-f} curve in Fig.
\ref{plot_fv1Dweakdis} and its inset, a small kink is observed for
$f\lesssim \mu$. Associated with this kink we find a transition to
a much more ordered state. We have illustrated the temporal
evolution of vortex displacements in this state in Fig.
\ref{plot_nuccomdisord}(b) for $f=0.08$. The 'staircase' structure
has vanished and the relative vortex displacements are greatly
reduced. In fact, in the above mentioned study Ref.
\cite{ChenPRB96} a very similar transition in the CDW dynamics was
found, and was shown to be of first order. We leave the precise
dependence of this transition on disorder and vortex interactions
in our channels for future studies.

We now turn to the incommensurate case. The  {\it v-f} curves with
$w \neq b_0$ in Fig. \ref{plot_fv1Dweakdis} all exhibit a finite
threshold instead of the vanishing threshold for the
incommensurate channels without disorder (Fig.
\ref{plot_fvord1d}). With disorder the defects that are present in
the channel, couple to the disorder in the CE, which causes a
pinning barrier $f_d$. This barrier has a distribution along the
channel $\{f_d\}$ and maximum value $f_d^{max}$. We now focus on
the curves with small defect density $c_d\lesssim 1/l_d$ for which
the defects are individually pinned. In this regime the threshold
force $f_s$ satisfies $f_s \lesssim f_d^{max}$.

\begin{figure}
\scalebox{0.35}{\includegraphics{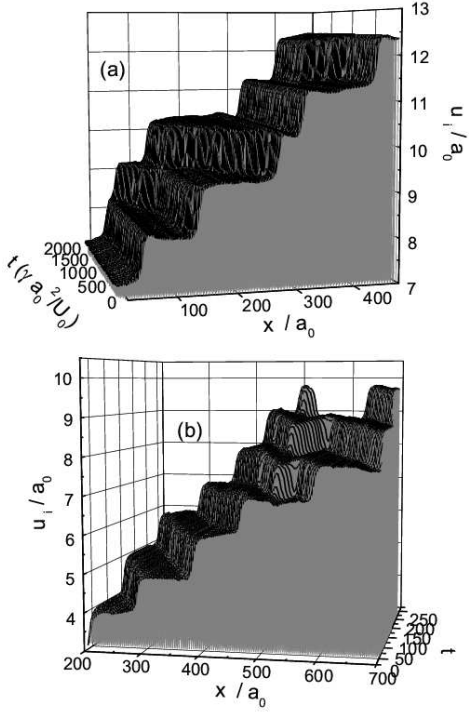}}

\caption{(a) Evolution of displacements for $w/b_0=0.99$ and
$f=0.013>f_s$ for $t>0$. Defects at $x\approx 50a_0$, $x\approx
150a_0$ and $x\approx 325a_0$ are initially pinned while the
others are mobile. A defect collision-release process occurs at
$x\approx150a_0$ and $t\approx 1500$. (b)Evolution of
displacements when $f$ is suddenly increased to $f=0.048$.
Nucleation is observed for $x\approx 500a_0$.}
\label{plot_uincomweakdis}
\end{figure}

The precise threshold behavior depends on the distribution of
barriers $\{f_d\}$, similar as for LJJ and CDW systems
\cite{MalomedPRB89_JosCDWrapid}. As an illustration, we show in
Fig. \ref{plot_uincomweakdis}(a) the evolution of displacements
for a channel with $w/b_0=0.99$ for a force just above threshold
$f(t>0)>f_s$. The static configuration at $t=0$ ($f<f_s$) shows
that the disorder breaks the periodicity of the 'soliton' chain.
For $t>0$, depinning starts with the defect at $x\simeq 270 a_0$
and proceeds via a 'collision-release' process between the moving
defect and its pinned neighbor. Thus, for $f \gtrsim f_s$ strong
local variations in the defect mobility exist and the overall
chain velocity depends strongly on the distribution $\{f_d\}$ (of
which we show an example below). However, as seen in the {\it v-f}
curves in Fig. \ref{plot_fv1Dweakdis}, for $f \gtrsim 2f_d^{max}$
these effects vanish and the mobility approaches $dv/df \simeq c_d
a_0 M_d$ with $M_d$ the defect mobility without disorder. Another
feature in the {\it v-f} curves for small defect densities is the
velocity upturn at a force $f\simeq f_n^{min}<\mu$. It is caused
by nucleation of {\it new defect pairs} in the incommensurate
chain. The start of such a process is illustrated in Fig.
\ref{plot_uincomweakdis}(b): at $x\simeq 500a_0$ the chain is
unstable against pair nucleation and the nucleated
interstitials/vacancies are formed 'on top of' the moving
incommensurate structure. This process only occurs at small defect
densities when the time between passage of {\em existing} defects
$\simeq 1/(v_dc_d)$ exceeds the nucleation time $R_n^{-1}$. For
$f\gtrsim \mu$, the structure of both defects disappears again.
The resulting dynamic state resembles that of the large velocity
profile shown in Fig. \ref{displ97ord}, but with additional
'roughness' due to the weak CE disorder.

The {\it v-f} curves in Fig. \ref{plot_fv1Dweakdis} at
$w/b_0=0.93$ and $w/b_0=1.1$, for which the defect density in the
chain is larger, exhibit a smaller threshold force. In this regime
the interaction between defects starts to become important and
$f_s$ is determined by {\it collective} pinning of the defects.
This situation was studied analytically for the case of Josephson
vortices in a disordered LJJ in \cite{VinokurJETP90_disordJosJun}.
We will not consider this situation explicitly but we note here
that, as the disorder and the typical pinning force on the defects
increases, the onset of the collective pinning regime shifts to
larger defect density, where defect interactions are stronger
\cite{VinokurJETP90_disordJosJun}.

\begin{figure}
\scalebox{0.4}{\includegraphics{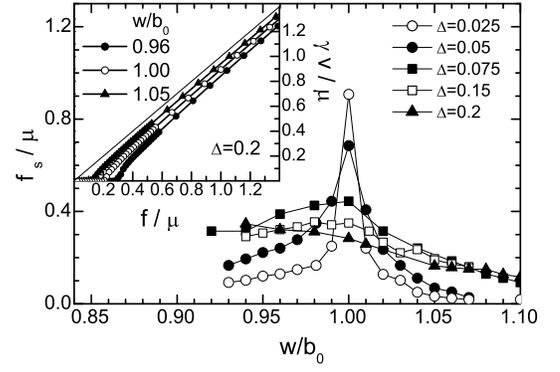}}
\vspace{0cm}\caption{Threshold $f_s$, obtained from a velocity
criterion $v \approx 0.025 \mu/\gamma$ and $L=1000a_0$, versus
$w/b_0$ for several disorder strengths. Data were averaged over
$5$ disorder realizations. Inset: {\it v-f} curves in the strong
disorder regime $\Delta=0.2$.} \label{plot_fs1DversuswDel}
\end{figure}

\begin{figure}
\scalebox{0.50}{\includegraphics{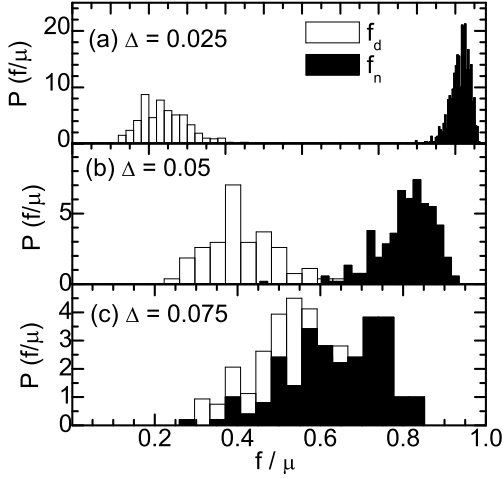}}
\vspace{0cm} \caption{Probability density of critical forces in
channels of length $L=100a_0\approx 5l_d$ and $w=b_0$ for a
commensurate chain ($\bullet$) and for a chain with one vacancy
($\circ$) for $\Delta=0.025$ (a), $\Delta=0.05$ (b) and
$\Delta=0.075$ (c).} \label{plot1Dpinstatistics}
\end{figure}

In Fig. \ref{plot_fs1DversuswDel} we show the dependence of $f_s$
on channel width, both for the weak disorder regime treated above
and for larger disorder. The data at $\Delta=0.025$ exhibit a
sharp peak at $w=b_0$, reflecting the gap between minimum
nucleation threshold and maximum defect pinning force. Larger
disorder however rapidly smears the peak, being eventually
completely suppressed for $\Delta \gtrsim 0.15$. The origin of
this behavior is a spontaneous nucleation of defects in the static
chain at larger disorder. This is conveniently illustrated via the
changes in the distribution of the individual defect pinning force
$\{f_d\}$ and that of the nucleation threshold $\{f_n\}$ for
increasing disorder strength, shown in Fig.
\ref{plot1Dpinstatistics}. The data were obtained by simulating
hundreds of short channels ($L=100a_0$) both with one vacancy,
yielding $\{f_d\}$, and without 'geometric' defects, yielding
$\{f_n\}$. While for $\Delta=0.025$ the distributions are
separated (formally, in infinite systems, such separation only
exist for bounded disorder, see the next section), for larger
disorder they start to overlap and they become nearly identical
for $\Delta=0.075$. This implies that nucleated defect pairs at
$w=b_0$ can remain pinned, while at incommensurability defects may
be nucleated before the 'geometrical' defects are released. In
other words, regardless of the matching condition the {\it static}
configuration contains both kinks and antikinks.

\begin{figure}
\scalebox{0.40}{\includegraphics{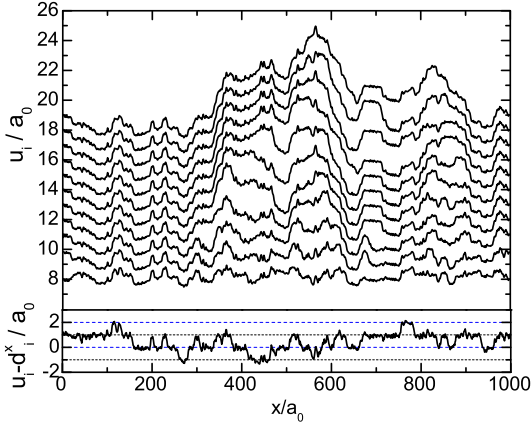}}
\caption{Lower panel: relative displacements $u_i-d^x_i$ for
$\Delta=0.15$ and $f<f_s\simeq 0.0165$ at commensurability
$w=b_0$. Upper panel: evolution of displacements at depinning,
$f=0.017$. The time increment between consecutive snapshots is
$\Delta t=10$ and for clarity each snapshot has been shifted up by
$a_0$.} \label{plot_Del15_1d_ucomdisord}
\end{figure}

While for bounded disorder static defects first appear at a
disorder strength defined as $\Delta_c$, the complete collapse of
the peak in $f_s$ is associated with the presence of a {\it
finite} density of disorder induced defects, of the order of the
inverse kinkwidth $\sim l_d^{-1}$. We define the disorder strength
at which this occurs as $\Delta_*$, here $\Delta_* \simeq 0.15$.
An example of the displacement fields for this disorder strength
is shown in Fig. \ref{plot_Del15_1d_ucomdisord} for $w=b_0$. The
lower panel shows the displacements for $f\lesssim f_s$, relative
to the displacements in the CE. Clearly, the static configuration
has numerous defects. In general, the approach to the critical
pinned state occurs by avalanches in which local nucleation and
repinning, i.e. nonpersisting nucleation events, drive the
rearrangements. The upper panel displays the evolution of
displacements above threshold, revealing a growth of 'mountains'
due to persistent nucleation, superimposed on a disordered
background.

The effect of large disorder on the shape of the {\it v-f} curves
is shown in the inset to Fig. \ref{plot_fs1DversuswDel}. All
curves now exhibit essentially linear behavior
\cite{fn_fvchainstrongdisorder}, except in a small regime just
above $f_s$. We note that also in this disordered regime a gradual
transition to a smoother displacement field occurs at larger
forces, similar to the dynamic transition found for CDW's. Going
back to Fig. \ref{plot_fs1DversuswDel} we should also mention the
overall asymmetry of $f_s$ with respect to $w/b_0=1$ and the
slight decrease of $f_s$ on increasing $w/b_0$ at large disorder.
These effects are unrelated to the competition between
commensurability pinning and disorder discussed so far, but simply
reflect the overal decrease of the edge potential for larger
width.

\subsection{Analysis of pinning forces and crossover to strong disorder}
\label{subdis1Dana}

Using the results of Sec.\ref{subdis1deqm}, we now analyze in more
detail the dependence of the pinning force on disorder and the
vortex interaction range. We focus on the {\it average} pinning
strength of isolated defects, which we derive here in a
semi-quantitative fashion (the formal calculation is deferred to
App.\ref{appD}). Our analysis applies to the case of weak
disorder, i.e. we assume that the defect shape is unaffected by
disorder \cite{VinokurJETP90_disordJosJun}). Extrapolation to
larger disorder provides a useful estimate for the crossover value
$\Delta_*$ at which the commensurability peak vanishes. We
conclude the section with a summary of previous results
\cite{BesselingEPL2003} for the threshold forces $f_n^{min}$ and
$f_d^{max}$ in the special case of bounded disorder.

The disorder correction Eq.(\ref{textformHdistotal}) to the energy
of the vortex chain consists of a term $H_a$, due to amplitude
fluctuations in the periodic potential, and a term $H_s$, due to
random coupling to the strain. We first evaluate the typical
pinning energy of a defect $\sqrt{\langle E_a^2 \rangle}$ due to
the amplitude fluctuations. The local fluctuations are assumed to
be uncorrelated on a length scale $a_0$, and have a variance
$\langle (\delta \mu/k_0)^2 \rangle$. Hence, for a defect in the
chain, which extends over a range $l_d$, the resulting random
potential has a variance $\langle E_a^2 \rangle \simeq \langle
(\delta \mu/k_0)^2 \rangle(l_d/a_0) \simeq \mu^2
\Delta^2l_da_0/4$. The typical pinning force on a defect is then
given by $\sqrt{\langle E_a^2 \rangle}/l_d$ which reduces to:
\begin{eqnarray}
\sqrt{\langle f_a^2 \rangle}\simeq  0.2 \mu \Delta g^{-1/4}
\label{rmsdefpinback}
\end{eqnarray}

The typical pinning energy $\sqrt{ \langle E_s^2 \rangle}$ of a
defect due to the term $H_s$ in Eq.(\ref{textformHdistotal}) is
estimated in a similar way. The mean square energy due to coupling
of a single fluctuation in $V_s$ to the strain of a defect is
$\sim \Gamma_s(0) (r_d/a_0)^2(2a_0/l_d)^2$ where $r_d$ is the
range of $\Gamma_s$, given below Eq.(\ref{textformHdistotal}), and
$a_0/l_d$ represents the strain. On the scale of a defect, there
are $l_d/r_d$ such fluctuations. Thus the associated random
potential for a defect has a variance $\langle E_s^2\rangle \sim
\Gamma_s(0) (r_d/a_0)^2(a_0/l_d)^2 (l_d/r_d)$. Taking
$\Gamma_s(0)\simeq \Gamma_l(0)$ and using Eq.(\ref{correlator}),
in which case $r_d=\lambda$, yields $\langle E_s^2\rangle \simeq
2C_{\alpha}(U_0\Delta\lambda)^2(\lambda/a_0)^{\alpha}/(l_da_0)$
(the factor $2$ comes from the refined calculation in
App.\ref{appD}) The typical pinning force $\sqrt{ \langle E_s^2
\rangle }/l_d$ due to random coupling to the strain is then given
by:
\begin{eqnarray}
\sqrt{ \langle f_s^2 \rangle } \simeq \sqrt{3C_{\alpha}} \mu
\Delta (g/3\pi )^{\frac{2+\alpha}{2}} g^{-3/4}
\label{rmsdefpinforw}
\end{eqnarray}

The ratio between the two characteristic defect energies is
$\sqrt{\langle E_s^2 \rangle}/\sqrt{\langle E_a^2 \rangle}\simeq 8
\sqrt{C_{\alpha}} (\lambda/a_0)^{\frac{\alpha+1}{2}}$, which shows
that, particularly for increasing $\lambda/a_0$, the dominant
pinning is due to random coupling to the strain. Henceforth we use
only this contribution. We next estimate the disorder strength
where the commensurability peak vanishes. As mentioned before,
this collapse occurs when the density of disorder induced defects
becomes $\sim l_d^{-1}$, in other words, when the typical energy
gain of a defect due to disorder becomes of the same magnitude as
its bare elastic energy $\int (dx/a_0)(\kappa_0/2)(\partial_x u)^2
\simeq \mu a_0 \sqrt{g}$. This leads to $\Delta_* \propto
C_{\alpha}^{-1/2} g^{-(2\alpha+1)/4}$. For the particular case of
random strains that are identical for all rows ($\alpha=2$),
$\Delta_*$ is given by:
\begin{eqnarray}
\Delta_*\simeq 3 g^{-5/4} \label{deltastarequation}
\end{eqnarray}
This can be compared to the numerical data in Fig.
\ref{plot_fs1DversuswDel}. Even though those results were obtained
for $\lambda/a_0=1$ ($g= 3 \pi$), formally outside the regime of
validity of our analysis, the predicted value $\Delta_* \simeq
0.18$ is in reasonable agreement with the data.


In the particular case of {\it bounded} random strains in the
CE's, the distribution of the nucleation force $\{f_n\}$ at
commensurability is bounded from below by $f_n^{min}$ and that of
the defect pinning force $\{f_d\}$ is bounded from above by
$f_d^{max}$ (at weak disorder). For completeness we give here the
previously derived results \cite{BesselingEPL2003} for these
extremal values: both occur due to disorder fluctuations on the
same length scale as that on which the displacement field $u(x)$
varies. For a defect this naturally corresponds to $l_d$. The
associated maximum defect pinning force is $f_d^{max}/\mu \propto
\Delta g^{3/2}$ (for uniform strains \cite{BesselingEPL2003}). For
nucleation, at $f\lesssim \mu$, the appropriate length scale is
$l_{san}$, the extent of a so called small amplitude nucleus
\cite{ButtikLandauerPRA81}. Due to the nonlinearity of the pinning
force, $l_{san}$ itself depends on the force, i.e. $l_{san}(f) >
l_d$ and it diverges for $f \rightarrow \mu$. As shown in detail
in \cite{BesselingEPL2003}, this leads to a minimum nucleation
threshold given by $1-(f_n^{min}/\mu) \propto [g^{3/2}
\Delta]^{4/3}$. From the condition $f_d^{max}=f_n^{min}$ one then
obtains the disorder strength $\Delta_c$ at which pinned defects
can first appear spontaneously in the system: $\Delta_c \simeq
g^{-3/2} < \Delta_*$.

\section{Wide channels with weak disorder}
\label{secweakdis2D}

We now consider how channels of larger width, in which vortices
have the $2$D degrees of freedom, behave in the presence of weak
edge disorder. Close to commensurability, the effects we find are
similar to that for a single chain. However, around 'half filling'
the importance of the transverse degrees of freedom of the channel
vortices become apparent.

\subsection{Behavior near commensurability}
\label{subweakdis2Dnearcom}

\begin{figure}
\scalebox{0.40}{\includegraphics{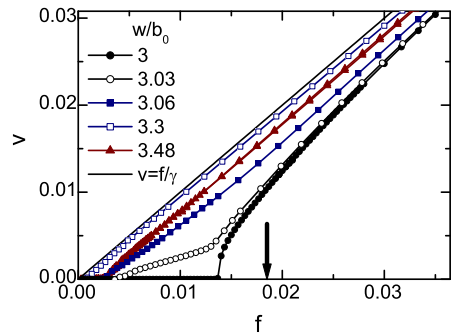}}
\vspace{0cm} \caption{{\it v-f} curves for weak disorder
($\Delta=0.05$) and several channel widths. The arrow indicates
the yield strength $f_s=\mu/3$ for $w/b_0=3$ and no disorder. All
data are computed for channel lengths $L > 400a_0$.}
\label{plot_fvwide_ranampi10}
\end{figure}

\begin{figure}
\scalebox{0.50}{\includegraphics{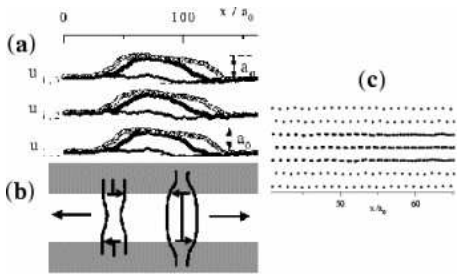}} \vspace{0cm}
\caption{(a): Time evolution of longitudinal displacements
$u_{i,j}$ of individual rows $j=1,2,3$ at depinning for $w/b_0=3$
and $\Delta=0.05$. (b) Square lattice representation of the
nucleated stacks of discommensurations. Small arrows indicate the
Burgers vector of the dislocations terminating each stack. The
large arrows indicate their propagation direction. (c) vortex
trajectories during nucleation of the vacancy stack between
$x=40a_0$ and $x=65a_0$ (up to the time corresponding to the
filled symbols in (a)).} \label{plot_w3nuc}
\end{figure}

For the commensurate case, $w/b_0=n$, weak CE disorder causes a
reduction of the threshold with respect to the ideal value
$f_s^0=\mu/n$, see the data in Fig. \ref{plot_fvwide_ranampi10}.
The reduction originates from defect formation at threshold, as
illustrated for $w=3b_0$ in Fig. \ref{plot_w3nuc}. In (a), three
snapshots of the displacements of individual rows inside the
channel are displayed. The first snapshot is for $f<f_s \simeq 0.7
f_s^0$ and yields the 'flat' profile. The subsequent snapshots for
$f>f_s$ reveal simultaneous nucleation and motion of a pair of
'oppositely charged' TSF's, each terminated at the CE's by a pair
of edge dislocations (see Fig. \ref{plot_w3nuc}(b)). The
macroscopic, stationary motion of the array is governed by
periodic repetition of this process at the least stable nucleation
site. In (c) we show the vortex trajectories during nucleation of
the TSF at the left in Figs. \ref{plot_w3nuc}(a),(b). Very similar
images were obtained in decoration experiments at the initial
stage of VL depinning in NbSe$_2$
\cite{MarchevskyPRB99_depinning}, implying that even for weakly
disordered VL's, defects may nucleate at depinning (see also
\cite{PardoPRL97_topdef}). We also note that in simulations of a
rapidly moving $2$D vortex lattice
\cite{FangohrPRB2001_nucleation} the same nucleation mechanism as
in Fig. \ref{plot_w3nuc}, but relative to the comoving frame, was
identified as source of velocity differences between chains.

For incommensurate channels close to commensurability, the TSF's
which are caused by the mismatch are pinned by the disorder. For
the weak disorder strengths considered here the random stress from
the CE's is not sufficient to break up the TSF's. Consequently, at
zero drive the array in the channel consists of a weakly
disordered superlattice of TSF's. The behavior of the transport
curves is shown in Fig. \ref{plot_fvwide_ranampi10} for
$w/b_0=3.03$ and $3.06$. It reveals features very similar to the
curves of a single chain (Fig. \ref{plot_fv1Dweakdis}). For $f_s
\lesssim f \lesssim \mu/n$, a low mobility regime in the {\it v-f}
curves develops due to the motion of the TSF's. Here $f_s$
corresponds either to the rms pinning force of individual TSF's or
to the collective pinning force for larger stack density. For
forces $f > \mu/n$ the curves approach linear flow behavior again.

\subsection{Behavior around 'half filling'}
\label{subweakdis2Dhalffilling}

\begin{figure}
\scalebox{0.40}{\includegraphics{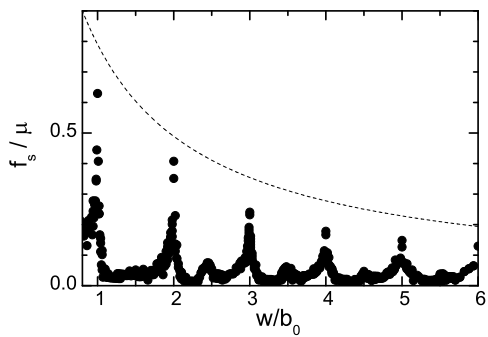}}
\vspace{0cm} \caption{Critical force versus channel width for a
disorder strength $\Delta=0.05$ using a velocity criterion $v
\approx 0.01 \mu/\gamma$. The dashed line represents the continuum
result $f_s=\mu b_0/w$.} \label{fcweakdisversusw}
\end{figure}

\begin{figure}
\scalebox{0.5}{\includegraphics{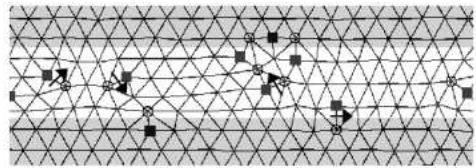}}
\vspace{0cm} \caption{Triangulation of the static ground state
structure for $w/b_0=3.48$ and $\Delta=0.05$. The arrows indicate
the Burgers vectors of the dislocations. Shown is a characteristic
segment of the total channel length $L=500 a_0$.}
\label{plot_weakdis_staticcoex}
\end{figure}

The dependence of $f_s$ on the channel width for $\Delta=0.05$ is
shown in Fig. \ref{fcweakdisversusw} for two disorder
realizations. For larger channel widths, smaller channel lengths
were used with $L \gtrsim 1000/(w/b_0)$. The data around matching
($w/b_0 \simeq n$) reflect the nucleation and pinning of TSF's as
discussed above: $f_s$ {\it at} commensurability is reduced
compared to the pure value $\mu/n$ and the commensurability peak
is considerably broadened, particularly for larger $n$, due to the
pinning of TSF's. The apparent discontinuity in the peak may be an
artifact of the finite channel length (see the discussion in
Sec.\ref{subdis1Dnum} on the distribution of nucleation sites). A
new, and robust feature, however, is that around 'half filling'
distinct maxima in $f_s$ appear. The origin of these maxima is
illustrated by the static structure near half filling. An example
is given for $w/b_0=3.48$ in for Fig.
\ref{plot_weakdis_staticcoex}. The triangulation shows that, in
addition to aligned dislocations with $\vec{b}\parallel \vec{x}$,
also misaligned dislocations appear with Burgers vector at an
angle of about $\pm 60^{\circ}$ with the CE's. These misaligned
defects are locally stabilized by the disorder in the CE's and
thus also pinned by the disorder which leads to the increase in
the threshold force. In addition, the projection of the driving
force along the glide direction is always smaller for misaligned
dislocations than for aligned dislocations. It is seen that the
misaligned dislocations separate regions with $n$ rows from
regions with $n \pm 1$ rows. Either of these regions may thus be
considered as a longitudinal stacking fault (LSF). In
Sec.\ref{secord2D} we mentioned that in absence of disorder LSF's
are metastable, the structure with a single integer number of rows
and a regular distribution of TSF's is energetically somewhat more
favorable. In presence of weak disorder, however, the LSF are
stable and have the lowest energy.

\begin{figure}
\scalebox{0.6}{\includegraphics{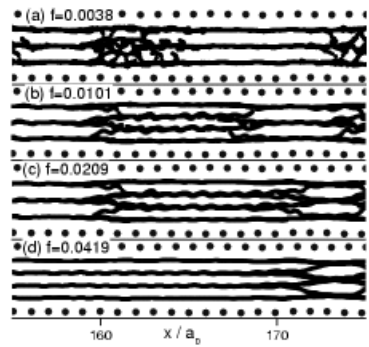}}
\vspace{0cm} \caption{Vortex trajectories for the channel segment
shown in Fig. \ref{plot_weakdis_staticcoex} ($w/b_0=3.48$,
$\Delta=0.05$) during motion over $\sim 4a_0$ at (a) $f=0.0038$,
(b) $f=0.0101$, (c) $f=0.0209$ and (d) $f=0.0419$.}
\label{plot_weakdis_dynacoex}
\end{figure}

As for the dynamics at $f>f_s$, the dislocation structure and flow
pattern are generally different from the static pattern of LSF's.
We illustrate this in Fig. \ref{plot_weakdis_dynacoex} where the
vortex trajectories at various (increasing) driving forces are
shown. For small drive (Fig. \ref{plot_weakdis_dynacoex}(a), $f
\simeq 0.25 \mu/(w/b_0)$), a region of plastic motion {\it within}
the channel is seen at about the location where the static pattern
shows a $4$ row structure. Vortex transport through these fault
zones occurs by repeated nucleation and annihilation of misaligned
defects. At a fixed driving force an LSF remains at the same
position, although its boundaries fluctuate over a distance of at
most $2-3 a_0$. This contrasts the situation in absence of
disorder where LSF's can move along the channel via a 'climb' like
process (see \cite{BesselingPRL99}).

For different driving forces, the location and amount of either
$n$ or $n \pm 1$-row regions or fault zones is different, as shown
in Figs. (b),(c) and (d). In this particular segment the $n=4$
region expands on increasing drive but at other locations the
reverse can occur. Moreover, after cycling the force, a different
structure can occur at the same drive, i.e. no unique structure
exists at a given force. This may also lead to small hysteresis in
the {\it v-f curves}. Overall, we see that the transverse degrees
of freedom in the channel, in combination with disorder, give rise
to an important new mechanism of yield strength enhancement. In
the following section we analyze these structures in more detail
in the context of strong disorder.


\section{Wide channels with strong disorder}
\label{secstrongdis2D}

The behavior of $f_s$ versus $w/b_0$ in Fig.
\ref{fcweakdisversusw} still exhibits considerable discrepancies
with the experimental data in Fig. \ref{plot_introFs}. Clearly,
the CE disorder underlying these experimental data (see
\cite{fn_fieldhistoryotheredge}) differs from the type of weak CE
disorder considered so far. Motivated by recent imaging
experiments \cite{BaarleAPL03}, which showed glassy vortex
configurations in the NbN edge material, we now consider the case
of strong CE disorder. As will be shown, in this regime the effect
of transverse degrees of freedom and the presence of misaligned
defects provide the main mechanism for the critical current
oscillations.

The simulations we discuss here in detail were performed at a
large disorder strength $\Delta=0.2$. The only remaining order in
the CE arrays is their preferred orientation with the principle
lattice vector along the CE's. The system sizes were typically
such that $wL \gtrsim 1500 a_0b_0$. We also allowed for quenched
defects between the two CE's (uncorrelated longitudinal strains in
the upper and lower CE arrays), but at these large disorder
strengths this is not essential.

\subsection{Static structures, yield strength and depinning}
\label{substrongdisstatic}

\begin{figure}
\scalebox{0.5}{\includegraphics{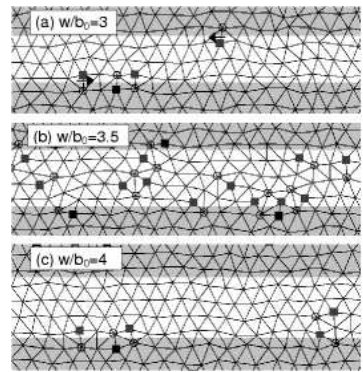}} \vspace{0cm}
\caption{Delauney triangulation of the ground state structure at
$\Delta=0.2$ for (a) $w/b_0=3$, (b) $w/b_0=3.5$ and (c) $w/b_0=4$.
The arrows in (a) indicate the Burgers vectors of the
dislocations. The aligned dislocations in (a) and (c) include a
disorder induced static dislocation pair (twisted bond) at the
lower CE, in (b) numerous misaligned dislocations are present.}
\label{plot_statstrongdistria}
\end{figure}

In Fig. \ref{plot_statstrongdistria} we show triangulations of the
static vortex configuration for channels of width $w/b_0=3$,
$w/b_0=3.5$ and $w/b_0=4$. As seen in Figs.(a) and (c), even in
the matching case dislocations are present due to the large random
stresses from the CE. While some of these may originate from
quenched 'phase slips' between upper and lower CE's, we checked
that, also without these 'phase slips', the matching structure at
this disorder strength always exhibits twisted bonds (defined by
two adjacent dislocations of opposite charge) at the CE's as well
as oppositely 'charged' TSF's terminated by defect pairs at the
CE. While most defects are thus situated along the CE with
$\vec{b} \parallel \vec{x}$, occasionally they can be located
inside the channel and have misoriented Burgers vector. Turning to
the mismatch case in (b), it is seen that a region of $4$ rows
(left) coexists with a $3$ row region (right). In between these
regions, there must be dislocations with misaligned Burgers
vectors. In addition, numerous other misaligned defects are
visible, rendering local destruction of the chain alignment with
the CE's (although less frequent, the latter can also occur in the
matching state, see Fig. (c)).

\begin{figure}
\scalebox{0.40}{\includegraphics{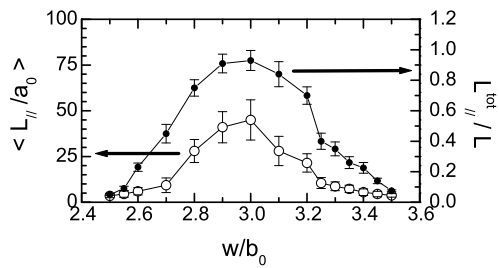}} \vspace{0cm}
\caption{The 'correlation' length $ \langle L_{//} \rangle$
($\circ$) and the fraction $L_{//}^{tot}/L$ ($\bullet$) of regions
with $n=3$ rows versus $w/b_0$.} \label{plot_Del20statLcor}
\end{figure}

To further characterize the disorder in the $f=0$ structures, we
analyzed, for the regime $2.5< w/b_0 < 3.5$, the average length of
domains without misaligned dislocations, $ \langle L_{//}
\rangle$, as well as the total fraction, $L_{//}^{tot}/L$, of
regions with $n=3$ rows. The results are shown in Fig.
\ref{plot_Del20statLcor}. As observed, both quantities are maximum
at $w/b_0=3$ and decay considerably away from matching: for
$|w/b_0-3| > 0.3$ the average length of 'correlated' $n=3$ regions
in the static structure is no more than $10 a_0$ and they make up
less than $\sim 50 \%$ of the channel. The remaining fraction
$1-(L_{//}^{tot}/L)$ contains misaligned dislocations and small
regions with $2$ ($w/b_0 \simeq 2.5$) or $4$ (for $w/b_0 \simeq
3.5$) aligned chains.

\begin{figure}
\scalebox{0.4}{\includegraphics{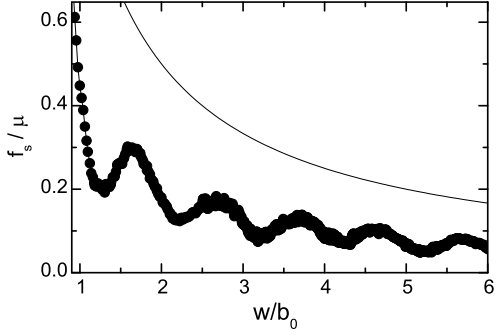}}
\vspace{0cm} \caption{Computed threshold force versus channel
width for strong disorder, $\Delta=0.2$. The data were obtained by
taking the friction force $f-\gamma v$ at a velocity criterion $v
\simeq 0.025 (\mu/\gamma)$ and subsequent averaging and smoothing
over $5$ disorder realizations. Drawn line: continuum result,
$f_s=\mu /(w/b_0)$.} \label{plot_fsversuswranampo40}
\end{figure}

In Fig. \ref{plot_fsversuswranampo40} we show the behavior of the
threshold force for $\Delta=0.2$. The modulation of $f_s$ with
channel width is still present but it has changed considerably
compared to the weak disorder case. The sharp maxima at integer
$w/b_0$ have vanished, very similar to the case of the 1D chain at
strong disorder, see Fig. \ref{plot_fs1DversuswDel}. Instead, we
now observe smooth oscillations, with maxima in $f_s$ for
$w/b_0\simeq n+0.65$ and minima for $w/b_0 \simeq n+0.15$. The
maxima, although occurring slightly above 'half filling', are of
similar nature as the local maxima at $w/b_0 \simeq n \pm 1/2$ for
weak disorder: they are related to numerous misaligned defects
present in the structure around mismatch. They enhance the flow
stress compared to that of the structures around matching with
predominantly aligned defects.

\begin{figure}
\scalebox{0.50}{\includegraphics{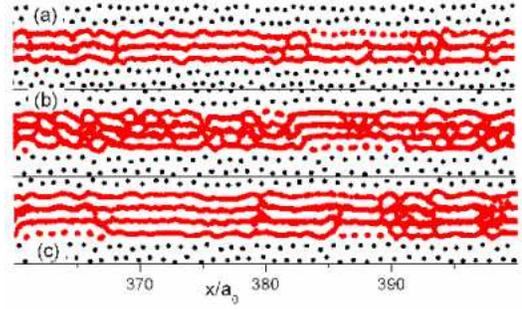}}
\vspace{0cm} \caption{Vortex trajectories at small velocity $v
\simeq 0.001$ for: (a) $w/b_0=3.1$ ($f=0.0053 \simeq 0.1 \mu$),
(b) $w/b_0=3.6$ ($f=0.0074 \simeq 0.14 \mu$) and (c) $w/b_0=4.1$ (
$f=0.0044 \simeq 0.08 \mu$), all during motion over $\sim 3a_0$.}
\label{plot_thresholdtrajec_Del20}
\end{figure}

The differences in threshold force are also reflected by the
vortex trajectories at small velocity. In Fig.
\ref{plot_thresholdtrajec_Del20} we show these trajectories for
channel widths $w/b_0=3.1$, $3.6$ and $4.1$, close to the extrema
in $f_s$. The first thing to notice is that the trajectories at
mismatch (Fig.\ref{plot_thresholdtrajec_Del20}(b), $w/b_0=3.6$)
are densely interconnecting, on a scale $\sim a_0$, i.e. the
motion is fully plastic and creation and annihilation of
misaligned defects occurs over nearly the full channel length. For
the 'near matching' cases in (a) and (c), the motion occurs mainly
in the form of integer chains. However, for the small velocity
considered here, the dynamics still exhibits a considerable amount
of plastic motion. This partly occurs due to vortices which remain
stuck at the CE's (also visible in (b)) and partly due to narrow
interconnecting regions. In fact, comparing the average length of
the regions without inter row switching in (a) with the data for
the static structure in Fig. \ref{plot_Del20statLcor}, it is seen
that near matching the structure at small $v$ is more disordered
than the corresponding static structure, reflecting nucleation of
misaligned defects in regions which were free of such defects at
$f=0$.

\subsection{Analysis of dynamic properties}
\label{substrongdisdynamic}

\begin{figure}
\scalebox{0.50}{\includegraphics{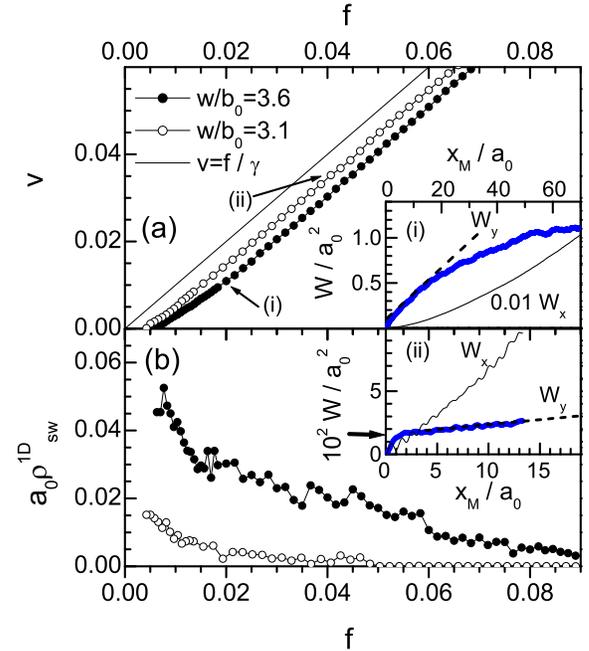}} \vspace{0cm}
\caption{(a) {\it v-f} curves around a minimum in flow stress
$w/b_0=3.1$ ($\circ$), and around a maximum $w/b_0=3.6$
($\bullet$). Inset: $W_x(x_{M})$ and $W_y(x_{M})$ for $w/b_0=3.6$
and $f=0.021$. The dashed line emphasizes the linear behavior of
$W_y$ in the regime $3a_0 < x_{M} < 10 a_0$. (b) The density of
switching points $\rho^{1D}_{sw}\equiv \Delta W_y/(\Delta
x_{M}a_0^2)$ versus force for $w/b_0=3.1$ ($\circ$) and
$w/b_0=3.6$ ($\bullet$). Inset: $W_x(x_{M})$ and $W_y(x_{M})$ for
$w/b_0=3.1$ and $f=0.041$. The arrow indicates the value $W_{y,c}$
referred to in the text.} \label{plot_fvDel20plusdif}
\end{figure}

We now consider in more detail the properties of the moving
structures at $f > f_s$. We first show in Fig.
\ref{plot_fvDel20plusdif}(a) two characteristic {\it v-f} curves
associated with a minimum ($w/b_0=3.1$) and a maximum
($w/b_0=3.6$) in flow stress. It is observed that the enhanced
threshold in the latter case also translates in a larger dynamic
friction, $f-\gamma v$, of the driven structure. In addition, the
latter curve exhibits a small positive curvature in the small
velocity regime $v \lesssim 0.01$. This nonlinearity is related to
the strong plastic nature of the motion in this regime.

A convenient way to characterize the amount of plasticity is to
calculate the mean square displacement of vortices form their
center of mass (M) positions \cite{KoltonPRL99_transfreeze}:
\begin{eqnarray}
W_{\alpha}(t)=\sum_i[\alpha_i(t+t_0)- \alpha_i(t_0)]^2/N_{ch},
\label{eq_Wydefinition}
\end{eqnarray}
where $\alpha=x-x_M$, $y-y_M$ denotes longitudinal and transverse
displacements, respectively. As shown by Kolton {\it et al} for a
'bulk' 2D vortex lattice \cite{KoltonPRL99_transfreeze},
$W_{\alpha}(t)$ can be characterized by
$W_{\alpha}=R_{\alpha}t^{\xi_{\alpha}}$. For example, when
$\xi_{y}=1$ we have normal transverse diffusion (caused by
'random' switching of vortices between chains) with $R_{y}=D_y$
the diffusion coefficient. However, in the channels $W_y$ will
become bounded at long times (large $x_M$) due to the finite
channel width. In the inset (i) to Fig.
\ref{plot_fvDel20plusdif}(a) we have illustrated this behavior for
$w/b_0=3.6$ and $f=0.021$. For $x_M \lesssim 15a_0$ (the point
$x_{M}=0$ was chosen in the steady state after transients had
disappeared), $W_y$ increases linearly as in usual diffusion, but
for larger displacements (times) $W_y$ levels off and eventually
saturates. In addition, even when all vortices remain in their
chain (no transverse diffusion), $W_y$ initially increases to a
value $W_{y,c}$ due to finite chain 'roughness'. Such behavior is
observed for $x_M \lesssim 2a_0$ in the inset (ii) to Fig.
\ref{plot_fvDel20plusdif}(b) where $W_{y}$ is shown for a more
coherent flow situation at $w/b_0=3.1$ and $f=0.041$. In practice,
we found that $W_{y,c}$ is always reached for $x_{M} <3a_0$, while
for the long time (large distance) behavior significant levelling
of $W_y$ occurs only when $x_{M} \gtrsim 10a_0$. The appropriate
regime we use to characterize real diffusion is therefore given by
$3a_0< x_{M} < 10 a_0$ (i.e. $\Delta x_{M}=7a_0$). Further, $D_y$
itself does not directly reflect the density of chain-switching
points along the channel. Defining the '1D' density of such
switching points as $\rho_{sw}^{1D} \simeq N_{sw}/L$, the {\it
rate} of switching increases linearly both with $\rho_{sw}^{1D}$
and with the average velocity: $t_{sw}^{-1} \simeq \rho_{sw}^{1D}
v$. Hence, the diffusion constant is given by: $D_y \simeq
a_0^2/t_{sw}=\rho_{sw}^{1D} v a_0^2$. Being interested in
$\rho_{sw}^{1D}$, we therefore divide out the intrinsic velocity
dependence of $D_y$ and calculate $W_y(t)/(vta_0^2)=\Delta
W_y/(\Delta x_{M}a_0^2) \equiv \rho_{sw}^{1D}$.

The results of $\rho_{sw}^{1D}$ versus $f$ are shown in Fig.
\ref{plot_fvDel20plusdif}(b) for the two cases $w/b_0=3.1$ and
$w/b_0=3.6$.  Around the minimum in flow stress ($\circ$), the
overall value of $\rho_{sw}^{1D}$ is considerably smaller than
around the maximum ($\bullet$), similar to what is seen in the
trajectories in Fig. \ref{plot_thresholdtrajec_Del20}. For both
cases, $\rho_{sw}^{1D}$ is clearly reduced on increasing the
force. This reflects both a decrease of the number of fault zones
in the moving structure as well as a suppression of switch events
(called 'transverse freezing' in \cite{KoltonPRL99_transfreeze})
within regions already organized in $n$ moving rows. Around
matching, $\rho_{sw}^{1D}$ smoothly vanishes at $ f \sim \mu
\approx 0.05$, indicating complete dynamic ordering into an $n=3$
row structure without transverse wandering. For $w/b_0=3.6$ an
ordering transition is also observed but it occurs at much larger
drive ($f \sim 3 \mu$, not shown) and the array orders into an
$n=4$ row configuration, with a reduced spacing $b < b_0$ between
the chains and average vortex spacing $a > a_0$ within the chains.
At the end of this section we will show how in this large drive
regime the number of rows changes with $w/b_0$.

The insets to Fig. \ref{plot_fvDel20plusdif} also show the
longitudinal mean square displacements. For strongly plastic flow
at mismatch (inset (i)), $W_x$ is large and $\xi_{x}$ is close to
$2$, as expected when some vortices remain stuck at the CE's. For
the more coherent situation in (ii), where transverse switching
has nearly ceased, $W_x$ is smaller and $\xi_x \gtrsim 1$. We
always find an exponent $1< \xi_{x}<2 $, similar to the results
for 2D VL's in \cite{KoltonPRL99_transfreeze}. Interestingly, even
without transverse wandering ($\rho^{1D}_{sw}=0$), $W_x$ increases
indefinitely (with $\xi \gtrsim 1$), indicating that the moving
integer chain structure still exhibits slip events and (local)
velocity differences between the chains.

\begin{figure}
\scalebox{0.50}{\includegraphics{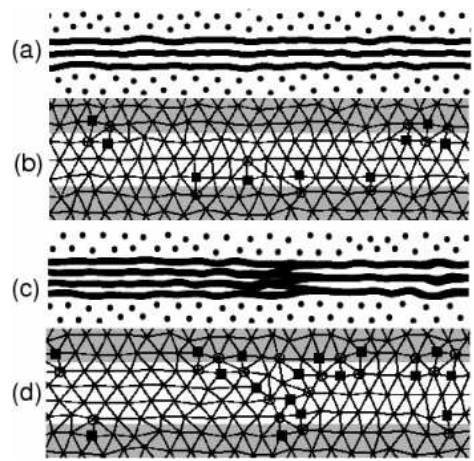}}
\vspace{0cm} \caption{(a) Flow trajectories at large drive (a)
during motion over $\sim 3a_0$ for $w/b_0=3.1$ and $f \simeq
2\mu$. (b) Delauney triangulation of one snapshot of (a). (c)
Trajectories for $w/b_0=3.55$ and $f \simeq 2 \mu$. (d)
Triangulation of a snapshot of (c).}
\label{plot_triatrajecDel20largev}
\end{figure}

We illustrate some more aspects of the structures at large drive
in Fig. \ref{plot_triatrajecDel20largev}, where the vortex
trajectories and triangulations of a single snapshot are displayed
for $w/b_0=3.1$ and $w/b_0=3.55$, both at $f \simeq 2 \mu$. For
$w/b_0=3.1$, we do not see any transverse wandering in Fig.
\ref{plot_triatrajecDel20largev}(a). Figure
\ref{plot_triatrajecDel20largev}(b) shows that in this case the
dynamic structure exhibits only dislocations with
$\vec{b}\parallel \vec{x}$, mainly located at the CE but also
occasionally between chains inside the channel. The latter
dislocations are possibly dynamically nucleated. They are
non-stationary in the comoving frame and lead to slip events and
the growth of $W_x$ as was discussed above. Turning to the
mismatch case (Figs. \ref{plot_triatrajecDel20largev}(c) and (d)),
it is seen that the dynamic structure consists of $n=3$ and $n=4$
row regions coexisting in the channel. At the driving force
considered here, $\rho^{1D}_{sw}$ has a finite but small value
$a_0\rho^{1D}_{sw} \simeq 0.007$, which is solely due to switching
of vortices in the fault zones separating the $3$ and $4$-row
regions. {\it Within} these regions transverse wandering is
absent. At yet larger forces the minority $3$-row regions vanish
and complete ordering into $4$ rows occurs, as for the case
$w/b_0=3.6$. For a given driving force, the $n$-row regions again
remain quasi-static during motion. The triangulation in (d)
exhibits the expected misaligned dislocations at the fault zone
but in general also aligned dislocations are present between the
chains within an $n$-row region (not shown in the figure).

It is also interesting to compare the velocities in the two
coexisting regions with $n$ and $n\pm 1$ rows. Denoting the vortex
velocity in an $n$-row region by $v_n$ and the longitudinal vortex
spacing there by $a_n$, flux conservation implies that $n
v_n/a_n=n' v_{n'}/a_{n'}$. We checked that, in both regions, the
average flux density $1/(a_n b_n)$, with $b_n=w/n$ the row
spacing, was equal to $1/(a_0b_0)$ within $\sim 4 \%$. Therefore
$a_{n'}=(n'/n)a_n$, and consequently the average vortex velocities
are equal, $v_n=v_{n'}$. However, the local {\it washboard
frequency}, $\nu_n=v_n/a_n$, is different in both regions. Indeed,
the spectrum of the velocity fluctuations in channels with dynamic
coexistence of $n$ and $n\pm1$ rows showed two shallow fundamental
peaks at frequencies $\nu_n/\nu_{n'}=n'/n$. We however note that,
both around matching (where a single peak occurs at $\nu \sim
v/a_0$) and around mismatch, the amplitude of the washboard
peak(s) decays on increasing the channel length $L$. In addition,
for large velocities the mixed $n / n\pm 1$ structures ultimately
anneal into a single $n$ or $n\pm 1$ domain, causing the collapse
of one of the peaks.

\begin{figure}
\scalebox{0.45}{\includegraphics{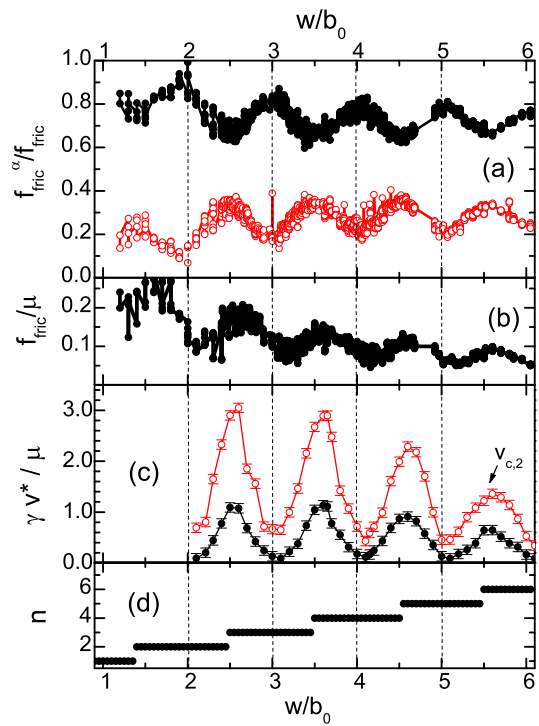}}
\vspace{0cm} \caption{(a) The contributions $f_{fric}^x/f_{fric}$
($\bullet$) and $f_{fric}^y/f_{fric}$ ($\circ$) to the dynamic
friction versus $w/b_0$ for a velocity $v=0.1 \mu/\gamma$. (b)
Total friction $f-\gamma v$ at $v=0.1 \mu/\gamma$. (c) Ordering
velocity $v_{c,1}$ ($\bullet$) determined from a criterion
$a_0\rho_{sw}^{1D} \simeq 0.01$, and $v_{c,2}$ ($\circ$) using a
criterion $a_0\rho_{sw}^{1D} \simeq 0.002$ (error bars were
estimated from different disorder realizations). (d) Number of
moving rows $n$ at large drive ($v \gtrsim v_{c,2}$) versus
$w/b_0$.} \label{plot_Del20dynpropversusw}
\end{figure}

The simulations also allow to explicitly show the influence of the
transverse degrees of freedom on the modulations of the dynamic
friction force (and, ultimately, the critical current).
Generalizing the expression for the friction force
Eq.(\ref{dissiprate}) in Sec. \ref{subord1Dfv} including
transverse fluctuations leads to:
\begin{eqnarray}
f&&=(\gamma/v)[\langle(\partial_t
u_x)^2\rangle_{i,t}+\langle(\partial_t
u_y)^2\rangle_{i,t}] \nonumber \\
&&=\gamma v+f_{fric}^x+f_{fric}^y \nonumber \\
&& \equiv \gamma v +f_{fric}, \label{eq_transvdissip}
\end{eqnarray}
where $f_{fric}^x$ and $f_{fric}^y$ denote the contribution to the
total friction due to longitudinal and transverse fluctuations,
respectively. Figure \ref{plot_Del20dynpropversusw}(a) displays
the numerical results for these quantities, normalized by the
total friction force $f-\gamma v$ as obtained from the ${\it v-f}$
curves. The sum of the two data is $1$ as it should be, confirming
the correct numerical evaluation of these quantities. The total
friction is also shown for clarity (Fig.(b)) and is essentially
the same as the data shown in Fig. \ref{plot_fsversuswranampo40}.
Clearly, the {\it relative} contribution of longitudinal
fluctuations to $f_{fric}$ {\it decreases} on approaching a
mismatch situation, while $f_{fric}^y/f_{fric}$ increases
accordingly. These qualitative features remain present also at
larger velocities, where permeation modes between chains are being
suppressed.

Finally, we consider the continuous modulation of structural
(dis)order in the moving arrays when varying the channel width. As
was shown in Fig. \ref{plot_fvDel20plusdif}(b), for a fixed
velocity, the density of switching points $\rho_{sw}^{1D}$ is
maximum around mismatch while for fixed channel width it decreases
with velocity. We can then define an ordering velocity $v_c$
operationally as the point at which $\rho_{sw}^{1D}$ is reduced
below a certain threshold. In Fig.
\ref{plot_Del20dynpropversusw}(c) the behavior of $v_c$ for
channel widths $w/b_0>2$ is shown for two criteria. The lower
curve ($v_{c,1}$), with $a_0\rho_{sw}^{1D}\approx 0.01$,
corresponds to a situation with $\sim 70\%$ of the channel length
'transversely frozen' into integer chain regions of length
$\gtrsim 10a_0$. The upper curve ($v_{c,2}$), with
$a_0\rho_{sw}^{1D} \approx 0.002$, corresponds to nearly fully
annealed arrays. As observed, $v_{c,1}$ increases smoothly by
about an order of magnitude between a matching and mismatching
situation, while $v_{c,2}$ increases by a factor $\gtrsim 5$
(except for $w/b_0 > 5$). Regardless of the criterion, the
amplitude of the modulation of $v_c$ is considerably larger than
the amplitude of the $f_s$ modulation, which we will further
discuss shortly.

For $v \gtrsim v_{c,2}$ the arrays all completely anneal into a
single $n$-row structure without permeation modes. Figure
\ref{plot_Del20dynpropversusw}(d) displays the number of rows $n$
of these structures versus $w/b_0$. The switching from $n$ to
$n+1$ rows is seen to occur at half integer channel widths for
$w/b_0>3$ but the transitions $1 \rightarrow 2$ and $2 \rightarrow
3$ take place below these points, at $w/b_0 \simeq 1.35$ and
$w/b_0 \simeq 2.4$, respectively. Around the transition points,
there are regions in which $n$ and $n\pm 1$ rows coexist at
smaller velocity $v \lesssim v_{c,2}$. The widths of these regions
were found to be $\Delta w/b_0 = \pm 0.05$.

\section{Discussion}
\label{secdiscuss}

\begin{figure}
\scalebox{0.40}{\includegraphics{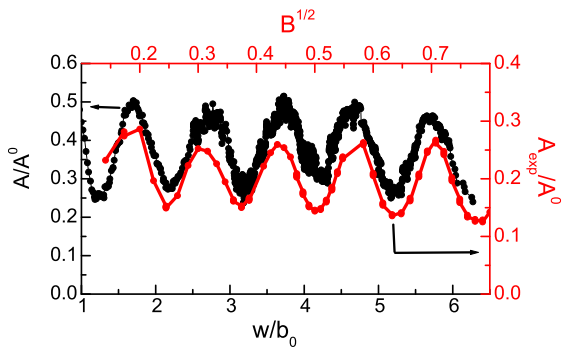}}
\vspace{0cm} \caption{Comparison between the experimentally
measured yield strength of an artificial flow channel and the
numerical results at strong disorder \cite{fn_expvcriterion},
represented through the parameter $A$ (see the text) normalized by
the value $A^0=(\pi\sqrt{3})^{-1}$ for the ideal lattice.}
\label{plot_sim_exp_compare}
\end{figure}

The results for $f_s$ versus $w/b_0$ at strong disorder in the
previous section show strong resemblance to the measurements of
the critical current versus field $B$ in the experimental channel
system, see Fig. \ref{plot_introFs}. For a more detailed
comparison, we combine in Fig. \ref{plot_sim_exp_compare} both
data, represented in terms of the parameter $A$ (see
Sec.\ref{secintro}): for the experiment, $A_{exp}$ is determined
from $A_{exp}=F_s w/(2 c_{66})$ and is plotted versus $B^{1/2}$,
for the simulations, $A=(f_s/\mu) (w/b_0) A^0$
(\cite{fn_expvcriterion}). As observed, the shapes of the
oscillations are in very reasonable agreement, although the
overall value of $A$ from the numerics is larger than the
experimental value \cite{fn_AexpAnumcomp}. The important
conclusion from the simulations here is thus that the maxima in
the measured critical current do not correspond to traditional
commensurability peaks, but are caused by enhanced plastic motion
and transverse deviations in mismatching channels with strong edge
disorder. Previously, we indeed obtained experimental evidence for
this mechanism via mode-locking (ML) experiments
\cite{KokuboPRL02} in which an rf-drive is superimposed on the
dc-drive. Interference between the former and collective modes of
deformation in the moving array leads to plateaus in the
current-voltage curves. The plateau (ML) voltage then directly
yields the number of moving rows $n$ and transitions from $n
\rightarrow n \pm 1$ rows were observed to coincide with maxima in
critical current \cite{fn_weffwetched}.

More detailed ML-experiments provided a wealth of additional
information on the dynamics of the arrays
\cite{KokuboPRB_chanMLvcB}. In particular, a minimum velocity was
required to observe the ML-phenomenon, which we identified as
ordering velocity. This ordering velocity exhibited a strong
upturn away from matching, similar to the behavior of $v_c$ in the
simulations. As proposed in \cite{KokuboPRB_chanMLvcB}, $v_c$ can
be estimated from a modified version of the dynamic ordering
theory of Koshelev and Vinokur (KV) \cite{Koshelevrecryst}.
Firstly, instead of the 2D random potential in
\cite{Koshelevrecryst}, for the channels $v_c$ is related to the
short range correlated random stress from the CE, with r.m.s.
amplitude $\sim \varepsilon_{ce} c_{66}$ and $\varepsilon_{ce}$
the random strain. Secondly, when thermal fluctuations can be
ignored compared to typical defect energies (see below), $v_c$ is
inversely proportional to the energy $k_BT_p$ for creation of
small defect pairs (see also \cite{Faleski}): $v_c/a \simeq
\sqrt{3/32\pi}(a_0/a)^2(\varepsilon_{ce}c_{66})^2a_0b_0/(\gamma
k_B \bar{T}_p)$, with $k_B \bar{T}_p \simeq A_p c_{66} a_0^2/2
\pi$ (per unit vortex length). The typical pairs referred to are
those with misaligned Burgers vector, which are responsible for
breakup of the chain structure. Hence, the increase of $v_c$ away
from matching implies a decrease in the pair formation energy,
which is accounted for by including the parameter $A_p\lesssim 1$
in $k_B \bar{T}_p$, while setting $A_p \equiv1$ at matching.
Experimentally, $A_p$ decreased to $\sim 0.1$ at mismatch. The
behavior of $v_c$ and $A_p$ in the simulations is analyzed using
the dimensionless form of the above formula for $v_c$:
\begin{eqnarray}
(\gamma v_c/\mu) \simeq 2.5 \varepsilon_{ce}^2(a_0/a)/A_p
\label{orderingvelocity}
\end{eqnarray}
The data for $v_{c,1}$ around matching (where $A_p=1$) yield as
measure for the random strain $\varepsilon_{ce} \simeq 0.19$ and,
near mismatch, a reduction of the defect pair creation energy by a
factor $A_p \simeq 0.1$. The latter is in very reasonable
agreement with the experiments, and the reduction of $k_BT_p$ near
mismatch also qualitatively agrees with the large number of
disorder induced fault zones observed in the static structures
near mismatch (Fig.\ref{plot_statstrongdistria}). However, the
result for $\varepsilon_{ce}$ is considerably larger than the
value $\varepsilon_{ce}=0.025$ found experimentally. This is also
manifest in the fact that in the experiments the pinning
frequency, defined as $f_s/(\gamma a)$, always exceeds the
ordering frequency $v_c/a$, while in the simulations $f_s \lesssim
\gamma v_c$.

Within the modified KV theory, the pinning frequency and the
ordering frequency are directly related via
\cite{KokuboPRB_chanMLvcB}: $v_c/a=\tau (f_s/\gamma a)^2$ with
$\tau=(\varepsilon_{ce}^2/2A^2)(wB)^2/(A_pc_{66}\rho_f)$.
Experimentally, $\tau$ was found to be independent of the matching
condition. The decrease in the defect pair creation energy
($\propto A_p$) then relates to the increase in the pinning
frequency (yield strength) away from matching ($\propto A$) as
$A_p \propto 1/A^{2}$. The numerical results for $A_p$ versus $A$
behave similarly and can be fitted by $A_p \propto 1/A^\varsigma$
with $\varsigma \sim 2-3$, but data collected over the full range
of $w/b_0$ show to much scatter to make a more detailed
comparison. Nevertheless, the KV model qualitatively accounts for
the enhanced amplitude of the $v_c$ modulation compared to that of
the yield strength.

As for the discrepancy between the experimental and numerical
values for $\varepsilon_{ce}$ or $v_c$, one should keep in mind
that experimentally the ordering velocity is determined from the
onset of an $n$-row ML plateau, i.e. it corresponds to the
velocity at which coherent $n$-row regions first appear, while
incoherent regions may still exist in other parts of the channel.
In the simulations this may thus correspond to $v_c$ determined
using a larger criterion for the density of switching points
$\rho^{1D}_{sw}$. In addition, the superimposed rf-drive in the
experiments may assist reordering of the structure, also leading
to smaller values of $v_c$. For future studies it is interesting
to directly compare numerical simulations of channels with mixed
rf-dc drive with the experiments and test which criterion best
represents the reordering phenomenon.

Additionally, the experimental results indicated that in the large
drive regime (where the ML amplitude saturates at a constant
value), the coherently moving fraction of vortices does not exceed
$\sim 40\%$ (at matching), while it was reduced on approaching
mismatch. This feature appears at odds with the simulations where
eventually the arrays all order into a completely transversely
frozen $n$-row domain at large drive, regardless of the matching
condition. At present we do not have a good explanation for this
discrepancy.

Finally we comment on the $T=0$ approximation made throughout this
study. When thermal fluctuations are important, not only does one
expect the dynamic ordering behavior to be changed (see
\cite{Koshelevrecryst,BesselingPRLDynMelt}), also the sharp
threshold behavior in the {\it v-f} curves will be smeared and
activated flow may occur. The relevant energy scale for both
phenomena is again the energy for formation of small defect pairs
$k_BT_p$. In the low magnetic field experiments in
\cite{KokuboPRB_chanMLvcB} this energy was $\sim 2$ orders of
magnitude larger than $k_BT$ and for comparison with these results
the $T=0$ approximation is justified. To compare with the
experiments near the melting field in \cite{BesselingPRLDynMelt},
it would be required to include thermal fluctuations in the
simulations.





\section{Summary}
\label{secsum}

We have presented a detailed study of the properties of vortices
confined to narrow flow channels with pinned vortices in the
channel edges. In the experimental system which motivated this
work \cite{Pruymboom,KokuboPRL02,KokuboPRB_chanMLvcB}, the
threshold force (yield strength) shows pronounced commensurability
oscillations when the natural vortex row spacing is varied through
integer fractions of the channel width. The analysis and
simulations presented in this paper show that in a mesoscopic
channel system the dependence of threshold on commensurability as
well as the dynamics of vortices in the channels drastically vary
with the amount of disorder in the confining arrays. At zero or
weak disorder the system behaves similar to 1D LJJ-systems and
defects at the CE's reduce the yield strength. At large disorder
the behavior involves transitions from quasi-1D to 2D structures,
where an increase in the amount of plastic deformations enhances
the yield strength similar to the situation in the classical peak
effect in superconductors.

We first presented a generalized sine-Gordon description for a 1D
vortex chain in an ideally ordered channel. In this case (or for
channels with multiple chains near commensurability) the threshold
force has sharp peaks at commensurate widths, whereas it is
essentially vanishing at incommensurability due to easy glide of
'aligned' defects, i.e. defects with Burgers vector along the CE.
The model was then extended to study the effects of weak disorder
in the confining arrays. Simulations and analytical results showed
that this disorder causes the sharp maxima in the threshold force
at matching to be lowered and broadened due to nucleation (at
matching) and pinning(at mismatch) of edge defects. Apart from
these defects, the arrays respond elastically in this regime, both
near threshold and at large drive. We studied numerically the
relevant edge defect dynamics and, using the sine-Gordon model, we
analyzed the crossover to strong defect pinning on increasing the
disorder strength.

For large disorder in the CE's, matching between the longitudinal
vortex spacings in and outside the channel becomes irrelevant and
the peaks in threshold force around matching completely vanish
with a 'saturated' value for $f_s$ of about $30 \%$ of the ideal
lattice strength. Around mismatch however, the arrays become
susceptible to formation of defects with Burgers vector
misoriented with the channel direction. Such defects either
locally break up the integer chain structure or exist at the
boundaries of $n$ and $n \pm 1$-row regions coexisting in the
channel. At large disorder, they are strongly pinned and cause the
threshold force to exceed that around matching. Approaching a
matching condition, the density of misaligned defects is reduced
and a smooth modulation of $f_s$ results, with minima near
matching. The depinning transition always involves plastic
deformations {\it inside} the channel, but the amount of
plasticity drastically increases away from matching. Using the
density of transverse switching points (obtained from the
transverse diffusion in the moving structures) as dynamic 'order
parameter', we study the evolution of the moving structures on
changing the channel width and the drive. The arrays reorder
(partially) into transversely frozen $n$-row regions at a velocity
$v_c$ which shows a similar modulation with commensurability as
the threshold force. Finally, we compared the modulations of $f_s$
and $v_c$ at strong disorder with the available experimental
results and with the dynamic ordering theory in
\cite{Koshelevrecryst} and find good qualitative agreement.

\begin{acknowledgments}
R.B. and P.K. are supported by the Nederlandse Stichting voor
Fundamenteel Onderzoek der Materie (FOM) and V.M.V. is supported
by U.S.DOE, Office of Science under contract $\#$W-31-109-ENG-38.

\end{acknowledgments}

\appendix

\section{Ordered channel for arbitrary field}
\label{appA}

In this appendix we calculate the edge potential and sine-Gordon
parameters in a symmetric, ordered channel with $w\simeq b_0$ for
arbitrary field. The interaction between a vortex at ${\bf
r}=(x,y)$ in the channel and the pinned vortices in the CE's at
${\bf R}_{n,m}$ is given by:
\begin{eqnarray}
V_{ce}({\bf r})=(2\pi)^{-2}\int d{\bf k} \sum_{n,m} V({\bf
k})e^{i{\bf k}\cdot({\bf r}-{\bf R}_{n,m})}, \label{Vedge}
\end{eqnarray}
where $V({\bf k})$ is the Fourier transform of the vortex-vortex
interaction. To obtain an expression valid over larger range of
fields than the (low field) London-regime, we use a generalization
of the London potential Eq.(\ref{VLondon}) as proposed by Brandt
\cite{Brandtelas}. This generalization accounts for the reduction
of superfluid density with field and an additional attractive
interaction due to overlapping vortex cores. The Fourier transform
of this interaction reads:
\begin{eqnarray}
V({\bf k})=2\pi U_0(1-b)\left[\frac{1}{|{\bf
k}|^2+\lambda'^{-2}}-\frac{1}{|{\bf k}|^2+\xi'^{-2}}\right],
\label{FTfullVBrandt}
\end{eqnarray}
where $b=B/B_{c2}$, $\lambda'=\lambda/(1-b)^{1/2}$ and $\xi'=\xi
\frac{C}{(2-2b)^{1/2}}$ the effective coherence length with $C\sim
1$. It is convenient to split Eq.(\ref{Vedge}) for the total
potential in terms of the contribution $V_m$ of row $m$ (see Fig.
\ref{plot_geometry}). Integrating over $k_y$ and using Poisson
summation yields
\begin{eqnarray}
V_m({\bf r})=\sum_l|V_m^l(y)|\cos{l k_0 (x-ma_0/2)} \label{ordVmb}
\end{eqnarray}
where $k_0=2\pi/a_0$ and the prefactors are:
\begin{eqnarray}
|V_m^l(y)|=k_0U_0(1-b)\left[ \frac{
e^{-\overline{lk_{0,\lambda'}}|y_m'|}}{\overline{lk_{0,\lambda'}}}-\frac{
e^{-\overline{lk_{0,\xi'}}|y_m'|}}{\overline{lk_{0,\xi'}}} \right]
\label{ordVmbprefac}
\end{eqnarray}
with $\overline{lk_{0,\lambda'}}=\sqrt{(lk_0)^2+(\lambda')^{-2}}$,
$\overline{lk_{0,\xi'}}=\sqrt{(lk_0)^2+(\xi')^{-2}}$ and
$y_m'=-y+m[b_0+((w-b_0)/2|m|)]$. First of all, we neglect in
Eq.(\ref{ordVmb}) the $l=0$ terms which represent uniform
($x$-independent) interaction. Secondly, when $\lambda\gtrsim
a_0$, as practically encountered in films,
$\overline{k_{0,\lambda'}}\approx k_0$. Then the $|l|>1$ terms can
be neglected, resulting in a sinusoidal potential $|l|=1$.
Further, in the summation over $m$ only the contributions from the
$m=\pm 1$ terms are significant. Next we employ the relation
$\xi^2/a_0^2=b\sqrt{3}/4\pi$ and rewrite
$\overline{k_{0,\xi'}}=(1/a_0)\sqrt{4\pi ^2+8\pi(1-b)/(\sqrt{3}C^2
b)}$. The resulting expression for the total edge potential is:
\begin{eqnarray}
V_{ce,0}({\bf
r})=&&-\left[\cosh(k_0y)-\frac{e^{(1-s(b))\pi\frac{w+b_0}{a_0}}}{s(b)}\cosh(k_0
y s(b)) \right]\nonumber\\
&& \times 2U_0(1-b)e^{-\pi\frac{w+b_0}{a_0}}\cos{k_0 x},
\label{ordVrfinal}
\end{eqnarray}
where $s(b)=\sqrt{1+(2-2b)/(\pi \sqrt{3} C^2 b)}$.

For a channel with $w=b_0$, the maximum $\mu(b)$ of the sinusoidal
pinning force $-\partial_x V_{ce,0}$ at $y=0$ is then given by:
\begin{eqnarray}
\mu(b) \simeq 2 \frac{U_0(1-b)f(b)}{12\pi a_0},
\label{muhighfield}
\end{eqnarray}
where we used $e^{-\pi \sqrt{3}}\simeq 1/24 \pi^2$ and
\begin{eqnarray}
f(b)=1-24\pi^3
\frac{e^{-\sqrt{\frac{\pi\sqrt{3}(2-2b)}{C^2b}+3\pi^2}}}
{\sqrt{\frac{\pi(2-2b)}{C^2\sqrt{3}b}+\pi^2}}.
\label{f(b)expression}
\end{eqnarray}
It can be checked that the associated shear modulus $c_{66}=\pi
\sqrt{3}\mu(b)/2a_0$ is very similar to the interpolation formula
of Brandt Eq.(\ref{c66}). Additionally, the edge potential
Eq.(\ref{ordVrfinal}) is harmonic for all fields. Hence the ideal
flow stress of a commensurate, ordered channel, is characterized
by Frenkels value $A^0=1/\pi\sqrt{3}$ for all fields
\cite{fn_nofluctuations}. In the low field limit $b\lesssim 0.2$
(and $\lambda/a_0 \gtrsim 1/\pi$) the above expressions reduce to
Eq.(\ref{OrdVrlowfield}), (\ref{mudef}) in Sec. \ref{subord1Dana}.

Using the field-dependent vortex interaction
Eq.(\ref{FTfullVBrandt}), one can derive the parameters in the
sine-Gordon description of Sec. \ref{subord1Dana} generalized for
higher field. The equation for the chain stiffness becomes:
\begin{eqnarray}
\kappa_q=U_0\pi(1-b)\left[
\frac{\lambda'/a_0}{\sqrt{1+\lambda'^2q^2}}-\frac{\xi'/a_0}{\sqrt{1+\xi'^2q^2}}\right].
\label{allfieldstiffness}
\end{eqnarray}
The reduced stiffness is obtained from $g(b)=\kappa_0(b)/(k_0
\mu(b)a_0^2)$:
\begin{eqnarray}
g(b)=\frac{3\pi}{f(b)} \sqrt{\frac{\sqrt{3}b}{4\pi
(1-b)}}\left(\frac{\lambda}{\xi}-\frac{C}{\sqrt{2}}\right)
\label{allfieldg}
\end{eqnarray}
Taking into account these refinements in
Eq.(\ref{semidiscretefel}) and Eq.(\ref{nonlocalordconteqm}), the
defect width in the nonlocal regime becomes:
\begin{eqnarray}
l_d^{nl}(b)=6 \pi^2 a_0/f(b).\label{ldnonlocb}
\end{eqnarray}
One can obtain the typical crossover field $b_{nl}$ (or typical
$\lambda/a_{0,nl}$) at which nonlocal behavior sets in for a chain
in an ordered channel by equating Eq.(\ref{ldnonlocb}) to the s-G
value for the kink width $2\pi a_0\sqrt{g(b)}$. Approximating
$f(b)\simeq (1-b)$, one finds $1-b=3\pi/ \sqrt{g(b)}$, which has
the approximate solution
\begin{eqnarray}
b_{nl}\simeq
\frac{1}{2}\left[1-\sqrt{1-(48\sqrt{3}\pi^3\xi^2/\lambda^2})\right]\label{bnonloc}
\end{eqnarray}
Hence, the nonlocal regime is absent for a channel in a material
with $\lambda/\xi \lesssim 50$ (and thickness $d\gtrsim \lambda$).
For $\lambda/\xi \gtrsim 60$, nonlocal behavior occurs for
$b>b_{nl}$ with $b_{nl}\lesssim 0.2$. This is to be compared with
the estimate $a_{0,nl} < \lambda/3\pi$ resulting from a simple
London interaction, see Sec. \ref{subord1Dana}.

\section{Solution to the dynamic sine-Gordon equation}
\label{appB}

For a displacement field of the form Eq.(\ref{persolansatz}),
expressed in modes with wave vector $mq=2\pi mc_d$ and amplitude
$h_m$, Eq.(\ref{fvdispl}) for the {\it v-f} curve attains the
form:
\begin{eqnarray}
{f}={\gamma} v\left[1 + 2k_0^2 \sum^M_{m=1} m^2|h_m|^2\right].
\label{coeffric}
\end{eqnarray}
The amplitudes $h_m$ are obtained by inserting
Eq.(\ref{persolansatz}) into the equation of motion
(\ref{ordconteqm}) with the wavelength dependent elasticity
$\kappa(q)$ from Eq.(\ref{kappaexpression}). Since $h$ can become
of the order of a lattice spacing $a_0$, one expands the 'sin'
term in (\ref{ordconteqm}) up to second order in $h$. Furthermore,
we assume that $h_m$ decays rapidly upon increasing $m$ and we
keep only the three lowest order contributions $h_m$ with $m \leq
3$. Collecting terms of equal wave number, one obtains the
following set of approximate equations:
\begin{eqnarray}
\frac{2}{\mu}(i\gamma vk_0+K_{1,q})h_1=&& -i +
ik_0^2(|h_1|^2+|h_2|^2)\nonumber\\
&&+k_0h_2-i(k_0^2/2)h_1^2 \label{hullset1}
\end{eqnarray}
\begin{eqnarray}
\frac{2}{\mu}(2i\gamma vk_0+K_{2,q})h_2=
k_0h_1+ik_0^2(h_1^*h_2-h_1h_2) \label{hullset2}
\end{eqnarray}
\begin{eqnarray}
\frac{2}{\mu}(3i\gamma vk_0+K_{3,q})h_3= k_0h_2+i(k_0^2/2)h_1^2,
\label{hullset3}
\end{eqnarray}
where $K_{m,q}=m^2 q^2\kappa(mq)$ and $h^*$ denotes the complex
conjugate of $h$. At small $v$, the real components of $h$ vanish
and the amplitudes describing the shape of the quasi-static
deformations are given by:
\begin{eqnarray}
|h_1|\approx
\frac{(2K_{1,q}/\mu)-\sqrt{(2K_{1,q}/\mu)^2+6k_0^2}}{3k_0^2}
\label{quasistat1}
\end{eqnarray}
\begin{eqnarray}
|h_2|\approx \frac{-k_0|h_1|}{2k_0^2|h_1|-(2K_{2,q}/\mu)}
\label{quasistat2}
\end{eqnarray}
\begin{eqnarray}
|h_3|\approx \frac{\mu(k_0|h_2|-k_0^2|h_1|^2/2)}{2K_{3,q}}.
\label{quasistat3}
\end{eqnarray}
For arbitrary $v$, the $h_m$'s in
Eq.(\ref{hullset1},\ref{hullset2},\ref{hullset3}) may be
determined by a mathematical program. However, the most important
effect of the coupling is that above a characteristic velocity
$v^*$ (see below), $|h_2|$ and $|h_3|$ decrease rapidly as
$|h_2|\sim|h_1|/v$ and $|h_3|\sim|h_1|^2/3v$, so that only the
first Fourier mode survives. This mode is described by ($v \gg
v^*$):
\begin{eqnarray}
h_1(v)&&= \frac{(2 \gamma v/\mu)-\sqrt{(2\gamma
v/\mu)^2+2}}{k_0}\nonumber\\
&&\approx -\frac{\mu}{2\gamma v k_0}, \label{largevelrel}
\end{eqnarray}
Since Eq.(\ref{largevelrel}) can be obtained by neglecting the
elastic force terms $\kappa_{n,q}$, it describes the deviation
from the average velocity of a single particle in a periodic
potential. Further analysis of
Eq.(\ref{hullset1},\ref{hullset2},\ref{hullset3}) shows that the
crossover velocity $v^*$ is determined by the amplitude of the
quasi-static result in
Eq.(\ref{quasistat1},\ref{quasistat2},\ref{quasistat3}):
\begin{eqnarray}
(2 \gamma v^* k_0/ \mu)^2 \sum_{m=1}^{3} (m|h_m|)^2  \approx 1.
\label{velcriterion}
\end{eqnarray}
Using the definition
\begin{eqnarray}
K_{eff}^2(c_d=\frac{q}{2\pi})\equiv\frac{\mu^2}{4}\left(\sum_{m=1}^{3}
(m|h_m|)^2\right)^{-1},\label{kappacdef}
\end{eqnarray}
we rewrite Eq.(\ref{velcriterion}):
\begin{eqnarray}
k_0 v^*=K_{eff}(c_d)/\gamma. \label{crossovervel}
\end{eqnarray}
Here $\gamma/K_{eff}(c_d)$ can be interpreted as the effective
relaxation time for the non linear deformations in the chain,
which is expressed through the relaxation times $\gamma/K_{m,q}$
of linear modes (phonons) by Eq.(\ref{kappacdef}). The velocity
dependence of $\sum (m|h_m|)^2$ may then be written as:
\begin{eqnarray}
\sum_{m=1}^{3} m^2|h_m(v)|^2= \frac{\mu^2}{4[K_{eff}^2+(\gamma
vk_0)^2]}. \label{endresult}
\end{eqnarray}
This has the correct small $v$ behavior (where higher modes play a
role) and large $v$ behavior (where $h_2\approx h_3\approx 0$).
First order perturbation using only $h_1$ to first order in
Eq.(\ref{hullset1}) yields the same functional form but with
$K_{eff}^2$ replaced by $K_{1,q}^2$, supporting the analytical
interpolation made in obtaining Eq.(\ref{endresult}). Finally,
using Eq.(\ref{coeffric}) one arrives at Eq.(\ref{fvresultana}) in
section \ref{subord1Dfv}.

\section{Elastic shear waves in commensurate, ordered channels}
\label{appC}

Starting from Eq.(\ref{elasheareqm}) we write $u$ as the sum of a
spatially uniform dc-component and a non-uniform, time-dependent
component $h(y,t)$. The result for $h(y,t)=u(y,t)-vt$ is the
following modified diffusion equation:
\begin{eqnarray}
\gamma \partial_t h(y,t)=(f-\gamma v)+c\partial^2_y h(y,t),
\label{comhytequation}
\end{eqnarray}
where $c=b_0^2k_0 \mu/2=c_{66}a_0b_0$. The boundary condition is
set by:
\begin{eqnarray}
\gamma\partial_t h(-w/2,t)=&&-(\mu/2)\sin[\omega_0 t+h(-w/2,t)] \nonumber\\
&&+(c/b_0^2)\Delta h(t), \label{boundarycondition}
\end{eqnarray}
and similarly for $y=+w/2$. Here the discrete character of the
array near the CE's is retained in $\Delta
h=h(-w/2+b_0,t)-h(-w/2,t)$. For the pinning term (last term on the
r.h.s.) we now use the large velocity expansion for the restoring
force from one CE: $(\mu/2)\sin[k_0 (v t+h)]\simeq
(\mu/2)\sin[\omega_0 t]$. Then
Eqs.(\ref{comhytequation}),(\ref{boundarycondition}) become
similar to those for heat diffusion in a rod with at both ends
heat sources that vary sinusoidally in time. In our case the
frequency is the washboard frequency $\omega_0=k_0 v$. By
separation of variables one finds:
\begin{eqnarray}
h(y,t)=&& h_e A_{e,v}(w)[f_v(y)\cos(\omega_0 t) +g_v(y)\sin(\omega_0t)]\nonumber \\
&&-\frac{(f-\gamma v)}{2c}y^2, \label{comuytsolution}
\end{eqnarray}
where $f_v(y)=-\cos(y/l_{\perp,v})\cosh(y/l_{\perp,v})$ and
$g_v(y)=\sin(y/l_{\perp,v})\sinh(y/l_{\perp,v})$ with a velocity
(frequency) dependent 'healing' length
$l_{\perp,v}=\sqrt{2c/\gamma \omega_0}=\sqrt{(\mu/\gamma v)}b_0$.
The factor $A_{e,v}(w)=1/\sqrt{f_v^2(w/2)+g_v^2(w/2)}$ normalizes
the displacement at the CE to $h_e$. The latter is obtained from
the boundary condition Eq.(\ref{boundarycondition}). In the limit
$w/b_0=n \geq 3$ and $\gamma v/\mu \gtrsim 0.25$, $h_e$ can be
approximated by:
\begin{eqnarray}
h_e \simeq \frac{\mu}{2k_0\gamma v}\sqrt{\frac{2\gamma v}{2\gamma
v+\mu}} \label{heexpression}
\end{eqnarray}

\begin{figure}
\scalebox{0.70}{\includegraphics{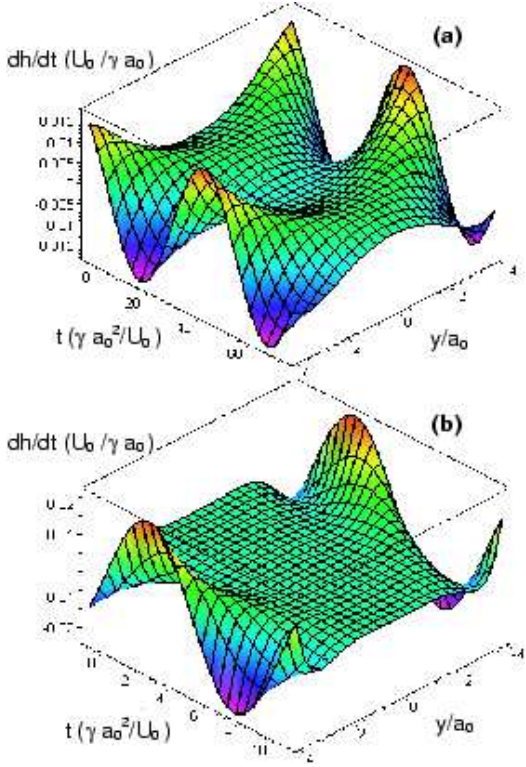}}

\vspace{0cm} \caption{Velocity profile $\partial_t h(y,t)$ versus
time from Eq.(\ref{comuytsolution}) in a channel with $w/b_0\simeq
9$. (a) for $\gamma v/\mu \simeq 0.4$ and (b) for $\gamma v/\mu
\simeq 2$.} \label{anawa0_8}
\end{figure}

\begin{figure}
\scalebox{0.30}{\includegraphics{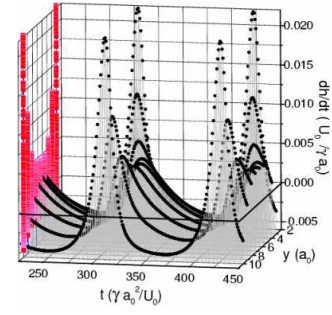}}

\vspace{0cm} \caption{Simulation result of the profile of the
fluctuating velocity component $\partial_t h$ versus time in a
channel with $w/b_0=9$ at $f=0.011$. The projection in the
$y-dh/dt$ plane further clarifies the $y$-dependence of $|h'|$.}
\label{vytcompurew9f0110}
\end{figure}

The solution (\ref{comuytsolution}) describes a periodic velocity
modulation $\partial_t h$ of each chain, with a $y$-dependent
amplitude $|h'|$ and phase shift. The latter reflects periodic
lagging and advancing of the inner rows with respect to the outer
ones. Both the decay of $|h'|$ away from the CE and the phase
shift strongly increase with velocity through $l_{\perp,v}$. In
Fig. \ref{anawa0_8} we have illustrated this behavior for small
velocity $\gamma v/\mu=0.4$ (a) and larger velocity $\gamma
v/\mu=2$ (b). Although the result in (a) is actually out of the
regime of validity of the high velocity expansion, the 'in-phase'
behavior at small $v$ is a feature that qualitatively agrees with
the simulations. For comparison we show in Fig.
\ref{vytcompurew9f0110} numerical results for the oscillating
velocity component for $w/b_0=9$ and small drive, $f\simeq
2f_s\simeq \mu/5$ ($\gamma v/\mu \approx 0.16$). In addition to
the small phase shift one observes that the modulation is highly
non-sinusoidal, which is not captured by our approximate solution.

Based on the above, one can also evaluate the dynamic friction
force $f- \gamma v$. Using Eq.(\ref{dissiprate}),(\ref{fvdispl}),
the expression for the elastic-continuum $u(y,t)$ would be
$f-\gamma v=(\gamma/vw)\int dy dt [\partial_t h(y,t)]^2$. However,
in the velocity regime where our solution applies, the length
$l_{\perp,v} \leq 2b_0$. Since we have a discrete number of vortex
rows, it is the two outer rows at $y= \pm w/2$ which give the
dominant contribution. Evaluating $f-\gamma v$ using
Eq.(\ref{heexpression}) then leads to
Eq.(\ref{widecomfvexpression}) in Sec. \ref{secord2D}.

\section{Disordered channel potential and pinning of defects}
\label{appD}

In this appendix we derive the disorder corrections to the channel
potential and the effect on pinning of the chain. For the vortex
interaction we use the London potential of Eq.(\ref{VLondon}).
Starting from Eq.(\ref{Vedgeformal}) we first split the potential
in contributions from row $m$ with transverse coordinate $y$:
\begin{eqnarray}
V_m(x)=(2\pi)^{-1}\int dk V(k,y) \rho_m(k) e^{ikx},
\label{disordVm}
\end{eqnarray}
where $k=k_x$ and
\begin{eqnarray}
V(k,y)=U_0\pi
\frac{e^{-y\sqrt{k^2+\lambda^{-2}}}}{\sqrt{k^2+\lambda^{-2}}}.
\label{Vmsk}
\end{eqnarray}
In  addition, $\rho_m(k)$ in Eq.(\ref{disordVm}) is related to the
longitudinal displacement field $d^x_m(x)$ via
\cite{GiamarchiPRB95}
\begin{eqnarray}
\rho_m(k)=a_0^{-1}\int dx \left[1-\partial_x
d_{m}+\delta\rho_m(x-d_{m})\right]e^{-ikx}, \label{rhomk}
\end{eqnarray}
which contains the microscopic density modulation
$\delta\rho_m(x-d_{m}(x))=\sum_{l\neq 0} \cos(lk_0[x-d_{m}(x)])$
as well as the density variations $\propto \partial_x d_{m}$ due
to long wavelength deformations. Inserting Eq.(\ref{rhomk}) in
Eq.(\ref{disordVm}) yields a total potential of the form
$V_m(x)=V_{l,m}(x)+ V_{p,m}(x)$ where
\begin{eqnarray}
V_{l,m}(x)=-a_0^{-1}\int dx' V(x-x',y)\partial_{x'} d_{m}(x'),
\label{Vml}
\end{eqnarray}
and
\begin{eqnarray}
V_{p,m}(x)=\frac{2}{a_0}\int dx' V(x-x',y) \cos [ k_0(x'-d_{m})].
\label{Vmpstart}
\end{eqnarray}
A constant offset arising from the 'DC'part of the density has
been omitted in Eq.(\ref{Vml}) and in Eq.(\ref{Vmpstart}) only the
$l= \pm 1$ components of the density are taken into account. The
latter represents a quasi-periodic potential with wavelength $\sim
a_0$. It is only significant for $y\lesssim 1.5b_0$, i.e. for
$m=\pm 1$, as in the pure case. Therefore $V_p$ can be written as
\begin{eqnarray}
V_p=-[\mu+\delta\mu(x)]\cos[k_0(x-d)]/k_0, \label{Vpformula}
\end{eqnarray}
where $d=(d_1+d_{-1})/2$ and
\begin{eqnarray}
\delta \mu(x)&&=\frac{k_0}{a_0^2}\int dk
[V(k_+,b_0)-V(k_-,b_0)]i k d(k)e^{ikx}\nonumber\\
&&\simeq \pi \sqrt{3} \mu \partial_x d, \label{deltamurx}
\end{eqnarray}
where $k_{\pm}=k_0\pm k$ and the second line is valid for
$k\lesssim 0.4 k_0$. The local strains thus also provide
variations in the potential height. A similar conclusion holds
when transverse shifts in the CE are included. In the simulations
the mean square strain is $\langle (\nabla \cdot {\bf d})^2
\rangle=\Delta^2/3$ and short range ($\sim a_0/2$) correlated
along $x$ with correlator $S(s,0)\equiv \langle
\partial_x d(x,y)
\partial_x d(x+s,y'=y) \rangle_x \simeq (\Delta^2/3)
e^{-(2s/a_0)^2}$. Accordingly, the amplitude fluctuations of the
periodic potential are characterized by:
\begin{eqnarray}
\Gamma_a(s)=\frac{\langle \delta \mu (x) \delta \mu
(x+s)\rangle_x}{k_0^{2}} \simeq (\mu \Delta a_0/2)^2
e^{-(\frac{2s}{a_0})^2} \label{Gamma_arepeat}
\end{eqnarray}

The non-local contributions Eq.(\ref{Vml}) add up to a total
contribution $V_l=\sum_{m\neq 0} V_{l,m}$ in $V_{ce}$. $V_l$ thus
originates from strains within a region $\sim \lambda$ around the
channel and will be smooth on the scale $a_0$ (we assume $\lambda
\gg a_0$). Transforming the sum into an integral and using
Eq.(\ref{Vmsk}), the correlator $\Gamma_l=\langle V_l(x)V_l (x+s)
\rangle$ can be written as \cite{fn_correlator}:
\begin{eqnarray}
\Gamma_l(s)=\frac{1}{a_0^4}\int_{CE}dy dy'\int dx A_{y,y'}(x)
S(s-x,y'-y), \label{Vrcorrelatorbasic}
\end{eqnarray}
The term $A_{y,y'}(x)=(2\pi)^{-1}\int dk V(k,y)V(k,y')\cos(kx)$
can be approximated by
\begin{eqnarray}
A_{y,y'}(x)\simeq U_0^2 \pi \lambda e^{-\frac{|y|+|y'|}{\lambda}}
e^{-(\frac{x}{\lambda})^2}.
\end{eqnarray}
In case the strains are uncorrelated between the rows $S(x,y'-y)
\sim \exp(-(2(x^2+(y'-y)^2)/a_0^2))$ and
Eq.(\ref{Vrcorrelatorbasic}) can be approximated by
\begin{eqnarray}
\Gamma_l(s) \simeq C_{\alpha}\Delta^2U_0^2(\lambda/a_0)^{1+\alpha}
e^{-(\frac{s}{\lambda})^2}, \label{repeatcorrelator}
\end{eqnarray}
with $\alpha=1$ and $C_1\simeq 1$. In case of uniform strains,
$S(x)$ is independent of $y'-y$. Then again the correlator is
given by the above formula, but with $\alpha=2$ and $C_2=
(4/3)\pi^{3/2}$.

To study the effect of the disorder on the pinning of vortices
inside the channel, we write the total energy as a sum of an
elastic and a pinning term:
\begin{eqnarray}
H(x,u)= H_{el}+H_p=a_0^{-1}\int dx \frac{\kappa_0}{2}(\partial_x
u)^2+H_{p}. \label{Htotal}
\end{eqnarray}
The dispersion of the elastic constant has been neglected and $u$
represents the displacements of vortices inside the channel. By
writing the density of the chain as
$\rho_c(x,u)=a_0^{-1}[1-\partial_x u+\sum_{l\neq 0}
\cos(lk_0(x-u))]$, the pinning term $H_p$ in Eq.(\ref{Htotal}) can
be expressed as:
\begin{eqnarray}
H_{p}=a_0^{-1}\int dx (V_p+V_l)(\delta \rho(x,u)-\partial_x u),
\label{Hpinbasic}
\end{eqnarray}
where only the lowest Fourier components of $\rho_c$ are retained,
i.e. $\delta \rho(x,u)=2\cos k_0(x-u)$. This expression can be
simplified as follows. Since we consider the limit where
$\partial_x u$ is nearly constant within $a_0$, the cross-term of
the quasi-periodic potential $V_p$ and $\partial_x u$ can be
neglected compared to the $V_l(x)\partial_x u$ term. The product
$V_l \delta \rho(x,u)$ is also oscillatory and can be neglected as
well \cite{fn_Vrdeltarho}. The other remaining term $V_p(x) \delta
\rho(x,u)$ can be written as $(\mu(x)/k_0) \cos[k_0(u-d)]$.
Shifting the argument via $k_0(u-d)=k_0 \tilde{u}$ and writing
$\tilde{u} \rightarrow u$, an extra term $\kappa_0\partial_x
d\partial_x u$ is generated in $H_{el}$ in Eq.(\ref{Htotal}) (and
also $u$-independent terms which can be neglected). The total
coupling to the strain $\partial_x u$ then consists of $V_{s}
\equiv V_l-\kappa_0\partial_x d$. The result for the total energy
is
\begin{eqnarray}
H=H_{SG}- \int  \frac{dx}{a_0}\left[\frac{\delta
\mu(x)}{k_0}\cos(k_0 u)- V_s(x)\partial_x u \right ].
\label{approxHtotal}
\end{eqnarray}
The potential $V_s$ has a correlator $\langle V_s(x)V_s(x+s)
\rangle=\Gamma_s(s)$ which is characterized by
\begin{eqnarray}
\Gamma_s(s)=\frac{16\pi^2
g^2}{3}\Gamma_a(s)+\Gamma_l(s)+\frac{2U_0\Delta^2\lambda}{a_0}V(s,b_0)
\label{Vscorrelator}
\end{eqnarray}
in which $g=\kappa_0/(2\pi \mu a_0)$ and the last term arises from
cross-correlations. In Eq.(\ref{approxHtotal}), $H_{SG}=\int dx
[(\kappa_0/2)(\partial_x u)^2-(\mu/k_0) \cos(k_0 u)]$ is the
original sine-Gordon energy functional of the pure model and the
remaining terms reflect the corrections due to disorder. Finally,
we denote the first correction, which is due to the amplitude
fluctuations, as $H_a$ and the second, coupling to the strain, as
$H_s$.

For weak disorder, we can now calculate the effect of disorder on
a defect in the chain by assuming that the shape of a defect at
$x$, $u_d(x'-x)=2a_0\arctan[\exp(\pm 2\pi (x'-x)/l_d)]/\pi$, is
unchanged by disorder \cite{VinokurJETP90_disordJosJun}. The
pinning energy of a defect due to the term $H_a$ is:
\begin{eqnarray}
E_a(x)=(a_0k_0)^{-1}\int dx' \delta \mu(x') \cos(k_0u_d(x'-x)).
\end{eqnarray}
The correlations of $E_a$ are given by \cite{fn_correlator}:
\begin{eqnarray}
\langle E_a(x) E_a(x+s) \rangle =\frac{1}{a_0^{2}} \int dp
A_{a}(p) \Gamma_a(s-p), \label{Ebcorrelator}
\end{eqnarray}
where
\begin{eqnarray}
A_a(p)=\int \frac{4(l_d/2\pi) d\tilde{x}}{\cosh^2(\tilde{x})
\cosh^{2}(\tilde{x}+\tilde{p})}\simeq l_d
e^{-(\frac{4p}{l_d})^2},\label{AbVformula}
\end{eqnarray}
with $\tilde{x}=2\pi x/l_d$. For $l_d \gg a_0$ we can approximate
$\Gamma_a(s)$ in Eq.(\ref{Ebcorrelator}) by $\Gamma_a(s)\simeq
(\mu\Delta)^2(\sqrt{\pi}a_0^3/8)\delta(s)$ leading to:
\begin{eqnarray}
\langle E_a(x) E_a(x+s) \rangle=(\sqrt{\pi}/8)\mu^2 \Delta^2 l_d
a_0 e^{-(\frac{4s}{l_d})^2}.\label{Ebcorresult}
\end{eqnarray}

The effect of the coupling to the strain is given by
$E_s(x)=(a_0)^{-1}\int dx' V_s(x') \partial_x u_d(x'-x)$ which has
the following correlator:
\begin{eqnarray}
\langle E_s(x) E_s(x+s) \rangle =\frac{1}{a_0^{2}}\int dp A_s(p)
\Gamma_s(s-p). \label{Efcorrelator}
\end{eqnarray}
Using $\partial_x u_d=(2a_0/l_d)\cosh^{-1}(2\pi x/l_d)$, $A_s$ is
given by:
\begin{eqnarray}
A_s(p)=\int \frac{(2a_0^2/\pi l_d) d\tilde{x}} {\cosh(\tilde{x})
\cosh(\tilde{x}+\tilde{p})}\simeq \frac {4a_0^2
e^{-(\frac{2p}{l_d})^2}}{\pi l_d}.\label{AfVformula}
\end{eqnarray}
For $l_d \gtrsim \lambda$ the final result is:
\begin{eqnarray}
\langle E_s(x) E_s(x+s) \rangle \simeq \frac{(U_0 \Delta
\lambda)^2}{l_d a_0}[C_l+4] e^{-(\frac{2s}{l_d})^2},
\label{Efcorresult}
\end{eqnarray}
where the term $C_l \simeq 2C_{\alpha}(\lambda/a_0)^{\alpha}$ in
square brackets is due to the nonlocal contribution $V_l$ and the
factor $4$ arises from the term $\sim \partial_x d$ in $V_s$.
Hence, for large $\lambda/a_0$ the nonlocal term dominates in the
coupling to the strain.


\begin{references}

\bibitem{Blatterbible}G. Blatter {\it et al.}, Rev. Mod. Phys. {\bf 66}, 1125,
(1994).

\bibitem{Gruner}G. Gr\"uner, Rev. Mod Phys. {\bf 60}, 1129 (1988).

\bibitem{Perssonbook}B.N.J. Persson, {\it Sliding Friction:
Physical Principles and Applications} (Springer-Verlag, Berlin,
1998).

\bibitem{Wigner}Y.P. Li, T. Sajoto, L.W. Engel, and D.C. Tsui,
Phys. Rev. Lett. {\bf 67}, 1630 (1991); M. Cha and H.A. Fertig,
Phys. Rev. B {\bf 50}, 14368 (1994).

\bibitem{Seshadri}R. Seshadri and R.M. Westervelt, Phys.\ Rev.\ Lett.\
{\bf 70}, 234 (1993); R. Seshadri and R.M. Westervelt, Phys.\
Rev.\ B {\bf 47}, 8620 (1993).

\bibitem{BaertperpinPRL95}M. Baert, V. V. Metlushko, R. Jonckheere, V. V. Moshchalkov, and
Y. Bruynseraede, Phys. Rev. Lett {\bf 74} 3269 (1995).

\bibitem{MartinperpinPRL99}J. I. Martín, M. Vélez, A. Hoffmann, I. K. Schuller, and J. L. Vicent,
Phys. Rev. Lett. {\bf 83}, 1022 (1999).

\bibitem{twinHTC}W. K. Kwok, U. Welp, G. W. Crabtree, K. G. Vandervoort,
R. Hulscher, and J. Z. Liu, Phys.\ Rev.\ Lett.\ {\bf 64}, 966
(1990); W. K. Kwok, U. Welp, V. M. Vinokur, S. Fleshler, J.
Downey, and G. W. Crabtree, Phys.\ Rev.\ Lett.\ {\bf 67}, 390
(1991);

\bibitem{JensenPRLPRB88}H.J. Jensen, A. Brass, and A. J. Berlinsky,
Phys. Rev. Lett. \ {\bf 60}, 1676 (1988); H. J. Jensen, A. Brass,
Y. Brechet, and A. J. Berlinsky, Phys. Rev. B {\bf 38}, 9235
(1988).

\bibitem{ShiPRL91_def}A.C. Shi and A.J. Berlinsky, Phys.\ Rev.\ Lett.\ {\bf 67}, 1926
(1991).

\bibitem{MarchevskyPRB99_depinning}M. Marchevsky, J. Aarts, and P.H. Kes, Phys. Rev. B {\bf
60}, 14601 (1999).

\bibitem{PardoPRL97_topdef}F. Pardo, F. De La Cruz, P.L. Gammel, C.S. Oglesby,
E. Bucher, B. Batlogg, and D.J. Bishop,  Phys.\ Rev.\ Lett.\ {\bf
78}, 4633 (1997).

\bibitem{pardonatureBragg}F. Pardo, F. de la Cruz, P.L. Gammel, E. Bucher, and D.J.
Bishop, Nature {\bf 396}, 348 (1998).

\bibitem{Larkin}A.I. Larkin, Zh. Eksp. Teor. Fiz. {\bf 58},
1466 (1970); A.I. Larkin and Yu. Ovchinikov, J. Low Temp. Phys.
{\bf 34}, 409 (1979).

\bibitem{Kes2DCP}P.H. Kes and C.C. Tsuei, Phys. Rev. Lett. {\bf
47}, 1930 (1981); Phys. Rev. B {\bf 28}, 5126 (1983).

\bibitem{WordenweberPRB86}R. W\"ordenweber and P.H. Kes, Phys. Rev. B {\bf 34}, 494 (1986).

\bibitem{TroyanovskiPRL02_PE}A.M. Troyanovski, M. van Hecke, N. Saha, J. Aarts, and P.H.
Kes, Phys.\ Rev.\ Lett.\ {\bf 89}, 147006 (2002).

\bibitem{MehtaReichOlson}A. P. Mehta, C. Reichhardt, C. J. Olson, and F.
Nori, Phys. Rev. Lett. {\bf 82}, 3641 (1999); C.J. Olson, C.
Reichhardt, and F. Nori, Phys.\ Rev.\ Lett.\ {\bf 81}, 3757
(1998).

\bibitem{RyuPRL96_2ddynord}S. Ryu, M.C. Hellerqvist, S. Doniach, A. Kapitulnik, and D.
Stroud, Phys.\ Rev.\ Lett.\ {\bf 77}, 5114 (1996).

\bibitem{HellerDanckwerts}M.C. Hellerqvist, D.Ephron, W.R. White,
M.R. Beasley, and A. Kapitulnik, Phys.\ Rev.\ Lett.\ {\bf 76},
4022 (1996); M. Danckwerts, A. R. Goñi, and C. Thomsen, Phys. Rev.
B {\bf 59}, 6624 (1999).

\bibitem{DaldiniPRL74}O. Daldini, P. Martinoli, J. L. Olsen, and G.
Berner, Phys. Rev. Lett. {\bf 32}, 218 (1974).

\bibitem{MartinoliPRB78}P. Martinoli, Phys. Rev. B. {\bf 17}, 1175
(1978).

\bibitem{matching2dperpin_exp}K. Harada, O. Kamimura, H. Kasai, T. Matsuda, A. Tonomura, and
V. V. Moshchalkov, Science {\bf 274}, 1167 (1996). M. Velez, D.
Jaque, J. I. Martín, M. I. Montero, I. K. Schuller, and J. L.
Vicent, Phys. Rev. B {\bf 65}, 104511 (2002).

\bibitem{RosseelLook}E. Rosseel, M. Van Bael, M. Baert, R. Jonckheere, V. V. Moshchalkov, and Y. Bruynseraede,
Phys. Rev. B {\bf 53}, 2983 (1996); L. Van Look, E. Rosseel, M. J.
Van Bael, K. Temst, V. V. Moshchalkov, and Y. Bruynseraede, Phys.
Rev. B {\bf 60}, 6998 (1999).

\bibitem{Reichhardtperpin}C. Reichhardt, C.J. Olson, and F. Nori,
Phys.\ Rev.\ Lett.\ {\bf 78}, 2648 (1997); C. Reichhardt, C.J.
Olson, and F. Nori, Phys. Rev. B {\bf 58}, 6534 (1998); C.
Reichhardt, G. T. Zimányi and N. Gronbech-Jensen, Phys. Rev. B
{\bf 64}, 014501 (2001).

\bibitem{Dewhughes87_shearinGB}D. Dew-Hughes, Phil. Mag. B {\bf 4}, 459 (1987).

\bibitem{Pruymboomthesis}A. Pruymboom. Ph.D. Thesis, Leiden
University (1988).

\bibitem{Gurevich}A. Gurevich, Phys.\ Rev.\ B {\bf 46},
3187 (1992); Phys.\ Rev.\ B {\bf 65}, 214531 (2002); A. Gurevich,
M. S. Rzchowski, G. Daniels, S. Patnaik, B. M. Hinaus, F. Carillo,
F. Tafuri, and D. C. Larbalestier, Phys.\ Rev.\ Lett.\ {\bf 88},
097001 (2002).

\bibitem{Pruymboom}A. Pruymboom, P.H. Kes, E. van der Drift, and S. Radelaar,
Phys.\ Rev.\ Lett.\ {\bf 60}, 1430 (1988); M.H. Theunissen, E. van
der Drift, and P.H. Kes, Phys. Rev. Lett. {\bf 77}, 159 (1996).

\bibitem{Brandtelas}E.H. Brandt, J. Low Temp. Phys. {\bf 26}, 709
(1977); {\it ibid.} {\bf 26}, 735 (1977); J. Low Temp. Phys. {\bf
28}, 263 (1977); {\it ibid.} {\bf 28}, 291 (1977); Phys. Stat.
Sol. (b) {\bf 77}, 551 (1976).

\bibitem{KokuboPRL02} N. Kokubo, R. Besseling, V.M. Vinokur, and P.H. Kes, Phys. Rev. Lett.
{\bf 88}, 247004 (2002).

\bibitem{KokuboPRB_chanMLvcB}N. Kokubo, R. Besseling, and
P.H. Kes, Phys. Rev. B {\bf 69}, 064504 (2004).

\bibitem{BesselingEPL2003} R. Besseling, T. Dr\"ose, V. M. Vinokur and P. H. Kes, Europhys. Lett. \textbf{62},
419 (2003).

\bibitem{BaarleAPL03}G.J.C. van Baarle, A.M. Troianovski, T. Nishizaki, P.H. Kes and J. Aarts,
Appl. Phys. Lett. {\bf 82}, 1081 (2003).

\bibitem{BesselingPRL99}R. Besseling, R. Niggebrugge and P.H. Kes, Phys.\ Rev.\ Lett.\ {\bf
82}, 3144 (1999).

\bibitem{Frenkel1926}J. Frenkel, Z.\ Phys.\ {\bf 37}, 572 (1926).

\bibitem{Koshelevrecryst}A.E. Koshelev and V.M. Vinokur, Phys.\ Rev.\ Lett.\ {\bf 73}, 3580
(1994).

\bibitem{OwenScalapinoPR67_LJJ}C.S. Owen and J. Scalapino, Phys. Rev. {\bf
164}, 538 (1967).

\bibitem{ButtikLandauerPRA81}M. B\"{u}ttiker and R. Landauer, Phys.\ Rev.\ A.
{\bf 23}, 1397 (1981).

\bibitem{BraunKivsharPhysRep98}O.M. Braun and Yu.S. Kivshar, Phys. Rep. {\bf 306}, 1
(1998).

\bibitem{Tinkham}M. Tinkham, {\it Introduction to
Superconductivity} McGRAW-HILL INTERNATIONAL EDITIONS, 1996.

\bibitem{BraunBraunMingaleev}O.M. Braun, Yu.S. Kivshar and I.I. Zelenskaya, Phys. Rev. B. {\bf
41}, 7118 (1990); O.M. Braun and Y.S. Kivshar, Phys.\ Rev.\ B {\bf
50}, 13388 (1994); S.F. Mingaleev, Y.B. Gaididei, E. Majernikova,
and S. Shpyrko, Phys. Rev. E {\bf 61}, 4454 (2000).

\bibitem{fn_discreteness}Practically, discreteness effects can be neglected for
$g\gtrsim 2$. One can then also consider the regime $\lambda/a_0
\lesssim 1$. The reduced stiffness is $g=V''(x=a_0)/V''_{ce,0}$
where $V''(x=a_0)\simeq(U_0/\lambda^2)\sqrt{\pi\lambda/(2x)}e^{-x/\lambda}(1+\lambda/x)$
and $V''_{ce}$ is obtained from the full ($\lambda$ dependent)
expression for $V_{ce,0}$ in App. \ref{appA} (Eq.(\ref{ordVmb})
and (\ref{ordVmbprefac})). The result is that the continuum
approach is valid for $\lambda/a_0 \gtrsim 0.3$. For the
frequently occurring geometry of a vortex chain confined by
vortices pinned in a square array one can also estimate the
continuum regime. In this case one replaces $(w+b_0)/2$ in
Eq.(\ref{OrdVrlowfield}) by $a_0/2$, yielding a periodic pinning
force with $\mu_{\Box}\simeq 10 \mu$. Correspondingly
$g_{\Box}\simeq \lambda/a_0$ and lattice discreteness is
unimportant for $\lambda/a_0\gtrsim 2$.

\bibitem{StrunzPRE98}T. Strunz and F. Elmer, Phys.\ Rev.\ E {\bf 58}, 1601 (1998).

\bibitem{Ziswiler}P. Ziswiler, V. Geshkenbein, and G. Blatter,
Phys.\ Rev.\ B {\bf 56}, 416 (1997).

\bibitem{CuleHwaPRB98}D. Cule and T. Hwa, Phys.\ Rev.\ B {\bf 57}, 8235 (1998).

\bibitem{GiamarchiPRB95}T. Giamarchi and P. Le Doussal, Phys. Rev. B {\bf 52}, 1242 (1995).

\bibitem{MalomedPRB89_JosCDWrapid}B.A. Malomed, Phys. Rev. B {\bf 39}, 8018
(1989).

\bibitem{VinokurJETP90_disordJosJun}V.M. Vinokur and A.E.
Koshelev, Sov. Phys. JETP {\bf 70}, 547 (1990).

\bibitem{FeigelVinCDW}M.V. Feigelman and V.M. Vinokur, Solid State
Communications, {\bf 45}, 595 (1983); {\it ibid.}, {\bf 45}, 599
(1983); V.M. Vinokur and M.B. Mineev, Sov. Phys. JETP {\bf 61}
1073 (1985).

\bibitem{FukuLeeRice}H. Fukuyama and P.A. Lee, Phys.\ Rev.\ B {\bf 17}, 535
(1978); P.A. Lee and T.M. Rice, Phys. Rev. B {\bf 19}, 3970
(1979); an early study of the competition between impurity and
commensurability pinning was reported in H. Fukuyama, J. Phys.
Soc. Japan {\bf 45}, 1474 (1978).

\bibitem{ChenPRB96}L. Chen {\it et al.}, Phys. Rev. B {\bf 54}, 12798 (1996).

\bibitem{KrugPRL95}J. Krug, Phys.\ Rev.\ Lett.\ {\bf 75}, 1795 (1995).

\bibitem{BalentstempPRL95}L. Balents and M.P.A Fisher, Phys. Rev. Lett. {\bf 75},
4270 (1995).

\bibitem{fn_convective term}The convective term $-\gamma v \partial_x u \simeq -f \partial_x u$
arises naturally when realizing that the driving force density on
a 'string' segment is the product of $f$ and the average vortex
density $\sim (1-\partial_x u$), see also \cite{CuleHwaPRB98}.

\bibitem{MyersPRB93}C.R. Myers and J.P. Sethna, Phys.\ Rev.\ B {\bf 47},
11171 (1993); D. Cule and T. Hwa, Phys.\ Rev. Lett.\ {\bf 77}, 278
(1996).

\bibitem{fn_fvchainstrongdisorder} A detailed study
of the {\it v-f} curves at strong disorder showed that the
friction force $f-\gamma v$ exhibited, besides the usual decrease
on increasing $v$ above threshold, a shallow minimum and then a
weak {\it increase} when further increasing $v$. These features
did not depend on system size or simulation time step and varied
little with commensurability. The observed behavior is in marked
contrast with analytical and numerical studies in standard CDW
models, which find a correction $f-\gamma v \propto v^{-1/2}$ at
large velocity for a $1$D system, see L. Sneddon, M.C. Cross, and
D.S. Fisher, Phys. Rev. Lett. {\bf 49}, 292 (1982) and also H.
Matsukawa, J. Ph. Soc. Jap. {bf 56}, 1522 (1987); {\bf 57} 3463
(1988). The difference could arise from the additional $u$
independent random force or the remaining (relatively small)
commensurability pinning which are present in our system, but
absent in the previous studies. This issue deserves further
investigation.


\bibitem{FangohrPRB2001_nucleation}H. Fangohr, S. J. Cox, and P. A. J. de Groot,
Phys. Rev. B {\bf 64}, 064505 (2001).

\bibitem{fn_fieldhistoryotheredge}The data in Fig. \ref{plot_introFs}
have been taken in a 'field down' experiment. Qualitatively
similar data were obtained in 'field cooled' experiments and both
data can be interpreted in terms of strong edge disorder. The
results of 'field up' measurements however are qualitatively
different and indicate much weaker CE disorder in this case, see
N. Kokubo {\it et al.}, in preparation.

\bibitem{KoltonPRL99_transfreeze}A.B. Kolton, D. Dominguez, and N.
Gronbech-Jensen, Phys.\ Rev.\ Lett.\ {\bf 83}, 3061 (1999).




\bibitem{fn_expvcriterion}For the numerical data, $f_s$ was obtained
from a criterion $\gamma v/\mu \approx 0.025$, i.e. $\gamma v a_0
/U_0 \approx 0.001$, while in the experiments a criterion
$v/a_0=1$ MHz was used for the critical current. The latter
corresponds to a dimensionless criterion $(v/a_0)(\gamma
a_0^2/U_0)\simeq (v/a_0)(20\pi \mu_0
\lambda^2B_{c2}/(\sqrt{3}B\rho_0))$ where $\rho_f \simeq 0.2 B
\rho_0/B_{c2}$ was used, with $\rho_0$ the normal resistivity and
$B_{c2}$ the upper critical field. Taking $\lambda=1.1$ $\mu$m,
$B\simeq 0.012 (w/b_0)^2$ T for the channel system under
investigation, $B_{c2}=1.55$ T and $\rho_0=2$ $\mu \Omega$m,
yields $(v/a_0)(\gamma a_0^2/U_0) \simeq 3.4 \cdot 10^{-9}
(v/a_0)/(w/b_0)^{2}=3.4 \cdot 10^{-3}/(w/b_0)^{2}$. Occasional
checks using the latter criterion however yielded results for
$f_s$ and $A/A^0$ essentially the same as in Figs.
\ref{plot_fsversuswranampo40} and \ref{plot_sim_exp_compare}.

\bibitem{fn_AexpAnumcomp} There are several possible explanations
for the discrepancy between $A_{exp}$ and the numerical result for
$A$. The experimental system consists of channels with walls of
finite hight at the CE's. Screening currents along these walls may
cause the average distance between the first mobile row and the
first pinned rows to be larger than that in the simulation (where
it is $b_0$ at matching), leading to an overall reduced value of
the shear interaction. In addition, the precise amount of disorder
in the experimental system is unknown (imaging studies only exist
on related geometries and outside the relevant field regime, see
B.L.T. Plourde, D.J. Van Harlingen, N. Saha, R. Besseling, M.B.S.
Hesselberth, and P.H. Kes, Phys. Rev. B {\bf 66}, 054529 (2002)
and M.V. Marchevsky, Ph.D. Thesis, Leiden University (1997)).
Further, differences in the amount of longitudinal and transverse
positional disorder may exist. Finally, compared to the
experiments, the simulations were performed for relatively small
ratio $\lambda/a_0$.

\bibitem{fn_weffwetched} In the experiments we found that the switching point $n \rightarrow n\pm 1$
together with the maxima and minima in $J_s$ can be shifted to
different magnetic field by changing the field history. This
implies a non-trivial relation between the effective width $w$ and
the physical channel width $w_{etched}$, depending on the amount
of screening currents along the channel walls, see N. Kokubo, R.
Besseling and P.H. Kes, Physica C {\bf 412-414}, 362 (2004).

\bibitem{Faleski}M.C. Faleski, M.C. Marchetti, and A.A. Middleton, Phys.\ Rev.\ B {\bf 54},
12427 (1996).

\bibitem{BesselingPRLDynMelt}R. Besseling, N. Kokubo and P.H. Kes,
Phys.\ Rev.\ Lett.\ {\bf 91}, 17702 (2003).




\bibitem{fn_nofluctuations}Note that this calculation neglects
fluctuation effects at high field.

\bibitem{fn_correlator}We employed the fact that for a function of the form
$R(x)=\int dx'f(\pm(x'-x))g(x')$ the correlator is given by
$\langle R_xR_{x+s}\rangle=\int dp A_f(p) \langle g(x)g(x+s \pm
p)\rangle$ with $A_f(p)=\int dx'f(x')f(x'+p)$.

\bibitem{fn_Vrdeltarho}One can calculate that the typical correction
$E_{b,l}$ to the pin energy of a defect due to a 'backward'
scattering term $\int V_l(x)\cos[k_0(x-u_d)]$ vanishes rapidly
with increasing $\lambda$, i.e. $\langle E^2_{b,l}\rangle \simeq
U_0^2 \Delta^2 g^{1/2} (\lambda/a_0)^{2+\alpha} e^{-(\pi
\lambda/a_0)^2}$.

\end{references}
\end{document}